%% file: thesis.tex
\documentclass[a4paper,12pt,twoside,openright]{report}
\usepackage[utf8]{inputenc}
\usepackage[T1]{fontenc}
\usepackage{fancyhdr}
\usepackage{enumerate}
\usepackage{color}
\usepackage{upgreek}
\usepackage{pdfpages}
\usepackage{mathtools}
\usepackage{txfontsb}
\usepackage{booktabs}
\usepackage{multirow}
\usepackage[english,brazil]{babel}
\usepackage{graphicx}
\usepackage[small]{caption}
\usepackage{bbm} %"black­board-style" cm fonts
\usepackage{siunitx}
\usepackage{mathrsfs}
\usepackage[all]{nowidow}
\usepackage[backend=biber,%
                natbib,%
                style=numeric-comp,%
                sorting=none,%
                firstinits,%
                sortlocale=en_US,%
                eprint,%
                isbn=false,%  if true, clear the issn for all entries
                maxnames=4,
                url=true]{biblatex}
\usepackage{tocloft}
\usepackage{subfig} %this must come after tocloft
\usepackage[%
    plainpages=false,
    pdfpagelabels,
    breaklinks,
    colorlinks,
    linktocpage]{hyperref}
\usepackage[toc,acronym,nomain,nogroupskip,nonumberlist]{glossaries}%nonumberlist
\makenoidxglossaries
\renewcommand{\finalnamedelim}{\ \&\ }
\newcommand{\eq}[1]{eq.~\eqref{#1}}
\newcommand{\Eq}[1]{Eq.~\eqref{#1}}
\input{glossary}
\input{symbols}
\DeclareMathOperator{\erfc}{erfc}

\hypersetup{colorlinks,
    citecolor = magenta,
    linkcolor = blue,
    urlcolor = blue, %commented
    hidelinks = true,
    pdfauthor = {Rafael Marcondes},
    pdftitle = {Interacting dark sector models in Cosmology and large-scale
        structure observational tests}
}
\allowdisplaybreaks[1]

\makeatletter
\def\cleardoublepage{\clearpage\if@twoside \ifodd\c@page\else
\hbox{}
\thispagestyle{empty}
\newpage
\if@twocolumn\hbox{}\newpage\fi\fi\fi}
\makeatother
\DeclareFieldFormat{doilink}{\href{http://dx.doi.org/\thefield{doi}}{#1}}
\DeclareBibliographyDriver{article}{%
  \usebibmacro{bibindex}%
  \usebibmacro{begentry}%
  \usebibmacro{author/translator+others}%
  \setunit{\labelnamepunct}\newblock
  \usebibmacro{title}%
  \newunit
  \printlist{language}%
  \newunit\newblock
  \usebibmacro{byauthor}%
  \newunit\newblock
  \usebibmacro{bytranslator+others}%
  \newunit\newblock
  \printfield{version}%
  \iffieldundef{doi}
    {}
    {%
      \newunit\newblock
      \printtext[doilink]{\usebibmacro{journal+issuetitle}%
      \newunit\newblock
      \usebibmacro{note+pages}}%
      }
  \newunit\newblock
  \iftoggle{bbx:isbn}
    {\printfield{issn}}
    {}%
  \newunit\newblock
  \usebibmacro{eprint}%
  \newunit\newblock
  \usebibmacro{addendum+pubstate}%
  \newunit\newblock
  \usebibmacro{pageref}%
  \usebibmacro{finentry}}

\title{Interacting dark energy models in Cosmology and large-scale structure
    observational tests}
\author{Rafael Marcondes}

\begin{document}
    \selectlanguage{brazil}
    \pagenumbering{roman}
    %\maketitle
    \begin{titlepage}
    %\end{titlepage}
        \thispagestyle{empty}
        \begin{center}
            {\large Universidade de São Paulo \\
                Instituto de Física} \\ \vskip 1.5 in
            {\Large Modelos de energia escura com interação em Cosmologia e
                testes observacionais com estruturas em grande escala } \\ \vskip 0.5 in

            {\large Rafael José França Marcondes} \\ \vskip 1 in
        \end{center}
        \begin{flushright}
            %%%% TESE
            \begin{minipage}{0.6\textwidth}
                \hfill Orientador: Prof.~Dr.~Elcio Abdalla \\ \vskip 0.15 in
                Tese de doutorado apresentada ao Instituto de Física para a obtenção do título de Doutor em Ciências \\ 
            \end{minipage} 
        \end{flushright}
        \vfill
        Banca examinadora: \\
        \\
        Prof.~Dr.~Elcio Abdalla (IF-USP) \\
        Profa.~Dra.~Ivone Freita Mota de Albuquerque (IF-USP)\\
        Prof.~Dr.~Laerte Sodré Júnior (IAG-USP)  \\
        Prof.~Dr.~Alberto Vazquez Saa (IMECC-Unicamp)\\
        Prof.~Dr.~Jailson Souza de Alcaniz (ON) \\
        \begin{center}
            São Paulo \\ 
            \the\year
        \end{center}

        %\newpage
        %\thispagestyle{empty}
        %\mbox{}
        \includepdf{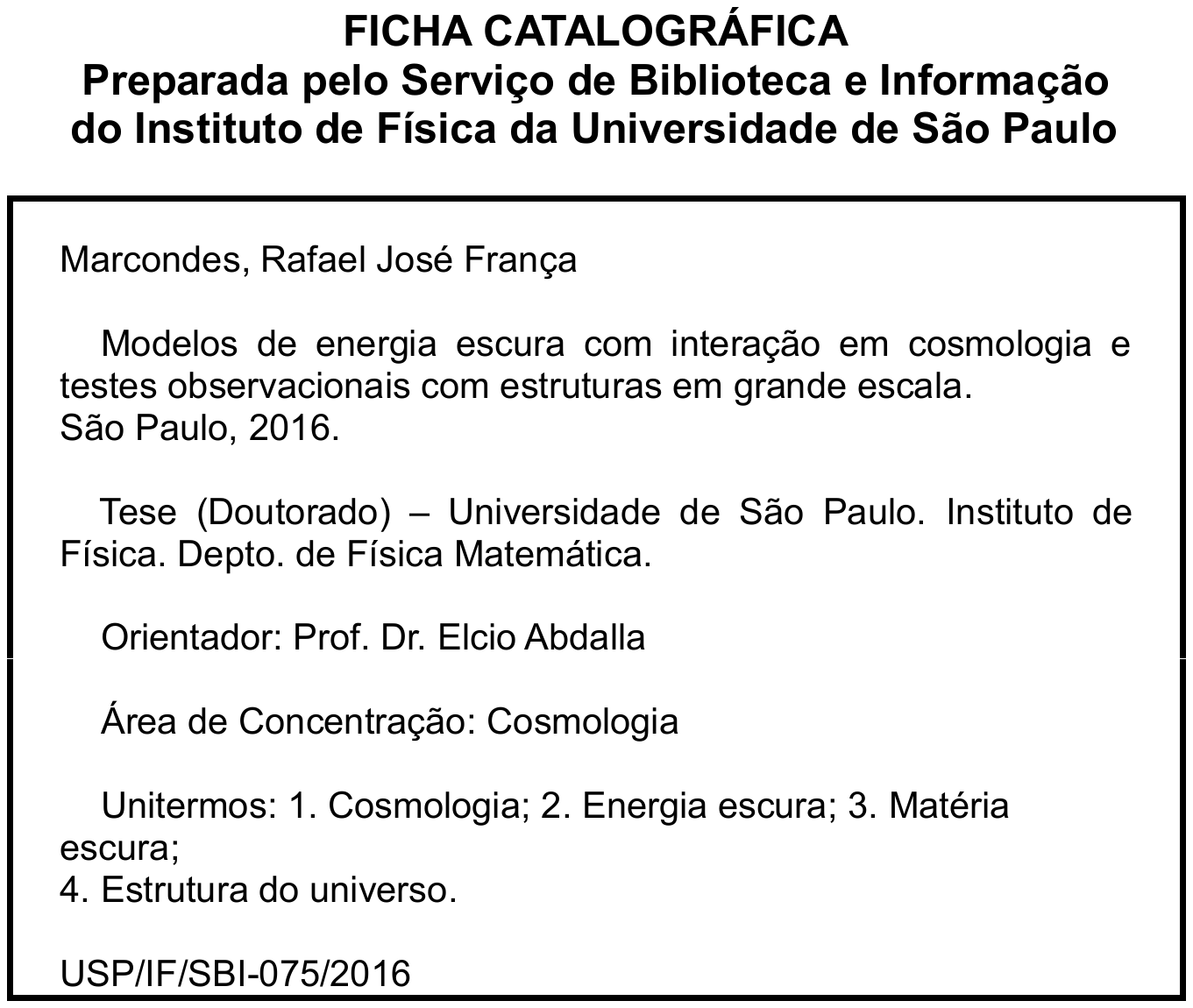}

        \newpage
        \thispagestyle{empty}
        \mbox{}
    %\maketitle
    %\begin{titlepage}

    \selectlanguage{english}
        \thispagestyle{empty}
        \begin{center}
            {\large University of São Paulo \\
                Institute of Physics} \\ \vskip 1.5 in
            {\Large Interacting dark energy models in Cosmology and
                large-scale structure observational tests} \\ \vskip 0.7 in

            {\large Rafael José França Marcondes} \\ \vskip 1 in
        \end{center}
        \begin{flushright}
            %%%% TESE
            \begin{minipage}{0.6\textwidth}
                \hfill Supervisor: Prof.~Dr.~Elcio Abdalla \\ \vskip 0.15 in
                Thesis presented to the Institute of Physics of the University
                of São Paulo in partial fulfillment of the requirements for the
                degree of Doctor of Science \\
            \end{minipage} 
        \end{flushright}
        \vfill
        Examination board: \\
        \\
        Prof.~Dr.~Elcio Abdalla (IF-USP) \\
        Prof.~Dr.~Ivone Freita Mota de Albuquerque (IF-USP) \\
        Prof.~Dr.~Laerte Sodré Júnior (IAG-USP) \\
        Prof.~Dr.~Alberto Vazquez Saa (IMECC-Unicamp) \\
        Prof.~Dr.~Jailson Souza de Alcaniz (ON) \\
        \begin{center}
            São Paulo \\ 
            \the\year
        \end{center}

    \end{titlepage}

    \newpage
    \thispagestyle{empty}
    \mbox{}

    \selectlanguage{english}

    \renewcommand{\abstractname}{Acknowledgements}
    \begin{abstract}
        \setcounter{page}{5}
        \thispagestyle{plain}
        I would like to start by thanking my supervisor Elcio Abdalla who warmly
        welcomed me from a different research area and handed me this challenging
        project.
        I thank him for his patience, guidance and for helping make this thesis
        possible.

        I would like to thank my collaborators Morgan Le Delliou, Gast\~ao Lima
        Neto and Bin Wang for the extensive discussions and advices.
        Professors Marcos Lima, Luis Raul Abramo and George Matsas were also
        important for my growth as a physicist and researcher.
        
        I am also very grateful to all my family -- Monica, Humberto,
        Gabriel, Wanessa and Igor -- for the unconditional support throughout
        this long journey
        and to all my colleagues and friends who contributed with their knowledge
        and friendship.
        A very special thank you to Ricardo Landim, Andr\'e Costa, Lucas Olivari,
        Riis Bachega, Elisa Ferreira, Maicon Siqueira, Rafael de Castro Lopes,
        Eduardo Marcondes, Jos\'e Ebram Filho, Daniela Pena, Fernando Henrique
        de S\'a, Jana\'ina Angeli, J\'essica Fernandes and Andreza Leite.

        Finally, I would like to show my most profound gratitude to
        my girlfriend Nathany for her love, dedication and for being so
        supportive and sweet.

    \end{abstract}

    \newpage
    \thispagestyle{empty}
    \mbox{} % manually inserted new page to make it obey the openright option. Needed because of tocloft

    \renewcommand{\abstractname}{Abstract}
    \begin{abstract}
        \setcounter{page}{7}
        \thispagestyle{plain}
        Modern Cosmology offers us a great understanding of the universe with
        striking precision, made possible by the modern technologies of the
        newest generations of telescopes.
        The standard cosmological model, however, is not absent of theoretical
        problems and open questions.
        One possibility that has been put forward is the existence of a coupling
        between dark sectors.
        The idea of an interaction between the dark components 
        could help physicists understand why we live in an epoch of the
        universe where dark matter and dark energy are comparable in terms of
        energy density, which can be regarded as a strange coincidence given that
        their time evolutions are completely different.
        
        Dark matter and dark energy are generally treated as perfect fluids.
        Interaction is introduced when we allow for a non-zero term 
        in the right-hand side of their individual energy-momentum tensor
        conservation equations. 
        We proceed with a phenomenological approach to test models of
        interaction with observations of redshift-space distortions.
        In a flat universe composed only of these two fluids, we consider
        separately two forms of interaction, through terms proportional to the
        densities of both dark energy and dark matter.
        An analytic expression for the growth rate approximated as $f =
        \Omega_{\dm}^{\gamma}$, where $\Omega_{\dm}$ is the percentage
        contribution from the dark matter to the energy content of the universe
        and $\gamma$ is the growth index, is derived in terms of the interaction
        strength and of other parameters of the model in the first case, while for
        the second model we show that a non-zero interaction cannot be
        accommodated by the index growth approximation.
        The successful expressions obtained are then used to compare the
        predictions with growth of structure observational data in a Markov Chain Monte
        Carlo code and we find that the current growth data alone cannot impose
        constraints on the interaction strength due to their large uncertainties.

        We also employ observations of galaxy clusters to assess their
        virial state via the modified \LI\ equation in order to detect signs of
        an interaction.
        We obtain measurements of observed virial ratios, interaction strength,
        rest virial ratio and departure from equilibrium for a set of clusters.
        A compounded analysis indicates an interaction strength of
        $0.29^{+2.25}_{-0.40}$, compatible with no interaction, but a compounded
        rest virial ratio of $0.82^{+0.13}_{-0.14}$, which means a $2\sigma$
        confidence level detection.
        Despite this tension, the method produces encouraging results while
        still leaves room for improvement, possibly by removing the assumption
        of small departure from equilibrium.

        \paragraph{Keywords:} Cosmology, Dark energy, Dark matter, Large-scale structure

    \end{abstract}

    \newpage
    \thispagestyle{empty}
    \mbox{} % manually inserted new page to make it obey the openright option. Needed because of tocloft

    \renewcommand{\abstractname}{\vspace{-10mm}Resumo}
    \begin{abstract}
        \selectlanguage{brazil}
        \setcounter{page}{9}
        \thispagestyle{plain}
        A cosmologia moderna oferece um ótimo entendimento do universo com uma
        precisão impressionante, possibilitada pelas tecnologias modernas das
        gerações mais novas de telescópios.
        O modelo cosmológico padrão, porém, não é livre de problemas do ponto de
        vista teórico, deixando perguntas ainda sem respostas.
        Uma possibilidade que tem sido proposta é a existência de um acoplamento
        entre setores escuros.
        A ideia de uma interação entre os componentes escuros 
        poderia ajudar os físicos a entender por que vivemos em uma época do
        universo na qual a matéria escura e a energia escura são comparáveis em
        termos de densidades de energia, o que pode ser considerado uma estranha
        coincidência dado que suas evoluções com o tempo são completamente
        diferentes.

        Matéria escura e energia escura são geralmente tratadas como fluidos
        perfeitos. 
        A interação é introduzida ao permitirmos um tensor não nulo
        no lado direito das equações de conservação dos tensores de
        energia-momento.
        Prosseguimos com uma abordagem fenomenológica para testar modelos de
        interação com observações de distorções no espaço de \emph{redshift}. 
        Em um universo plano composto apenas por esses dois fluidos,
        consideramos, separadamente, duas formas de interação, através de termos
        proporcionais às densidades de energia escura e de matéria escura.
        Uma expressão analítica para a taxa de crescimento aproximada por $f =
        \Omega_{\dm}^{\gamma}$, onde $\Omega_{\dm}$ é a contribuição percentual da
        matéria escura para o conteúdo do universo e $\gamma$ é o
        índice de crescimento, é deduzida em termos da interação e
        de outros parâmetros do modelo no primeiro caso, enquanto para o segundo
        caso mostramos que uma interação não nula não pode ser acomodada pela
        aproximação do índice de crescimento.
        As expressões obtidas são então utilizadas para
        comparar as previsões com dados observacionais de crescimento de
        estruturas em um
        programa para Monte Carlo via cadeias de Markov.
        Concluímos que tais dados atuais por si só não são capazes de
        restringir a interação devido às suas grandes incertezas.

        Utilizamos também observações de aglomerados de galáxias para analisar
        seus estados viriais através da equação de \LI\ modificada a fim de
        detectar sinais de interação.
        Obtemos medições de taxas viriais observadas, constante de interação,
        taxa virial de equilíbrio e desvio do equilíbrio para um conjunto de
        aglomerados.
        Uma análise combinada indica uma constante de interação
        $0.29^{+2.25}_{-0.40}$, compatível com zero, mas uma
        taxa virial de equilíbrio combinada de $0.82^{+0.13}_{-0.14}$, o que
        significa uma detecção em um intervalo de confiança de $2\sigma$.
        Apesar desta tensão, o método produz resultados encorajadores enquanto
        ainda permite melhorias, possivelmente pela remoção da suposição de
        pequenos desvios do equilíbrio.

        \paragraph{Palavras-chave:} Cosmologia, Energia escura, Matéria escura,
        Estrutura do universo
    \end{abstract}

    \selectlanguage{english}
    \newpage
    \thispagestyle{empty}
    \mbox{}

    \printnoidxglossary[type=acronym, style=index] %index check if there is a non bold version
                            % long3col, long, super, super3col are nice but give errors

    \newpage
    \thispagestyle{empty}
    \mbox{} % manually inserted new page to make it obey the openright option. Needed because of tocloft

    \tableofcontents

    \chapter{Introduction}
    \pagenumbering{arabic}
    Cosmology is a branch of astronomy that studies the universe in large scale.
    The theory of \gls{gr} underlies the standard cosmological model, assuming 
    that the physics is the same everywhere and at all times, based on both
    laboratory results and inferences from what can be observed.
    However, the validity of this extrapolation is limited.
    At very high energy scales, \gls{gr} fails to describe the early
    universe, which may be dominated by quantum gravity effects.

    As far as observations are concerned, the universe looks the same no matter
    at what direction we look.
    Also, there is no reason to believe that we live in a privileged part of
    the universe.
    Combining these two ideas, we can translate this situation into a scenario
    well characterized by homogeneity and isotropy at scales larger than a
    hundred megaparsecs.
    The assumptions of homogeneity and isotropy are the fundamental principles
    of cosmology.
    With this in mind, cosmology attempts to track down the history and
    evolution of the universe as a whole, from the beginning until the present
    epoch.
    Component species, growth of structures and the future of the universe are
    also between the subjects of interest to cosmologists.

    Current observational facts, like the luminosity distances of
    \glsentryfull{sneIa} \citep{Riess1998, perlmutter1999}, offer us strong
    evidences of an accelerating expansion going on.
    This expansion traces back to a hot and dense phase, where all the content
    of the universe were concentrated in a very small region, then started to
    expand. This picture is known as the ``Big Bang''.
    If the \gls{efe} from \gls{gr} are correct for very large scales, such an
    accelerated expansion occurring today could only be sustained by a
    component with negative pressure dominating the universe, as the Friedmann
    equations point out \citep{Friedmann1999}. 
    We call this component \gls{de}.
    A simple cosmological constant can also play that role and is completely
    equivalent, from a cosmological point of view, to the existence of a
    \gls{de} fluid for which the sum of its pressure and energy density is
    exactly zero.

    At the scales of structures, e.g.~galaxies and galaxy clusters, observed
    galaxy rotation curves differ from the prediction of classical mechanics if
    the velocity profiles are due to the gravitational field of the matter that
    we can see (e.g.~baryons).
    When computing the mass considering the luminosity profiles and the
    mass-to-light ratio, the amount of luminous matter is not sufficient to
    match the observed profile \citep{Zwicky1933,Zwicky1937}. 
    One possibility for this discrepancy is the existence of a dark,
    non-baryonic component which does not interact with the other components
    except gravitationally.
    This \gls{dm} is supposed to permeate the galaxy, extending to the
    galaxy's halo.

    The most currently accepted cosmological model is called \gls{lcdm}.
    The standard cosmological model comprises the Big Bang and the subsequent
    expansion, the universe being composed today, in most part, of dark energy
    (or the cosmological constant $\Lambda$), cold (non-relativistic) dark
    matter, baryonic matter and a small amount of radiation.
    This model describes the universe quite well, in good agreement with the
    most recent and precise observations (chapter~\ref{ch:LCDM}), but still
    leaves some open questions, which we further detail in
    section~\ref{sec:LCDMproblems}.
    Perturbation theory is introduced in chapter~\ref{ch:perturbations}.
    This is done in a general way that goes beyond the standard model, allowing
    an interaction between dark energy and dark matter, which may enable us to
    address some of the questions not answered by the \gls{lcdm} model.
    Still in this chapter, the two-point functions are defined.
    Chapter~\ref{ch:rsd} follows with a brief study of the \glspl{rsd}.
    Then, in chapter~\ref{ch:coupling}, we work further on the interacting
    models and develop the equations describing the growth of structures, which
    we later employ in order to try to constrain the interaction and other model
    parameters by comparing their predictions with \gls{lss} data obtained from
    \gls{rsd} measurements.
    The work presented in this chapter has been submitted to the Journal of
    Cosmology and Astroparticle Physics \citep{Marcondes2016}.
    Next, in chapter~\ref{ch:nvclusters} we present a study of an interacting
    dark energy model based on the equilibrium states of galaxy clusters, trying
    to evaluate the effect of the interaction by their deviation from the virial
    theorem. 
    This interesting approach has been published in the Monthly Notices of the
    Royal Astronomical Society \citep{LeDelliouetal2015}.
    We conclude in chapter~\ref{ch:final} summarizing the results of the two
    approaches and briefly discuss some of the ideas that we may follow in the
    next works, aiming to improve those results, particularly for the \gls{lss}
    data in view of upcoming observations that will be provided by the newest
    and most advanced telescopes currently under construction.

    \chapter{The \texorpdfstring{\glsentrytext{lcdm}}{LCDM} model}
    \label{ch:LCDM}

    The universe is expanding.
    The obvious conclusion is that the distance between two distant galaxies was
    smaller in the past, possibly all the way back to a hot and dense state.
    This picture is reinforced by the perception of \citet{Hubble_1929} that the
    more distant the celestial objects are, the faster they recede from us.
    This became famous as the Hubble's law,
    \begin{align}
        \label{eq:HubblesLaw}
        \dot r = H_0 r,
    \end{align}
    where $\dot r$ is the recessional velocity, $r$ is the proper distance of
    the object from us and $H_0$ is the Hubble constant, the current value of
    the time varying expansion rate.  This recessional velocity is measured by
    the redshift $z \equiv (\lambda_{\obs} - \lambda_{\emit})/\lambda_{\emit}$ of the
    object, i.e., the relative shift of the spectral emission or absorption
    lines of that source's light.  The observed wavelength $\lambda_{\obs}$ differs
    from the emitted wavelength $\lambda_{\emit}$ due to the motion relative to the
    observer.  For small redshifts, the velocity is measured by $\dot r = c z$,
    where $c$ is the speed of light.  Since galaxies are, in general, receding,
    wavelengths are stretched, going towards the red side of the spectrum, hence
    the name \textit{red}shift.

    A convenient way to study the universe is to separate the expansion from
    other dynamics.  We write the proper distance $r$ of an object in terms of a
    universal time-only varying scale factor $a(t)$ times a coordinate distance
    $x$.  Since the wavelengths scale as this expansion parameter, $\lambda_{\obs}
    / \lambda_{\emit} = a_{\obs} / a_{\emit}$, the scale factor relates to the redshift
    by $1 + z = a(t_{\obs}) / a(t_{\emit})$.  The proper velocity is then $u \equiv
    \dot r = \dot a x + a \dot x$, the dot representing differentiation with
    respect to the cosmic time $t$.

    The second term is the peculiar velocity.
    When a galaxy does not have peculiar velocity and, therefore, has fixed
    comoving coordinate x, we can write the proper velocity as $\dot r = \dot a
    x = (\dot a/a) a x = H r$, which is similar to \eq{eq:HubblesLaw} but valid
    for all cosmic times. 
    Here we introduced the Hubble rate $H(t) \equiv \dot a(t) /a(t)$. 
    We use the index 0 to refer to the value of a quantity in the current
    days ($t_0$).
    The scale factor is normalized to 1 today [$a(t_0) \equiv a_0 \equiv 1$] and
    $H_0$ in \eq{eq:HubblesLaw} means
    $H(t_0)$.\footnote{\citet{Planck2015} constrains the Hubble parameter to $H_0 = \SI{67.74(46)}{\km\per\second\per\mega\parsec}$ through indirect
        (model-dependent) measurements, which is in tension with direct
        (model-independent) measurements by up to $3.4 \sigma$; a value of $H_0
        = \SI{73.24(174)}{\km\per\second\per\mega\parsec}$ is measured by
        \citet{Riessetal2016}.
        This tension has attracted attention of physicists lately.
        For more details, see \citet{BernalVerdeRiess2016} and other references
        therein.}
    In the absence of peculiar velocity, a galaxy is said to follow the Hubble
    flow.

    We adopt the $(-,+,+,+)$ convention for the metric signature.
    With the considerations of homogeneity and isotropy, the line element in
    such a smooth, expanding universe is 
    \begin{align}
        \ud s^2 = - \ud t^2 + a^2(t) \, \ud \ell^2 ,
    \end{align}
    in units with $c = 1$;
    $\ud \ell$ is the three-dimensional space element.
    It gives the proper spatial separation between two events occurring at a
    time $t$. 
    For a flat universe, the spatial element can be written simply as the
    Euclidean line element $\ud \ell^2 = \ud x^2 + \ud y^2 + \ud z^2$, but we
    also like to express it in hyperspherical coordinates, where it can be
    generalized to a form that includes a possible spatial curvature. 
    With the usual transformation $x = r \sin \theta \cos \varphi$, $y = r \sin
    \theta \sin \varphi$ and $z = r \cos \theta$, and then a new change of
    coordinate $r = \left\lbrace R \sinh \chi, R \chi, R \sin
    \chi \right\rbrace$, according to the curvature $K$ being negative,
    zero or positive corresponding to open, flat or closed universes,
    respectively, the line element in the new coordinates is
    \begin{align}
        \label{eq:FLRWmetric}
        \ud s^2 = - \ud t^2 + a^2(t) R^2 \left[ \ud \chi^2 + S_{K}^2(\chi) \, \ud \Omega_{\chi}^2 \right], 
    \end{align}
    where the function $S_{K}$ assumes
    \begin{align}
        S_{K}(\chi) = \begin{cases}
            \sinh \chi, & \text{for } K < 0; \\
            \chi, & \text{for } K =0; \\
            \sin \chi , & \text{for } K > 0;
              \end{cases}
    \end{align}
    and $\Omega_{\chi}$ is the solid angle defined by $\ud
    \Omega_{\chi}^2 = \ud \theta^2 + \sin^2 \! \theta \, \ud \varphi^2$.
    In these coordinates, $\chi$ is an angle coordinate, $R$ is a constant with
    units of length and can be interpreted as the comoving radius of curvature
    of the space.
    
    The function $S_{K}$ can still be written in a unified fashion as
    \begin{align}
        S_{K}(\chi) = \frac{1}{\sqrt{ - K}} \sinh \left( \sqrt{ - K } \chi \right) ,
    \end{align}
    the flat universe being recovered by taking the limit $K \rightarrow 0^{-}$.
    The \gls{flrw} metric
    \citep{Friedmann1999,Lemaitre1931,Robertson1935,Walker1937} is then obtained
    by equating $\ud s^2 = g_{\mu\nu} \, \ud x^{\mu} \, \ud x^{\nu}$ and
    \eq{eq:FLRWmetric}, so in the general coordinates we have the non-zero
    metric components
    \begin{gather}
        \label{eq:FLRWcurvcomponents}
        \begin{gathered}
            g_{tt} = -1, \qquad 
            g_{\chi\chi} = a^2(t) R^2, \\
            g_{\theta\theta} = a^2(t) R^2 S_K^2(\chi), \qquad
            g_{\varphi\varphi} = a^2(t) R^2 S_K^2(\chi) \sin^2 \! \theta 
        \end{gathered}
    \end{gather}

    The evolution of $a$ with $t$ in this \gls{flrw} universe is determined by
    the content of the universe.
    We describe the fluids by the total energy density $\bar\rho(t)$ and
    total pressure $\bar p(t)$, so the total energy-momentum tensor
    is\footnote{Greek letters are used for indices running through
        \numlist{0;1;2;3}.}
    \begin{align}
        \label{eq:totalEMtensor1}
        \bar{\mathcal{T}}_{\mu\nu} = \bar p \bar g_{\mu\nu} + \left( \bar p +
            \bar \rho \right) \bar u_{\mu} \bar u_{\nu},
    \end{align}
    where $\bar u^{\mu} = \left(1, \vecnotation{0}\right)$ is the four-velocity of the
    fluid, comoving with the Hubble flow.
    The bars indicate that these are background (unperturbed) quantities.
    Since the distinction will be important when we treat inhomogeneities later,
    we prefer to introduce this notation already to avoid confusion.

    With \eq{eq:totalEMtensor1} and the metric $\bar g_{\mu\nu}$ given by
    \eq{eq:FLRWcurvcomponents}, we write the time-time component of the
    \gls{efe} as
    \begin{align}
        \label{eq:FriedmannEq}
        H^2(t) = \frac{8 \uppi G}{3} \bar \rho(t) - \frac{K}{a^2(t) R^2},
    \end{align}
    where $G$ is the Newton's gravitational constant.
    This is the well-known Friedmann equation \citep{Friedmann1999}.
    The second term in the right-hand side is the contribution from the curvature.
    The constant $R$ can be arbitrarily set to \num{1} and $K$ normalized to
    \numlist[list-final-separator = { or }]{-1;0;1} simultaneously.
    This freedom of choice is possible with a redefinition of the radial
    coordinate $r$ and of the scale factor $a$.

    We can think of the curvature as a fluid component with energy density $\bar \rho_{K}
    (t) \equiv -3K (8\uppi G )^{-1} a^{-2}(t)$ and then write the Friedmann equation as
    \begin{align}
        \label{eq:Friedmannrho}
        H^2(t) = \frac{8 \uppi G}{3} \bigl[ \bar \rho_{\dm}(t) + \bar \rho_{\baryons}(t) + \bar \rho_{\de}(t) + \bar \rho_{\rad}(t) + \bar \rho_{K}(t) \bigr],
    \end{align}
    also expressing each component of the total $\bar \rho(t)$ explicitly: dark matter,
    baryonic matter, dark energy and radiation (photons and neutrinos).
    The $K=0$ flat universe has an energy density equal to $3 H^2(t)/8 \uppi G
    \equiv \bar \rho_{\critical}(t)$, which is called the critical energy
    density.\footnote{The value of the critical energy density today is
        $\bar \rho_{\critical,0} = \num[separate-uncertainty =
        false]{1.87847(23)e-29} \, h^2 \, \si{\gram\per\cubic\cm}$
        \citep{KOliveReview}, where $h$ is the Hubble constant $H_0$ in units of
        $\SI{100}{\km\per\second\per\mega\parsec}$.}
    Normalizing both sides of \eq{eq:Friedmannrho} by $\bar \rho_{\critical}(t)$
    and defining the dimensionless density parameters $\Omega_{\indi} (t) \equiv
    \bar \rho_{\indi}(t)/ \bar \rho_{\critical}(t)$, where $\indi$ stands for each of the
    components mentioned above, \eq{eq:Friedmannrho} becomes
    \begin{align}
        \label{eq:sumdenspar}
        1 = \Omega_{\matter}(t) + \Omega_{\rad}(t) + \Omega_{\de}(t) + \Omega_{K} (t).
    \end{align}
    The density parameters of matter, radiation and dark energy amount to 1 in a
    universe without curvature.
    In general $\Omega_{K} = 1 - \Omega$, where $\Omega \equiv
    \Omega_{\matter} + \Omega_{\rad} + \Omega_{\de}$ is the total density
    parameter.
    $\Omega_{\de}$ is also called the vacuum density.
    Note, by our definitions, that $\Omega_{K}$ and $K$ have opposite signs, $\Omega_{K} = - K/a^2 H^2 = - K/\dot a^2$.

    The spatial part of the \gls{efe} gives, using \eq{eq:FriedmannEq},
    another important equation referred to as the second Friedmann equation:
    \begin{align}
        \label{eq:2ndFriedmann}
        \frac{\ddot a}{a} = -\frac{4\uppi G}{3} ( \bar \rho + 3 \bar p),
    \end{align}
    where $K$ has been canceled out.
    From \eq{eq:2ndFriedmann} it follows that a current accelerated
    expansion restricts the effective \gls{eos} parameter---the ratio between
    the pressure and the energy density $w(t) \equiv \bar p(t)/\bar\rho(t)$---of the universe as a
    whole to be less than \num{-1/3} so that $\ddot a > 0$.

    We know from the energy-momentum tensor conservation law in an expanding
    universe and from their \gls{eos} that matter (both dark and baryonic) and radiation evolve with $a^{-3}$ and $a^{-4}$, respectively. 
    Still under the assumption that the conservation applies for each species,
    the \gls{de} density is proportional to $a^{-3 \left(1+w_{\de} \right)}$, where
    $w_{\de} = \bar p_{\de} / \bar \rho_{\de}$ is the dark energy \gls{eos} parameter.
    The density simplifies to a constant in the case $w_{\de} = -1$
    (cosmological constant),\footnote{Several models for dark energy which are
        different in nature have been proposed.
        They are characterized by the \gls{eos} parameter, which can be
        different from \num{-1} and still constant ($w$CDM models) or can vary
        with time (dynamic dark energy), e.g.~quintessence described by a
        self-interacting scalar field \citep{Wetterich1988668,PeeblesRatra1988,Caldwell1998}.
        \emph{Planck}'s \glsentryfull{cmb} data alone do not constrain $w$ too much
        because of degeneracy with other parameters, but combining them with
        \glsentryfull{wmap} polarization data \citep{Bennet2013}, \glsentrylong{sneIa},
        \glsentrylong{bao} and other data, \citet{Planck2015} found $w_{\de} =
        -1.019^{+0.075}_{-0.080}$ at \nfpercent\ confidence limit.} in which we
    may want to write the density as
    $\bar \rho_{\Lambda} \equiv \bar \rho_{\de}(t) = \bar \rho_{\de,0}$.
    The curvature energy density goes with $a^{-2}$.
    The energy densities as functions of $a$ rather than $t$ are thus given by
    \begin{subequations}
        \label{eq:evolutions}
        \begin{align}
            \label{eq:energyevolution}
            \bar \rho_{\matter}(a) &= \bar \rho_{\matter,0} \, a^{-3}, & 
            \bar \rho_{\de}(a) &= \bar \rho_{\de,0} \, a^{-3(1+w_{\de})}, \\
            \bar \rho_{\rad}(a) &= \bar \rho_{\rad,0} \, a^{-4}, &
            \bar \rho_{K}(a) &= \bar \rho_{K,0} \, a^{-2}.
        \end{align}
    \end{subequations}
    We may want to write \eq{eq:Friedmannrho} normalized by $\bar \rho_{\critical,0}$ instead of $\bar \rho_{\critical}(t)$, and also express the evolution in terms of the redshift rather than the scale factor.
    In this case we have
    \begin{align}
        \label{eq:Friedmann}
        \frac{H^2(z)}{H_0^2} = \Omega_{\matter,0} (1 + z )^3 + \Omega_{\rad,0}
        (1 + z )^4 + \Omega_{K,0} (1 + z)^2 + \Omega_{\de,0} (1 + z
        )^{3(1+w_{\de})}.
    \end{align}

    Recent observational data from the \gls{wmap} \citep{Hinshaw2013,Bennet2013}
    indicated, for a six-parameter \gls{lcdm} fit, a \gls{dm} density parameter
    $\Omega_{\dm,0} = \num{0.233 \pm 0.023}$, a baryonic density parameter
    $\Omega_{\baryons,0} = \num{0.0463 \pm 0.0024}$, the total matter density
    parameter being $\Omega_{\matter,0} = \num{0.279 \pm 0.025}$, and a \gls{de}
    component with $\Omega_{\Lambda} = \num{0.721 \pm 0.025}$.
    Curvature and radiation density parameters are assumed to be zero in this
    simple six-parameter model.\footnote{Radiation density is not actually zero but rather small, at the order of $\Omega_{\rad,0} \sim \num{8e-5}$ \citep{Dodelson_2003}. Although negligible today, radiation was important and dominated the universe at early epochs, when $a$ was small, as is evident from eqs.~\eqref{eq:evolutions}.}
    Placing limits on deviations from this simple model with a seven-parameter model
    allows non-zero curvature $\Omega_{K,0} = -0.037^{+0.044}_{-0.042}$.
    The precisions on these parameters are further improved by combining
    \gls{wmap} data with other \gls{ecmb} measurements from the \gls{act} and
    the \gls{spt}, \glsentrylong{bao} data, and direct measurements of the Hubble constant \citep{Bennet2013}. 
    Newer results from the \emph{Planck} satellite \citep{Planck2015} give the
    slightly different values $\Omega_{\matter,0} = \num{0.3089 \pm 0.0062}$,
    and $\Omega_{\Lambda} = \num{0.6911 \pm 0.0062}$.
    All uncertainties correspond to \sepercent\ confidence limits.

    \section{Cosmic probes and observational evidences for the \texorpdfstring{\glsentrytext{lcdm}}{LCDM} model}

    %\citet{LiMiao2011}
    In this section we discuss some of the most important cosmological
    probes---\gls{sneIa}, light element abundances, the \gls{cmb}, \gls{bao},
    \glsentryfullpl{rsd} and galaxy clusters---and comment on the Hubble diagram and the
    uniformity of the \gls{cmb}, two of the main and most convincing
    observational facts that settle the Big Bang model on a firm basis.

    \subsection{\glsentrylong{sneIa}}
    White dwarf stars that accrete mass from a companion star can eventually
    reach the Chandrasekhar limit, in which their masses become so big that the
    electron degeneracy pressure cannot continue counterbalancing the
    gravitational collapse \citep{Wolfgang2000}.
    The variable star which results from the thermonuclear explosion of a white
    dwarf is a supernova. 
    The classification of the supernovae is based on their spectral properties.
    In contrast to those of the Type II, Type I supernovae do not show any
    Balmer lines of hydrogren in their spectrum.
    The type Ia differs from its siblings Ib and Ic by the presence of a strong
    ionized silicon absorption line at wavelength \SI{6150}{\angstrom}
    \citep{Schneider_2006}.

    The acceleration of the expansion of the universe was first discovered by
    \citet{Riess1998} by measuring luminosity distances of \gls{sneIa}, used as
    standard candles.
    \citet{perlmutter1999} later confirmed the discovery with analysis of
    nearby and high-redshift supernovae.
    Once the absolute magnitude of the \gls{sneIa} is determined, one can
    obtain the observational distance modulus
    \begin{align}
        \mu_{\obs} = m - M,
    \end{align}
    where $m$ and $M$ are the apparent and absolute magnitudes, respectively.
    The theoretical distance modulus of a supernova (labelled by $i$), on the
    other hand, can be calculated as
    \begin{align}
        \mu_{\text{th}}(z_i) \equiv 5 \log_{10}  d_L(z_i) + 25.
    \end{align}
    Models can then be constrained through the dependence of the luminosity
    distance $d_L$ on the cosmological parameters by comparing $\mu_{\obs}$ and
    $\mu_{\text{th}}$.

    \subsubsection{The Hubble diagram}
    The Hubble diagram is still the most direct evidence of expansion.
    Plotting the redshift velocity versus the luminosity distance $d_L$ for
    distant galaxies reveals the linear increase at low redshifts.
    At higher redshifts, the luminosity distance is more sensitive to the
    contents of the universe through the Hubble rate:
    \begin{align}
        \label{eq:lumdist}
        d_L(z) = \frac{1 + z}{H_0 \sqrt{\lvert \Omega_{K}\rvert}} \, S_{K} \! \left[ H_0 \sqrt{ \lvert \Omega_{K} \rvert } \int_0^{z} \frac{\ud \tilde z}{H(\tilde z)} \right].
    \end{align}
    It is clear, from eqs.~\eqref{eq:Friedmann} and \eqref{eq:lumdist}, how the
    Hubble factor is nearly constant for small $z$, but affects strongly the
    luminosity distance at high redshifts, besides departing from the constant
    proportionality between velocity and distance.

\begin{figure}[tb]
        \centering
        \includegraphics[width=0.6\textwidth,trim=0.8cm 1cm 2.2cm 2cm,clip=true]{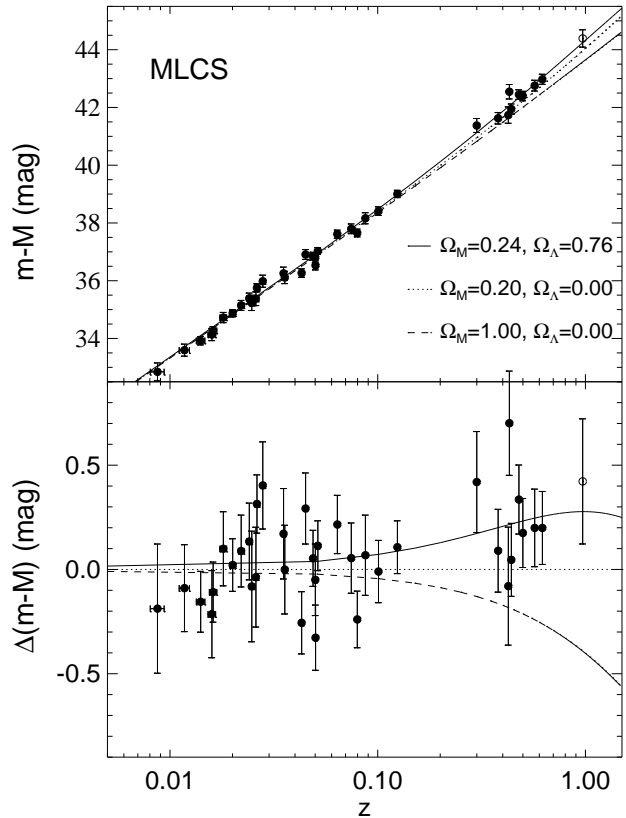}
        \caption{\acrlong{sneIa} diagram from \citet{Riess1998}. The upper panel
            shows apparent magnitudes (as an indicator of distance) versus
            redshift, for low- and high-redshift \gls{sneIa} samples. The bottom
            panel shows the residual magnitudes, thus elucidating the preference
            for a $\Lambda$-dominated universe supported by the high-redshift
            \gls{sneIa}.}
        \label{fig:Hdiagram}
    \end{figure}

    The major challenge in the construction of the Hubble diagram is to
    determine the distances of objects whose intrinsic brightness is unknown.
    With the use of standard candles one can determine the difference in the
    distances of these objects from us by their apparent brightness.
    It is possible to analyze the Hubble diagram at large redshifts with
    \gls{sneIa}, which are too distant but are bright enough they can still be
    detected. 
    The redshifts of these objects allow us to distinguish between flat matter
    dominated, open, and flat universe with a cosmological constant $\Lambda$
    \citep{Riess1998, perlmutter1999}, as shown in Figure~\ref{fig:Hdiagram}.
    Current high-redshift data favor a universe dominated by some form of dark
    energy or cosmological constant, with a best-fit of about \SI{70}{\percent}
    of the total energy density for this component.

    \subsection{The light element abundances}
    A distinct confirmation of the Big Bang is the prediction of light element
    abundances by the \gls{bbn}.
    Light elements started to form when the universe cooled down sufficiently so
    that protons and neutrons could combine into nuclei, and then nuclei and
    electrons could combine into atoms.
    Before that, in a hot universe with temperature $T$ of order
    $\SI{1}{\MeV}/\kB$ (where $\kB$ is the Boltzmann constant), the intense
    radiation prevented atoms from being produced when those particles collided.
    The atoms would be destroyed by high energy photons right after being formed.
    When the temperature went down way below the typical nuclei binding
    energies, the lightest elements started to form. 
    With knowledge of the conditions of the early universe and of the nuclear
    cross-sections of the relevant processes, one can calculate the expected
    amount of those elements in the primordial universe. 
    These predictions are in good agreement with current estimates of light
    element abundances \citep{Dodelson_2003}, hence serving as a good argument
    in favor of the Big Bang theory.\footnote{See also \citet{BBN} for more
        recent predictions of the light element abundances with the results by
        the \emph{Planck} satellite mission.}

    \subsection{The \glsentrydesc{cmb}}
    \label{ss:CMB}
\begin{figure}[tb]
        \centering
        \includegraphics[width=0.8\textwidth]{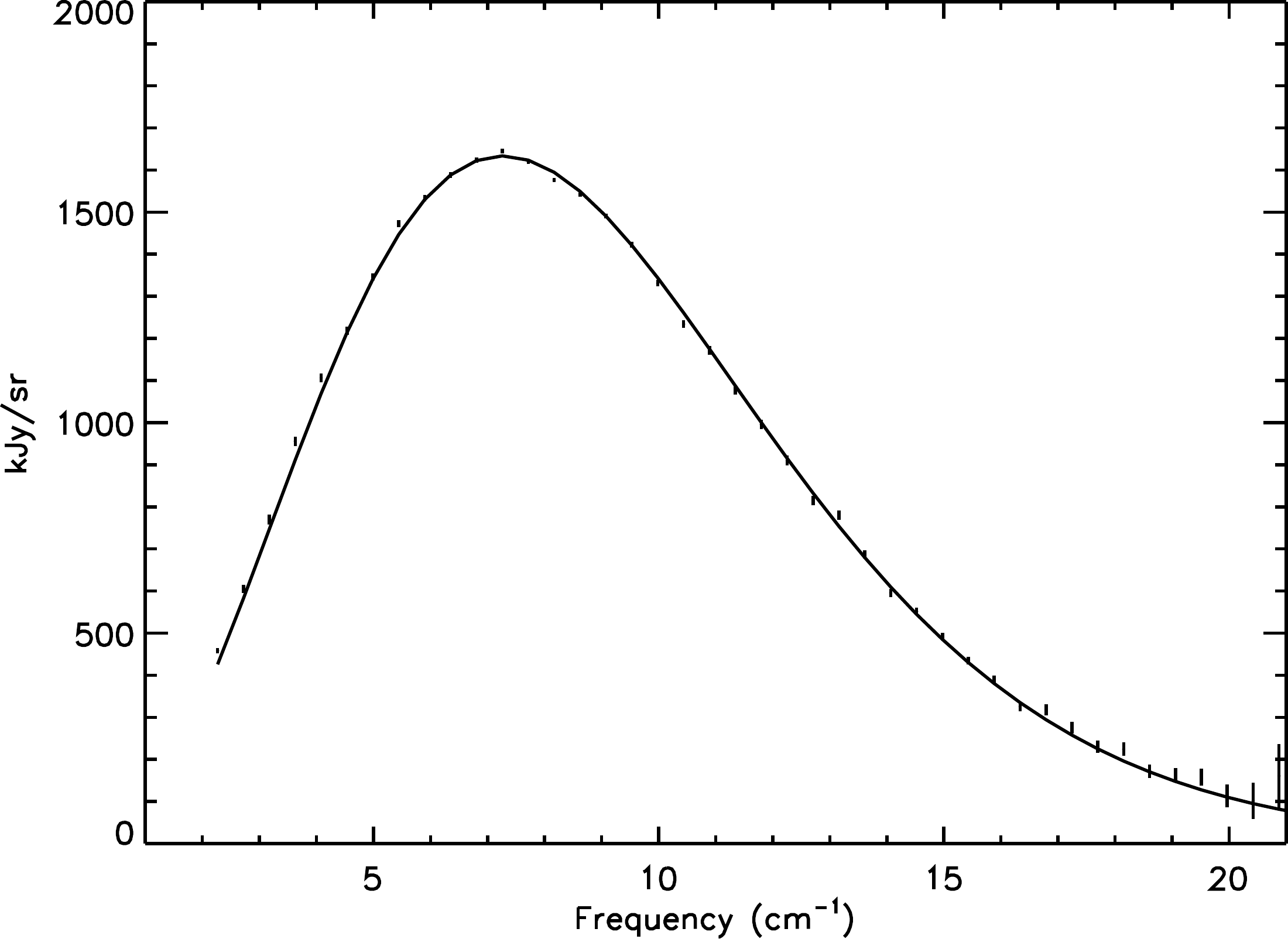}
        \caption{Comparison of the intensity of radiation observed with
            \gls{cobe}'s \gls{firas} radiometer with a black-body spectrum with
            temperature \SI{2.728}{\kelvin}, from \citet{Fixsen1996}. The
            intensity is in units of kiloJansky per steradian ($\SI{1}{\jansky}
            = \SI{e-26}{\watt\per\square\meter\per\hertz}$). The error bars
            indicate $1\sigma$ experimental uncertainty in intensity.}
        \label{fig:blackbody}
    \end{figure}

    The high degree of uniformity of the \gls{cmb} is the most compelling
    evidence of the universe starting with a Big Bang.
    At the epoch when the universe was hot enough for electrons to be bounded
    into atoms, the collisions of photons with free electrons had maintained a
    thermal equilibrium between radiation and matter, making the distribution of
    the number density of photons follow a black-body spectrum.
    At some time later, as the universe was expanding, the matter cooled down
    and became less dense.
    The radiation then decoupled from the matter to start a free expansion (we
    call this moment ``last scattering''), but the form of its spectrum was kept
    unchanged.
    The cosmic temperature when this last interaction of photons with matter
    took place was about \SI{3000}{\kelvin}, at a redshift \num{1100}. 
    Photons have travelled freely since then.
    In \citeyear{Penzias:1965wn}, \citet{Penzias:1965wn} discovered such cosmic
    radiation which later would be reported to have a temperature of about
    \SI{3}{\kelvin}.
    The cosmological implications of this \gls{cmb} were treated by
    \citet{Dicke:1965zz} in a companion article.
    More details about the history of the \gls{cmb} discovery can be found in
    \citeauthor{Weinberg_2008}'s \emph{Cosmology} \citep{Weinberg_2008}.
    More recently, observations with the \gls{firas} radiometer of the
    \gls{cobe} revealed an almost exact black-body spectrum in the wavelength
    range of \SIrange{0.5}{0.05}{\cm} \citep{Mather:1993ij}.
    Figure~\ref{fig:blackbody} shows these observations compared with the
    black-body spectrum. 
    The value of the redshifted \gls{cmb} temperature today has been determined with
    high precision by \citet{Fixsen2009} using \gls{wmap} data to recalibrate
    the \gls{firas} data. The new, reviewed value is \SI[separate-uncertainty =
    false]{2.72548 \pm 0.00057}{\kelvin}.

    Anisotropies in the \gls{cmb} (small temperature fluctuations) were discovered by the \gls{cobe} satellite in 1989.
    The \gls{wmap} measured precisely the temperature power spectrum and probed several cosmological parameters with high accuracy.
    A best-fitting value of $\Omega_{\Lambda} \approx \num{0.72}$ for the \gls{de}
    density parameter corresponding to a cosmological constant was found with
    the 9-year data.

    \subsection{Baryon acoustic oscillations}
    Acoustic waves propagating in the early universe from the end of inflation until decoupling left a characteristic imprint on the anisotropies of the \gls{cmb} and on the late-time matter power spectrum \citep{Einseinstein2005}.
    These \acrlong{bao} provide a standard ruler for the length scale of clustering of baryonic matter, which is about \SI{150}{\mega\parsec} today \citep{LiMiao2011}.
    The apparent size of the \gls{bao} measured from astronomical observations hints at the expansion history of the universe through measurements of the Hubble rate $H(z)$ and yields measurements of the angular diameter distance $d_{\text{A}}$, which is given by
    \begin{align}
        d_{\text{A}} \equiv \frac{\ell}{\theta} = \frac{1}{1 + z} \int_0^{z}
        \frac{\ud \tilde z}{H(\tilde z)}, \qquad \text{(flat universe)}
    \end{align}
    the physical size $\ell$ of an object (the \gls{bao} characteristic length, for instance) divided by the angle $\theta$ it subtends, and is related to the luminosity distance by $d_{\text{A}} = a^2 d_L$.

    \subsection{Redshift-space distortions and the growth factor}

    As we will see in chapters~\ref{ch:rsd} and \ref{ch:coupling}, \glspl{rsd}
    can be used to distinguish between dark energy and modified gravity models
    or even to help improve the constrains on model parameters.
    Measuring the growth rate of structures---the rate of change of the growth
    factor $D$, which is the solution of the dynamic equations governing
    gravitational instability of matter perturbations:
    \begin{align}
        f(z) = \frac{\ud \ln D}{\ud \ln a}.
    \end{align}
    Measurements of $f(z)$ from observations are dependent on the bias model,
    i.e., on how galaxies are supposed to trace the matter field.
    \glspl{rsd}, however, can provide a measurement of an observable that is independent on the bias model \citep{SongPercival2009}.

    \subsection{Galaxy clusters}
    Galaxy clusters are some of the largest structures in the universe, containing from
    hundreds to thousands of galaxies in a radius from approximately
    \SIrange{1}{5}{\mega\parsec} \citep{LiMiao2011}.
    Clusters are observed through a variety of techniques, e.g.~X-ray imaging
    and spectroscopy and gravitational lensing.
    N-body simulations can predict the number density $n(z, M)$ of dark matter
    halos of mass $M$ as a function of the redshift $z$ and of the halo mass
    $M$.
    These predictions can be compared to cluster surveys to provide constraints
    on the expansion history of the universe.

    In chapter~\ref{ch:nvclusters} we adopt a different
    approach to test an interacting model assessing the virial equilibrium
    states of clusters through the \LI\ equation.

    \section{Classical problems in Cosmology}
    \label{sec:LCDMproblems}

    The evidences we have seen so far constitute a solid ground for the
    \gls{lcdm} model with the Big Bang.
    \gls{lcdm} by itself, however, fails to answer some questions raised based
    on observational facts.
    Some of them have been explained by inflation remarkably well.
    Because of this success, \gls{lcdm} and inflation constitute, together, the
    current standard model of cosmology.

    In this section we present some of these classical problems---the first two
    of which have already been solved by inflation---and introduce, in
    section~\ref{ss:coincidenceproblem}, what motivates most this work, a puzzle
    whose solution can be the existence of a non-minimal coupling between
    \gls{de} and \gls{dm}: the coincidence problem.

    \subsection{The horizon problem}
    \label{ss:horizon}
    It is common to rewrite the line element of \eq{eq:FLRWmetric} in terms of a new time coordinate, the conformal time $\tau$, defined by $\ud \tau = \ud t/ a(t)$, such that the scale factor can be factored out as a common term to both time and spatial parts:
    \begin{align}
        \ud s^2 = a^2(\tau) ( - \ud \tau^2 + \ud \ell^2 ).
    \end{align}
    Note that with this definition, the geodesic of photons $\ud s^2 = 0$ gives
    $\ud \tau = \ud \ell$.
    The conformal time $\tau$, which has dimension of length, is the maximum
    comoving distance light could have traveled since the beginning of the
    universe.

    We discussed in section~\ref{ss:CMB} how smooth the \gls{cmb} is. 
    In fact, it is so smooth that it defies the principle of causality.
    Photons from the \gls{cmb} share the same temperature to one part in \num{e5}. 
    If this uniformity requires that photons were interacting so that larger
    disturbances in the temperature field could be washed out until an
    equilibrium was reached, how could it be that even photons that are
    separated by any distance larger than $\tau$ have the same temperature?
    They have never been in causal contact, no information could have ever
    propagated through that distance, no interaction could have happened to put
    them in equilibrium.

    An explanation for this is provided by the theory of inflation, which
    affirms that the universe passed through a period of incredibly fast
    expansion right after the Big Bang.
    The idea gained popularity after \citet{GuthPRD1981}, who realized that
    inflation could also address some other cosmological
    puzzles.\footnote{\citeauthor{GuthPRD1981}'s first version of inflation had
        a serious problem, explained in \citet{HawkingMoss1982} and
        \citet{GuthWeinberg1983}, and was replaced by a new model by
        \citet{Linde1982} and \citet{Albrecht1982}, known as slow-roll
        inflation.}
    The solution suggests that those particles are not in contact today but
    could have been in contact for some time before.
    That is, they are now separated by a comoving distance larger than the
    comoving Hubble radius $1/aH$, but smaller than the conformal time.
    The requirement is that $aH$ must have increased during inflation, thus
    implying $\ddot a > 0$, namely an accelerated phase of expansion
    \citep{Dodelson_2003,Baumann2009}.

    \subsection{The flatness problem}
    Back then when the scale factor was of order \num{3e-4}, at redshift about
    \num{3000}, the universe underwent a transition phase. 
    From a previously radiation dominated era, it became dominated by both
    radiation and matter, and then matter surpassed radiation in energy density.
    The scale factor, which had been increasing with $t^{1/2}$, then started to
    evolve with $t^{2/3}$ during the matter-dominated era until near the
    present.\footnote{At redshift $z \sim 0.4$ ($a \sim 0.7$) the universe
        became dominated by dark energy or cosmological constant and the scale
        factor dependence with time became exponential.}
    This can be seen by integrating \eq{eq:Friedmann} (in terms of $a$)
    considering only the dominant component in the right-hand side.

    Let us now turn our attention to the evolution of the curvature term
    $\Omega_{K}(t) = -K/\dot a^2$ of \eq{eq:sumdenspar}.
    Since $\dot a \propto t^{-1/3}$, $\Omega_{K}$ increased at the same rate as
    $a$, i.e., with $t^{2/3}$.
    Thus if $\lvert \Omega_{K} \rvert < 1$ today,\footnote{Since this
        \gls{de}-dominated phase began only recently, considering only the long
        matter-dominated era is sufficient to evince the flatness problem.  One
        can conclude that taking into account this latest regime of the scale
        factor worsens the problem, requiring an even finer tuning of the initial
        curvature.} as observations indicate, in the time the scale factor
    increased by a factor of order \num{e4}, then $\Omega_{K}$ must have increased by
    the same factor, which means $\lvert \Omega_{K} \rvert < \num{e-4}$ at that
    time of matter-radiation equality.
    The temperature of the primordial plasma of quarks and photons, which were in a thermal equilibrium, followed the same evolution of the temperature of a gas of photons, that is, scaling with $a^{-1}$ as a consequence of the redshift, since the temperature is proportional to the photon energy or frequency. 
    At that time, this temperature was of order \SI{e4}{\kelvin}.
    Before this, in the radiation-dominated era, $a$ was increasing as $t^{1/2}$ and then $\Omega_{K}$ was proportional to $t$. Since $T \propto a^{-1} \propto t^{-1/2}$ during that period, we can equivalently state that $\Omega_{K}$ was increasing as $T^{-2}$.
    The observed helium abundance coincides with this temperature being of order \SI{e10}{\kelvin} at the beginning of this period \citep{Weinberg_2008}. 
    If the temperature has decreased by a factor \num{e6}, the curvature must have increased by \num{e12} in the same period.
    In order for $\lvert \Omega_{K} \rvert$ to be smaller than \num{e-4} at $T \sim \SI{e4}{\kelvin}$, it could not have been greater than \num{e-16} at $T \sim \SI{e10}{\kelvin}$.
    At earlier times, $\lvert \Omega_{K} \rvert$ must have been even smaller.

    This fine tuning of the curvature density parameter potentially poses a problem.
    It would be good if we could explain why the universe was so flat at the
    beginning, although one can argue that there is no impediment even for
    $\Omega_{K}$ being exactly zero.
    However, it is more natural to expect that some mechanism could have been
    responsible for flattening the universe independently of its initial
    curvature, hence avoiding the need of specific assumptions.
    This problem is also solved by inflation.
    A sufficient preceding period of inflation is enough to guarantee that the
    curvature is negligible at the beginning of the radiation-dominated era.
    The condition required is the same that solves the horizon problem.
    If the universe has expanded during inflation more or less exponentially by
    a factor $e^{\mathcal{N}}$, where $\mathcal{N}$ is the number of
    $e$-foldings, the condition is
    \begin{align}
        e^{\mathcal{N}} > \frac{a_{\indI} H_{\indI} }{a_0 H_0},
    \end{align}
    where the subscript $\indI$ refers to the cosmic time when inflation ends
    and the radiation dominance begins.

    \subsection{The cosmological constant problem}
    The cosmological constant $\Lambda$ was first proposed by
    \citet{Einstein1917} as a free parameter to accommodate a static universe
    solution for his field equations as
    \begin{align}
        \bar{\mathcal{R}}_{\mu\nu} - \frac{1}{2} \bar{g}_{\mu\nu} \bar{\mathcal{R}} - \Lambda \bar{g}_{\mu\nu} = - 8 \uppi G \bar{\mathcal{T}}_{\mu\nu},
    \end{align}
    where $\bar{\mathcal{R}}_{\mu\nu} =
    \bar{\mathcal{R}}^{\rho}{}_{\mu\rho\nu}$ is the Ricci tensor,
    given by the contraction of Riemann tensor in the first and third indices,
    $\bar{\mathcal{R}} = \bar{\mathcal{R}}^{\nu}{}_{\nu}$ is the Ricci scalar (the
    contracted Ricci tensor) and $\bar{\mathcal{T}}_{\mu\nu}$ is the energy-momentum
    tensor, which describes the content of the universe. 
    
    With a non-zero cosmological constant, the Friedmann equation
    \eqref{eq:FriedmannEq} is written as
    \begin{align}
        \label{eq:FriedmannCC}
        H^2(t) = - \frac{K}{a^2(t)} + \frac{8 \uppi G}{3}\bar\rho(t) + \frac{\Lambda}{3},
    \end{align}
    where now the total density $\bar\rho(t)$ accounts for the matter (and
    radiation) content only, not including a dark energy field (i.e., the term
    $\Lambda$ is equivalent to $8 \uppi G \bar\rho_{\de}$).

    Cosmological observations constrain the effective vacuum energy density to
    be no greater than \SI{e-47}{\giga\eV^4}.
    On the other hand, by summing the zero-point energies of all modes (up to a
    cutoff) of some field describing the empty space gives $\bar\rho_{\vac} =
    \SI{2e71}{\giga\eV^4}$, a discrepancy of \num{118} orders of magnitude
    \citep{Weinberg1989}.\footnote{This discrepancy can vary depending on the
        chosen cutoff scale. For instance, taking the cutoff at the
        Planck scale gives $\bar\rho_{\vac} \simeq \SI{e74}{\giga\eV^4}$
        \citep{Amendola,Montani}, thus yielding a divergence of \num{121} orders
        of magnitude.} It is true that what really should be smaller than
    \SI{e-47}{\giga\eV^4} is the effective vacuum energy density, which is
    composed of the vacuum energy density $\bar\rho_{\vac}$, from the
    energy-momentum tensor in vacuum $\bar{\mathcal{T}}^{(\vac)}_{\mu\nu} = -
    \bar\rho_{\vac} \bar{g}_{\mu\nu}$, and the cosmological constant
    contribution $\Lambda/8\uppi G$, but then we should have the two terms
    canceling to 118 decimal places:
    \begin{align}
        \left \lvert \bar\rho_{\vac} + \frac{\Lambda}{8\uppi G} \right \rvert < \SI{e-47}{\giga\eV^4}. 
    \end{align}
    One could argue that the effective energy density $\lvert \bar\rho_{\vac} +
    \Lambda/8\uppi G \rvert$ is exactly zero (or equivalently the effective
    cosmological constant $\Lambda_{\eff} = \Lambda + 8\uppi G \bar\rho_{\vac}$
    is exactly zero), with an explanation yet to be given presumably by a
    theory of quantum gravity, but current cosmological observations point to a
    non-zero, although extremely small and fine-tuned, value for
    $\Lambda_{\eff}$ \citep{Padmanabhan2003}.

    \subsection{The cosmic coincidence problem}
    \label{ss:coincidenceproblem}
    We already know that dark matter and dark energy evolved completely
    differently with the expansion of the universe. 
    While the energy density of dark matter has decreased as $a^{-3}$, the dark
    energy's remained constant.
    Surprisingly, both components contribute to the energy content of the
    universe by similar (of the same order) amounts today \citep{Fitch1998,Zlatev1999}.
    One could expect totally different orders of magnitude, specially when
    noting that in the standard model the \gls{dm}-\gls{de} density ratio
    $\dmderatio$ should just cross the value \num{1} at some time, without
    being forced towards it. 
    Thus the fact that those two components have similar densities just now
    can be seen as a coincidence.

    Some physicists have proposed the existence of a mechanism that drives the
    ratio $\dmderatio$ close to \num{1}
    \citep{delCampo2006,HeWangAbdalla2011,Poitras2014}.
    In general, allowing an interaction in the dark sector, i.e., between
    \gls{dm} and \gls{de}, through a non-zero term in the right-hand side of
    the energy-momentum tensor conservation equation for these components can
    help alleviate the coincidence problem.
    In this case, despite the flux between the dark components, the total
    energy density is still conserved.
    The effect of such an interaction is that when the right conditions are
    satisfied we have $\dmderatio \sim 1$ for a longer period of time.
    Therefore, it becomes more reasonable to find a ratio of this order from
    observations \citep{FerreiraPavon2013}.

    Chapters~\ref{ch:coupling} and \ref{ch:nvclusters} present our works on 
    interacting dark sector models of cosmology which aim to
    provide a solution to this problem or at least alleviate it.

    \chapter{Fluid inhomogeneities and space-time perturbations}
    \label{ch:perturbations}
    Cosmology as we have seen so far is well described by the \gls{flrw} metric,
    as long as the scales are large enough so that homogeneity still applies.
    In smaller scales, the assumption is obviously invalid, since matter tends
    to clump in structures like clusters of galaxies, galaxies and planetary
    systems.
    We believe that \glsentrylongpl{lss} result from density fluctuations of a pressureless
    cold dark matter fluid amplified by the gravitational attraction.
    In this chapter we introduce perturbations to the metric and the fluids in a
    general interacting-dark sectors cosmology, which obviously also applies to
    the non-interacting standard model when the coupling is zero.
    Later, we define the two-point functions that describe the density
    fields statistically.

    \section{The perturbed metric and field equations}
    According to \gls{gr}, the matter or energy contents of the universe define
    the space-time geometry, which in turn determines the geodesic lines that
    particles and bodies will follow.
    This is encoded in the \gls{efe}
    \begin{align}
        \mathcal{R}_{\mu\nu} - \frac{1}{2} g_{\mu\nu} \mathcal{R} = -8 \uppi G
        \mathcal{T}_{\mu\nu},
    \end{align}
    where $\mathcal{T}_{\mu\nu}$ is the total energy-momentum tensor,
    \begin{align}
        \mathcal{R}_{\mu\nu} = \partial_{\nu}
        {\Gamma^{\lambda}}_{\mu\lambda} - \partial_{\lambda}
        {\Gamma^{\lambda}}_{\mu\nu} +
        {\Gamma^{\rho}}_{\mu\lambda}
        {\Gamma^{\lambda}}_{\nu\rho} -
        {\Gamma^{\rho}}_{\mu\nu}
        {\Gamma^{\lambda}}_{\lambda\rho}
    \end{align}
    is the Ricci tensor and $\mathcal{R}$ is the Ricci scalar, given by the
    space-time metric $g_{\mu\nu}$ and the Christoffel symbols
    \begin{align}
        {\Gamma^{\mu}}_{\nu\lambda} = \frac{1}{2} g^{\mu\rho}
        \left(\partial_{\nu} g_{\lambda\rho} + \partial_{\lambda} g_{\rho\nu} -
            \partial_{\rho} g_{\nu\lambda} \right).
    \end{align}
    If we want to study perturbations to the functions that describe
    the fluids, we have to consider also the perturbations to the metric.

    Perturbations are in general denoted by a $\delta$ preceding the quantity's symbol,
    while a bar denotes the unperturbed part. One exception is the metric, whose
    perturbation is usually denoted by $h_{\mu\nu}$, so the total metric tensor
    is
    \begin{align}
        g_{\mu\nu} = \bar g_{\mu\nu} + h_{\mu\nu}.
    \end{align}
    The perturbation to the inverse of a general matrix $M$ is $\delta
    M^{-1} = - M^{-1} \left(\delta M \right) M^{-1}$, so $h^{\mu\nu} = - \bar
    g^{\mu\rho} \bar g^{\nu\sigma} h_{\rho\sigma}$.
    The perturbed Christoffel symbols are
    \begin{align}
        \delta {\Gamma^{\mu}}_{\nu\lambda} = \frac{1}{2} \bar g^{\mu\rho} \left( -2
            h_{\rho\sigma} \bar{\Gamma}^{\sigma}{}_{\nu\lambda} +
            \partial_{\lambda} h_{\rho\nu}
            + \partial_{\nu} h_{\lambda\rho} - \partial_{\rho} h_{\nu\lambda}
        \right),
    \end{align}
    which lead to the perturbed Ricci tensor 
    \begin{align}
        \label{eq:pertRicci}
        \delta \mathcal{R}_{\mu\nu} = \partial_{\nu} {\delta \Gamma^{\lambda}}_{\mu\lambda} 
        - \partial_{\lambda} { \delta \Gamma^{\lambda}}_{\mu\nu}
        + {\delta \Gamma^{\rho}}_{\mu\lambda} \bar{\Gamma}^{\lambda}{}_{\nu\rho} 
        + {\delta \Gamma^{\lambda}}_{\nu\rho} \bar{\Gamma}^{\rho}{}_{\mu\lambda} 
        - {\delta \Gamma^{\rho}}_{\mu\nu} \bar{\Gamma}^{\lambda}{}_{\lambda\rho} 
        - {\delta \Gamma^{\lambda}}_{\lambda\rho} \bar{\Gamma}^{\rho}{}_{\mu\nu} 
    \end{align}
    and the perturbed \gls{efe}
    \begin{align}
        \label{eq:pertEFE}
        \delta \mathcal{R}_{\mu\nu} 
        - \frac{1}{2} \left( h_{\mu\nu} \bar g^{\lambda\rho} 
                + \bar g_{\mu\nu} h^{\lambda\rho} \right) \bar{\mathcal{R}}_{\lambda\rho} 
            - \frac{1}{2} \bar g_{\mu\nu} \bar g^{\lambda\rho} \delta \mathcal{R}_{\lambda\rho}
            &= - 8 \uppi G \delta \mathcal{T}_{\mu\nu} ,
    \end{align}
    up to first order in perturbations.

    \subsection{The perturbed \glsentryshort{flrw} metric}
    We write the perturbed \gls{flrw} in the Newtonian gauge
    \citep{Weinberg_2008} and restrict ourselves to scalar perturbations only.
    It will be convenient to use the conformal time, which we already introduced
    in section~\ref{ss:horizon}.
    The line element is
    \begin{align}
        \label{eq:perturbedFLRW}
        \ud s^2 = a^2(\tau) \left[ - \left(1 + 2 \psi \right) \ud \tau^2 +
            \left(1 - 2\phi \right) \delta_{ij} \, \ud x^i \ud x^j \right],
    \end{align}
    with $\phi = \phi(x^{\mu})$ and $\psi = \psi(x^{\mu})$ being small
    perturbations, satisfying $|\phi|, |\psi| \ll 1$.
    Assuming there is no anisotropic stress, we have $\psi = \phi$.
    The non-zero components of the unperturbed and perturbed metric parts are
    \begin{align}
        \begin{aligned}
            \bar g_{00} &= -a^2,  \qquad & 
            h_{00} &= - 2 a^2 \phi, \\
            \bar g_{ij} &= a^2 \delta_{ij}, \qquad &
            h_{ij} &= - 2 a^2 \phi \delta_{ij}.
        \end{aligned}
    \end{align}

    \section{Fluid perturbations and evolution equations}
    \label{sec:evequations}
    The unperturbed energy-momentum tensor assumes the perfect fluid form due to
    its rotational and translational invariance,
    \begin{align}
        \label{eq:totalEMtensor}
        \bar{\mathcal{T}}_{\mu\nu} = \bar p \bar g_{\mu\nu} + \left( \bar p +
            \bar \rho \right) \bar u_{\mu} \bar u_{\nu},
    \end{align}
    where $\bar p$ is pressure, $\bar \rho$ is the energy density and $\bar
    u^{\mu} = (a^{-1}, 0, 0, 0)$ is the four-velocity of the
    comoving fluid.
    It follows from the \gls{efe} that the total energy-momentum tensor must
    be conserved, $\Nabla_{\mu} \bar{\mathcal{T}}^{\mu}{}_{\nu} = 0$.
    Noting that the non-zero components of the energy-momentum tensor are
    $\bar{\mathcal{T}}^{0}{}_{0} = - \bar\rho$ and
    $\bar{\mathcal{T}}^{i}{}_{j} = \bar p \delta^i_j$, the $\nu = 0$
    conservation equation reads
    \begin{align}
        \label{eq:totalcons}
        \bar\rho' + 3 \mathcal{H} \left(1 + w \right) \bar\rho = 0,
    \end{align}
    where we have substituted $\bar p = w \bar \rho$ and $\mathcal{H} \equiv
    a'/a$ is the Hubble rate in terms of the conformal time, the prime denoting
    derivatives with respect to $\tau$.

    Treating the components as perfect fluids, we can also write
    \eq{eq:totalEMtensor} for each matter component $\indi$ separately, but the conservation of
    the energy-momentum tensor need not to apply individually.
    In fact, an interaction between dark matter and dark energy is included
    by allowing a non-zero tensor $\bar Q^{\indi\,}{}_{\nu}$ in the right-hand
    side of the conservation equations, $\Nabla_{\mu}
    \bar{\mathcal{T}}_{\!\indi\,}{{}^{\mu}}_{\nu}
    = \bar Q^{\indi\,}{}_{\nu}$, as long as the total energy-momentum
    conservation still applies, i.e., $\sum_{\indi} \bar Q^{\indi\,}{}_{\nu} = 0$.
    In this case, \eq{eq:totalcons} for the fluid $\indi$ reads
    \begin{align}
        \label{eq:fluidconsQ}
        \bar \rho_{\indi}' + 3 \mathcal{H} \left( 1 + w_{\indi} \right)
        \bar\rho_{\indi} = a^2 \bar Q^{\indi\,}{}^0 = - \bar Q^{\indi}_0
    \end{align}

    The perturbed energy-momentum tensor of the fluid $\indi$ has the components
    \begin{gather}
        \begin{aligned}
            \delta \mathcal{T}_{\!\indi\,} {{}^0}_0 &= -\delta \rho_{\indi}, \quad & \quad
            \delta \mathcal{T}_{\!\indi\,}{{}^i}_j &= \delta p_{\indi} \, \delta^i_j, \\ 
            \delta \mathcal{T}_{\!\indi\,}{{}^i}_0 &= - a^{-1} \left(\bar
                \rho_{\indi} + \bar p_{\indi} \right) \delta u_{\!\indi\,}{}^i, \quad & \quad
            \delta \mathcal{T}_{\!\indi\,}{{}^0}_i &= a^{-1} \left(\bar
                \rho_{\indi} + \bar p_{\indi} \right) \delta u_{\!\indi\,}{}_i, \\
            \delta \mathcal{T}_{\!\indi\,}{}_{00} &= a^2 \left( \delta
                \rho_{\indi} + 2 \bar\rho_{\indi} \phi \right), \quad & \quad
            \delta \mathcal{T}_{\!\indi\,}{}_{ij} &= a^2 \left( \delta p_{\indi}
                - 2 \bar p_{\indi} \phi \right) \delta_{ij},
        \end{aligned}    
    \end{gather}    
    and the perturbed energy-momentum conservation equations lead to the
    evolution equations for the density contrast and velocity perturbations
    \begin{subequations}
        \label{eq:dynamicalequations}
        \begin{align}
            \label{eq:dyndelta}
            - \delta_{\indi}' - \left[3 \mathcal{H} \left( c_{\!s\,\indi}^2 -
                    w_{\indi} \right) -
                \frac{\bar Q^{\indi}_{0}}{\bar\rho_{\indi}} \right]
            \delta_{\indi}  - \left(1 + w_{\indi} \right) \bigl(\theta_{\indi} -
            3 \phi'\bigr) &= \frac{\delta Q^{\indi}_{0}}{\bar\rho_{\indi}}, \\
            \label{eq:dyntheta}
            \theta_{\indi}' + \left[\mathcal{H} \left(1 -3 w_{\indi} \right) -
                \frac{\bar Q^{\indi}_{0}}{\bar\rho_{\indi}} +
                \frac{w_{\indi}'}{1+w_{\indi}} \right] \theta_{\indi} - k^2
            \phi - \frac{c_{\!s\,\indi}^2}{1+w_{\indi}} k^2 \delta_{\indi} &= \frac{i
                k^i \delta Q^{\indi}_{i}}{\bar\rho_{\indi} \left(1+w_{\indi}\right)},
        \end{align}
    \end{subequations}
    from $\nu = 0$ and $\nu = i$, respectively, using $\delta u_{\!\indi\,}{}_0 =
    -a \phi$, from the condition $g_{\mu\nu} u_{\!\indi\,}{}^{\mu}
    u_{\!\indi\,}{}^{\nu} = -1$.
    The density contrast (or overdensity) is defined as the relative density
    perturbation $\delta_{\indi} \equiv \delta \rho_{\indi}/\bar\rho_{\indi}$,
    $\theta_{\indi} \equiv a^{-1} i k^j \delta u_{\!\indi\,}{}_j$ is the
    divergence of the velocity perturbation in Fourier space, where $k^i$ are
    the components of the wavevector and $k^2 = \vecnotation{k} \cdot
    \vecnotation{k}$;
    $\delta Q^{\indi}_{\mu}$ are the perturbations to the exchange of
    energy-momentum in the perturbed conservation equations and
    $c_{\!s\,\indi}^2 \equiv \delta p_{\indi}/\delta\rho_{\indi}$ is the sound
    speed of the fluid $\indi$.
    Another useful equation is obtained from the perturbed time-time field
    equation,
    \begin{align}
        \label{eq:PoissonFullEq}
        k^2 \phi + 3\mathcal{H}^2 \phi &= -3 \mathcal{H} \phi' -4
        \uppi G a^2 \bar\rho \delta .
        %\left(1 + 3\mathcal{H}^2/k^2 \right) k^2 \phi &= -3 \mathcal{H} \phi' -4
        %\uppi G a^2 \bar\rho \delta .
    \end{align}
    This is the relativistic Poisson equation in Fourier space. It relates the
    potential $\phi$ in the metric to the total density perturbation $\delta
    \rho = \bar \rho \delta = \sum_{\indi} \bar\rho_{\indi} \delta_{\indi}$.

    \subsection{Growth function and growth rate}
    Structures form in the universe in the Newtonian regime of \gls{gr}, on
    spatial scales much smaller than the horizon, i.e., $k \gg \mathcal{H}$, and
    with negligible time variation of the gravitational potential.
    This allows us to discard the second term in the left-hand side of
    \eq{eq:PoissonFullEq} and also the term proportional to $\phi'$.
    Additionally, the sound speed of dark energy can be supposed large enough so
    that \gls{de} perturbations are smoothed out on sub-horizon scales
    \citep{Maartens2013}.
    The Poisson equation thus reduces to
    \begin{align}
        k^2 \phi = - 4 \uppi G a^2 \bar \rho \delta.
    \end{align}
    Considering the universe composed of matter and dark energy only, without
    interaction, we can also use \eq{eq:Friedmannrho} and write
    \begin{align}
        \label{eq:PoissonMatter}
        k^2 \phi = - \frac{3}{2} \mathcal{H}^2 \Omega_{\matter}
        \delta_{\matter}.
    \end{align}
    We now take the time derivative of \eq{eq:dyndelta} to replace
    $\theta_{\matter}'$ in \eq{eq:dyntheta} and, with \eq{eq:PoissonMatter}, get
    the evolution equation for the matter perturbations
    \begin{align}
        \label{eq:matterSOE}
        \delta_{\matter}'' + \mathcal{H} \delta_{\matter}' - \frac{3}{2}
        \mathcal{H}^2 \Omega_{\matter} \delta_{\matter} = 0.
    \end{align}
    
    In order to solve \eq{eq:matterSOE}, one needs to know the evolution of $a$
    or $\mathcal{H}$ with time.
    It is interesting to note, nonetheless, that this equation for
    $\delta_{\matter}$ does not involve derivatives with respect to spatial
    coordinates nor dependence on $\vecnotation{x}$.
    This allows us to decompose the solution separating the spatial and time
    dependences.
    Therefore, the general solution of \eq{eq:matterSOE} will have the form
    \begin{align}
        \label{eq:deltamodes}
        \delta_{\matter}(\tau, \vecnotation{x}) =
        \varepsilon_{\matter}^{\tplus}(\vecnotation{x}) D_{\matter}^{\tplus}(\tau) +
        \varepsilon_{\matter}^{\tminus}(\vecnotation{x}) D_{\matter}^{\tminus} (\tau),
    \end{align}
    a linear combination of the two particular solutions, the growing mode
    $D_{\matter}^{\tplus}$ and the decaying mode $D_{\matter}^{\tminus}$.
    At late times, the decaying mode becomes irrelevant as the increasing
    solution dominates.
    The functions $\varepsilon_{\matter}^{\tplus}$ and
    $\varepsilon_{\matter}^{\tminus}$ correspond to the density contrast field
    at some time that can be arbitrarily chosen according to the normalization
    of $D_{\matter}^{\tplus}$ and $D_{\matter}^{\tminus}$.
    For example, we can take $\varepsilon_{\matter}^{\tplus}(\vecnotation{x})$ to be the
    current density perturbation divided by the growth function today
    $\delta_{\matter,0}(\vecnotation{x})/D_{\matter,0}^{\tplus}$, so that
    (neglecting the decaying mode)
    \begin{align}
        \label{eq:backprop}
        \delta_{\matter}(z, \vecnotation{x}) =
        \frac{\delta_{\matter,0}(\vecnotation{x})}{D_{\matter,0}^{\tplus}}
        D_{\matter}^{\tplus}(z)
        \equiv \delta_{\matter,0}(\vecnotation{x}) \mathcal{D}_{\matter}(z;0),
    \end{align}
    with the last equality defining the backward propagation function
    $\mathcal{D}_{\matter}(z;0) \equiv \frac{D_{\matter}^{\tplus}(z)}
    {D_{\matter,0}^{\tplus}}$ for the evolution of the matter perturbation from
    redshift zero to $z$, with the implicit assumption that the evolution
    remains linear until the present epoch.

    It is also convenient to define the linear matter growth rate
    \begin{align}
        f(z) \equiv \frac{\ud \ln \delta_{\matter}}{\ud \ln a} \quad \text{or} \quad
        f(z) \equiv \frac{\ud \ln D_{\matter}^{\tplus}(z)}{\ud \ln a} 
    \end{align}
    and analyze \eq{eq:matterSOE} in terms of $f(z)$
    to simplify the study of \acrlongpl{rsd} and growth of structures, as we are
    going to do in chapters~\ref{ch:rsd} and \ref{ch:coupling}.
    In terms of the growth rate, the linearized continuity equation
    $\delta_{\matter}' + \theta_{\matter} = 0$ [\eq{eq:dyndelta}] is then
    \begin{align}
        \label{eq:flincont}
        \mathcal{H} a f \delta_{\matter} + \theta_{\matter} = 0.
    \end{align}

    \subsubsection{Solutions of the growth function}
    The solution of \eq{eq:matterSOE} can be found following an argument based
    on the Birkhoff's theorem, which says that different parts of the universe
    can be imagined to evolve as independent homogeneous universes
    \citep{Peebles1993}.
    We change back to the cosmic time in order to use this method.
    The local expansion parameter for an observer at $\vecnotation{x}$ differs
    from the mean background parameter $\bar a$ by a small quantity $\epsilon$ as
    \begin{align}
        a(t, \vecnotation{x}) = \bar a(t) \left[ 1 - \epsilon(t,
            \vecnotation{x}) \right].
    \end{align}
    That is, $a$ is the scale factor of a homogeneous universe with parameters
    slightly different from those of the homogeneous universe characterized by
    the scale factor $\bar a$.
    \Eq{eq:energyevolution} implies that
    \begin{align}
        \rho_{\matter} a^3 = \bar \rho_{\matter} \left( 1 + \delta_{\matter} \right) \bar a^3 \left(1 - \epsilon \right)^3 = \bar \rho_{\matter} \bar a^3 
    \end{align}
    and then 
    \begin{align}
        \delta_{\matter} = 3 \epsilon,
    \end{align}
    up to first order in $\delta_{\matter}$ and $\epsilon$.
    The perturbation to the matter density fluid is the fractional difference of the
    densities of the two slightly different universes and is three times the
    fractional difference of the expansion parameters.

    Let us suppose we have a family of functions $a(t, \alpha)$, labelled by the
    parameter $\alpha$, that are solutions to the scale factor in the Friedmann
    equations of different homogeneous universes.
    Then 
    \begin{align}
        \label{eq:deltaparalpha}
        \epsilon = - \frac{\delta a}{a} = -
        \frac{\delta \alpha}{a} \frac{\partial a}{\partial \alpha} \qquad
        \text{and} \qquad
        \delta = - 3 \frac{\delta \alpha}{a} \frac{\partial a}{\partial \alpha}.
    \end{align}
    The second Friedmann equation
    \begin{align}
        \ddot a = - \frac{4 \uppi G}{3}  \bar \rho_{\matter} a
            + \frac{\Lambda}{3} a 
    \end{align}
    integrated in $a$ gives
    \begin{align}
        \label{eq:defX}
        \dot a^2 = X(a), \qquad \text{with} \qquad
        X(a) \equiv \frac{8\uppi G}{3} \bigl( \bar \rho_{\matter} a^3 \bigr)
        \frac{1}{a} + \frac{\Lambda}{3} a^2 + \mathcal{K}.
    \end{align}
    This is the first Friedmann equation.
    Comparison with~\eq{eq:FriedmannEq} reveals the constant of
    integration $\mathcal{K}$ is related to the spatial curvature through
    $\mathcal{K} = - K/R^2$.
    Integrating \eqref{eq:defX} in time gives
    \begin{align}
        \label{eq:tsolution}
        t = \int \frac{\ud a}{X^{1/2}} + t_c,
    \end{align}
    where $t_c$ is a second constant of integration.
    $t_c$ and $\mathcal{K}$ can be thought as parameters distinguishing the
    neighbouring universes.
    Differentiating \eq{eq:tsolution} with respect to $\mathcal{K}$ and
    $t_c$, keeping $t$ fixed, we get
    \begin{gather}
        \frac{\ud t}{\ud \mathcal{K}} = \frac{\partial a}{\partial \mathcal{K}} \frac{\partial}{\partial a} \int \frac{\ud a}{X^{1/2}}  + \frac{\partial X}{\partial \mathcal{K}} \frac{\partial}{\partial X} \int \frac{\ud a}{X^{1/2}} + \frac{\ud t_c}{\ud \mathcal{K}} \nonumber \\
        0 = X^{-1/2} \frac{\partial a}{\partial \mathcal{K}} - \frac{1}{2} \frac{\partial X}{\partial \mathcal{K}} \int \frac{\ud a}{X^{3/2}} + 0 \Rightarrow \nonumber \\
        \Rightarrow \frac{\partial a}{\partial \mathcal{K}} = \frac{X^{1/2}}{2} \int \frac{\ud a}{X^{3/2}};
    \end{gather}
    and
    \begin{gather}
        \frac{\ud t}{\ud t_c} = \frac{\partial a}{\partial t_c} \frac{\partial}{\partial a} \int \frac{\ud a}{X^{1/2}} + \frac{\partial X}{\partial t_c} \frac{\partial}{\partial X} \int \frac{\ud a}{X^{1/2}} + \frac{\ud t_c}{\ud t_c} \nonumber \\
            0 = X^{-1/2} \frac{\partial a}{\partial t_c} + 0 + 1 \Rightarrow \frac{\partial a}{\partial t_c} = - X^{1/2},
    \end{gather}
    thus giving, from \eq{eq:deltaparalpha} with $\alpha = \mathcal{K}$,
    \begin{align}
        \label{eq:growthdelta}
        \delta^{\tplus}(t, \vecnotation{x}) = - \frac{3}{2} \delta \mathcal{K}
        \frac{X^{1/2}}{a} \int \frac{\ud a}{X^{3/2}}
    \end{align}
    and, with $\alpha = t_c$,
    \begin{align}
        \label{eq:decaydelta}
        \delta^{\tminus}(t, \vecnotation{x}) = 3 \delta t_c \frac{X^{1/2}}{a}.
    \end{align}
    $\delta^{\tplus}$ and $\delta^{\tminus}$ are the growing and decaying mode
    of the matter density field, corresponding to the terms
    $\varepsilon_{\matter}^{\tplus}(\vecnotation{x}) D_{\matter}^{\tplus}(t)$
    and
    $\varepsilon_{\matter}^{\tminus}(\vecnotation{x})
    D_{\matter}^{\tminus}(t)$ in \eq{eq:deltamodes}, respectively.
    One can verify that they satisfy $\ddot \delta_{\matter} + 2 H \dot
    \delta_{\matter} - \frac{3}{2} H^2 \Omega_{\matter} \delta_{\matter} = 0$,
    which is equivalent to \eqref{eq:matterSOE} but expressed in terms of the
    cosmic time, and are indeed the solutions to the matter density contrast
    equation.

    One interesting case is the simple \gls{eds} cosmology---a matter-only
    universe with the flat \gls{flrw} metric.
    In this cosmology, $X = \frac{8 \uppi G}{3} \frac{\bar \rho_{\matter}
        a^3}{a} \propto a^{-1}$ giving $D_{\matter}^{\tplus} \propto a \propto
    t^{2/3}$ from \eq{eq:growthdelta} and $D_{\matter}^{\tminus} \propto
    a^{-3/2} \propto t^{-1}$ from \eq{eq:decaydelta}, the time dependence coming
    from \eq{eq:defX}.
    The growth rate $f(a) = \frac{\ud \ln D_{\matter}^{\tplus}}{\ud \ln a}$ in
    the \gls{eds} universe is constant and equal to \num{1}.

    \section{The matter correlation function and power spectrum}
    \label{ss:corrpower}
    We now define a quantity of extreme importance for confronting
    theory and observations.
    In practice, in order to compare theory with observations, one cannot
    compare a map of galaxies generated by simulations to the actual observed
    distribution of galaxies.
    Instead, these tests are done by comparing their statistical properties.
    The key quantity is the two-point correlation function
    $\xi(\vecnotation{r})$, or autocorrelation function of the density
    field,\footnote{In this section $\vecnotation{r}'$ is a point of space just as
        $\vecnotation{r}$, not to be confused with $\ud \vecnotation{r}/\ud
        \tau$.}
    \begin{align}
        \label{eq:defxicorr}
        \xi_{\matter}(\vecnotation{r}) \equiv \bigl\langle
        \delta_{\matter}(\vecnotation{r}') \delta_{\matter}(\vecnotation{r}' +
        \vecnotation{r})
        \bigr\rangle = \frac{1}{V} \int_{V} \delta_{\matter}(\vecnotation{r}')
        \delta_{\matter}(\vecnotation{r}' + \vecnotation{r}) \, \ud \vecnotation{r}',
    \end{align}
    which is an average of the product of the density contrast at two points
    separated by $\vecnotation{r}$ over some volume $V$ (see \citet{Peebles1980}
    and references therein).
    Isotropy of the universe actually implies that $\xi_{\matter}$ depends only on the
    modulus $r$ of the vector $\vecnotation{r}$. We may then write $\xi_{\matter}(r)$
    instead.
    Of course the correlation function also depends on $\tau$, as the
    inhomogeneities evolve with time. However, we omit this dependence in this
    section for simplicity of notation.

    In Fourier space, the density contrast is the Fourier transform of
    $\delta_{\indi}(\vecnotation{r})$ (now denoting both the matter field
    $\delta_{\matter}$ and the galaxy field $\delta_{\gal}$),
    \begin{align}
        \label{eq:Fmode}
        \hat \delta_{\indi}(\vecnotation{k}) \equiv \int \ud \vecnotation{r}
        \, e^{- i \vecnotation{k} \cdot \vecnotation{r}}
        \delta_{\indi}(\vecnotation{r}).
    \end{align}
    We leave the normalization factor in the inverse Fourier transform,
    \begin{align}
        \label{eq:invFmode}
        \delta_{\indi}(\vecnotation{r}) = \left(2 \uppi \right)^{-3} \int \ud
        \vecnotation{k} \, e^{i \vecnotation{k} \cdot \vecnotation{r}}
        \hat\delta_{\indi}(\vecnotation{k}).
    \end{align}
    The covariance of two Fourier modes is
    \begin{align}
        \bigl\langle \hat\delta_{\indi}(\vecnotation{k}_1)
        \hat\delta_{\indi}(\vecnotation{k}_2) \bigr\rangle =
        \int \ud \vecnotation{r}_1 \int \ud \vecnotation{r}_2 \, e^{- i
            \vecnotation{k}_1 \cdot \vecnotation{r}_1} e^{- i \vecnotation{k}_2
            \cdot \vecnotation{r}_2} \bigl\langle
        \delta_{\indi}(\vecnotation{r}_1) \delta_{\indi}(\vecnotation{r}_2)
        \bigr\rangle .
    \end{align}
    Changing the variable of integration $\vecnotation{r}_1$ to
    $\vecnotation{r} = \vecnotation{r}_1 - \vecnotation{r}_2$ makes the
    integrand $e^{-i \vecnotation{k}_1 \cdot \vecnotation{r}_1} \times e^{-i
        \vecnotation{k}_2 \cdot \vecnotation{r}_2} \xi_{\indi}(r)$ equal to
    $e^{-i\vecnotation{k}_1 \cdot \vecnotation{r}} e^{-i (\vecnotation{k}_1 +
        \vecnotation{k}_2) \cdot \vecnotation{r}_2} \xi_{\indi}(r)$, which upon
    integration in $\vecnotation{r}_2$ gives the three-dimensional Dirac delta
    function $\Dirac(\vecnotation{k}_1 + \vecnotation{k}_2)$, expressing the
    hypothetical translational invariance or statistical homogeneity.
    Hence,
    \begin{align}
        \label{eq:defPsk}
        \bigl\langle \hat\delta_{\indi}(\vecnotation{k}_1)
        \hat\delta_{\indi}(\vecnotation{k}_2) \bigr\rangle = \left(2 \uppi\right)^3
        \Dirac(\vecnotation{k}_1 + \vecnotation{k}_2) \int
        \ud \vecnotation{r} \, e^{-i \vecnotation{k}_1 \cdot \vecnotation{r}}
        \xi_{\indi}(r).
    \end{align}
    The remaining integral in \eq{eq:defPsk} is the Fourier transform (evaluated at
    $\vecnotation{k}_1$) of the correlation function $\xi_{\indi}(r)$.
    We define it as the matter power spectrum
    \begin{align}
        \mathcal{P}_{\!\indi}(k) \equiv \int \ud \vecnotation{r} \, e^{-i \vecnotation{k} \cdot \vecnotation{r}}
        \xi_{\indi}(r)
    \end{align}
    and may alternatively express it, after integrating the angular part, as
    \begin{align}
        \label{eq:altPS}
        \mathcal{P}_{\!\indi}(k) = 2\uppi \int_0^{\infty} \ud r \, r^2 \frac{\sin(k r)}{k
            r} \xi_{\indi}(r).
    \end{align}
    The power spectrum $\mathcal{P}_{\!\indi}$ depends only on $k$, the modulus of
    $\vecnotation{k}$, thus reflecting the statistical isotropy.
    Naturally, the correlation function is the inverse Fourier transform of the
    power spectrum:
    \begin{align}
        \xi_{\indi}(r) = (2 \uppi)^{-3} \int \ud^3 \vecnotation{k} \, e^{- i
            \vecnotation{k} \cdot \vecnotation{r}} \mathcal{P}_{\!\indi}(k) =
        \int_0^{\infty} \ud k \, k^2 \frac{\sin(k r)}{k r}
        \mathcal{P}_{\!\indi}(k).
    \end{align}

    The correlation function and the power spectrum are equivalent descriptions
    of the statistical properties of the inhomogeneities.
    The statistical properties of the matter density field are completely
    characterized by the two-point correlation function or the power spectrum if
    the fluctuations are Gaussian (which means that the phases of the Fourier modes
    $\hat\delta_{\matter} (\vecnotation{k})$ are uncorrelated and random as a
    consequence of the central limit theorem).
    The requirement is that the initial perturbations produced during inflation are Gaussian,
    since the linear evolution preserves the phases.
    Indeed, the primordial fluctuations have been shown highly Gaussian
    \citep{Leistedt2014,Planck2013NG}.
    However, in case non-Gaussianities are eventually detected, the three-point
    correlation function (or equivalently its Fourier space counterpart, the
    bi\-spectrum) and higher order moments may be necessary to describe
    completely the statistical properties of the density field.

    \subsection{The galaxy correlation function, number density and bias}
    \label{ss:corrnumbias}
    The matter density field cannot be directly observed since its composed
    mostly of dark matter.
    Instead, we can directly see the galaxies and study their discrete
    distribution, which is expected to trace the underlying matter field.
    This idea was introduced by \citet{Kaiser1984} in \citeyear{Kaiser1984} to
    explain the properties of Abell clusters, despite the already known fact
    that clustering properties of galaxies vary with their morphology
    \citep{Dressler1980,Postman1984}, so they cannot all be good tracers of the
    mass distribution \citep{WMAP1yearmethodology2003}.
    This situation is eased, however, by the galaxy distribution, initially very
    biased when they were formed at high density regions of the matter
    fluctuation field, becoming less and less biased with time as its
    gravitational evolution takes place.

    The galaxy overdensity is defined in terms of a mean galaxy number density
    $\bar n_{\gal}$ rather than an energy density,
    \begin{align}
        \delta_{\gal}(\tau, \vecnotation{r}) \equiv \frac{n_{\gal}(\tau,
            \vecnotation{r}) - \bar n_{\gal}(\tau)}{\bar
            n_{\gal}(\tau)}.
    \end{align}
    The relation between the galaxy and the total matter distribution is made by
    the galaxy bias $b(\tau,\vecnotation{r})$,
    \begin{align}
        \label{eq:generalbias}
        \delta_{\gal}(\tau,\vecnotation{r}) = b (\tau,\vecnotation{r})
        \delta_{\matter}(\tau,\vecnotation{r}).
    \end{align}
    The bias is a consequence of the non-linear nature of galaxy formation.
    Several different and complicated biasing schemes have been introduced in the literature.
    The simplest form of bias is a constant $b$, so the galaxy density contrast $\delta_{\gal}$ is linearly biased, $\delta_{\gal} = b \delta$.
    This assumption is justified by the indication that, on sufficiently large scales, galaxy bias is scale independent \citep{Hoekstra2002,Verde2002}.
    The galaxy velocity field, on the other hand, follows exactly the matter
    velocity field, $\vecnotation{v}_{\gal} = \vecnotation{v}_{\matter}$.
    The correlation function and the power spectrum of galaxies will then be
    related to their matter counterparts by
    \begin{align}
        \label{eq:biastwopointfunctions}
        \xi_{\gal} = b^2 \xi_{\matter} \quad \text{and} \quad \mathcal{P}_{\!\gal} = b^2
        \mathcal{P}_{\!\matter}.
    \end{align}

    The probability of finding a galaxy centered in a random volume element
    $\ud V$ is $\bar n_{\gal} \, \ud V$.
    The galaxy two-point correlation function $\xi_{\gal}(r) = \bigl\langle
    \delta_{\gal}(\vecnotation{r}') \delta_{\gal}(\vecnotation{r}' +
    \vecnotation{r}) \bigr\rangle$ can be interpreted as the excess probability,
    compared to that of a random distribution, that a pair of galaxies can be
    found at a distance $r$, as this probability can be written as $\ud P = \bar
    n_{\gal}^2 \bigl[ 1 + \xi_{\gal}(r) \bigr] \, \ud V_1 \ud V_2$.

    \chapter{Redshift-space distortions}
    \label{ch:rsd}

    The distances of galaxies from us are usually inferred through the measured
    redshift, by the conversion of the redshift velocity to distance via
    Hubble's law.
    However, this is not exact. As we have seen in chapter~\ref{ch:LCDM}, the
    formation of structures due to gravitational instabilities induce galaxies
    to have peculiar velocities that distort the uniform Hubble flow.
    What we actually measure with the redshift is the sum of the two
    contributions to the velocity of galaxies---the Hubble expansion velocity
    and the line-of-sight projection of the peculiar velocity,
    \begin{align}
        c z = H r + \vecnotation{v} \cdot \vers{r}.
    \end{align}
    Therefore, the direct interpretation of the distance as $c z / H$ is
    contaminated by the extra term $v/H \equiv \vecnotation{v} \cdot \vers{r} /
    H$ due to the peculiar velocity. 
    The displacements of galaxies relative to their true positions
    in the redshift space when they possess peculiar velocity along the line of
    sight are what we call \acrlongpl{rsd} (\glspl{rsd}).

    Peculiar velocities constitute a powerful cosmic probe since they are
    related to the growth rate of structures [\eq{eq:flincont}].
    Redshift surveys have been used to constrain cosmological parameters.
    \citet{Peacock2001Nature} measure the amount of matter in the universe from
    clustering in the \gls{2df}.
    \citet{WMAP1yearmethodology2003} also use the \gls{2df} data to complement
    their analysis of the \gls{cmb} for parameter estimation from the \gls{wmap}
    observations with the matter power spectrum of the nearby universe.
    In this chapter, we will see how the \glspl{rsd} look like in redshift space
    and how they affect the statistical properties that we studied in
    section~\ref{ss:corrpower}, which is essential to learn how to compare 
    theoretical predictions with observations of \gls{rsd}.

    \section{Seeing the distortions in redshift space}
    \label{sec:distortionsinRS}
\begin{figure}[tb]
        \centering
        \includegraphics[width=\textwidth]{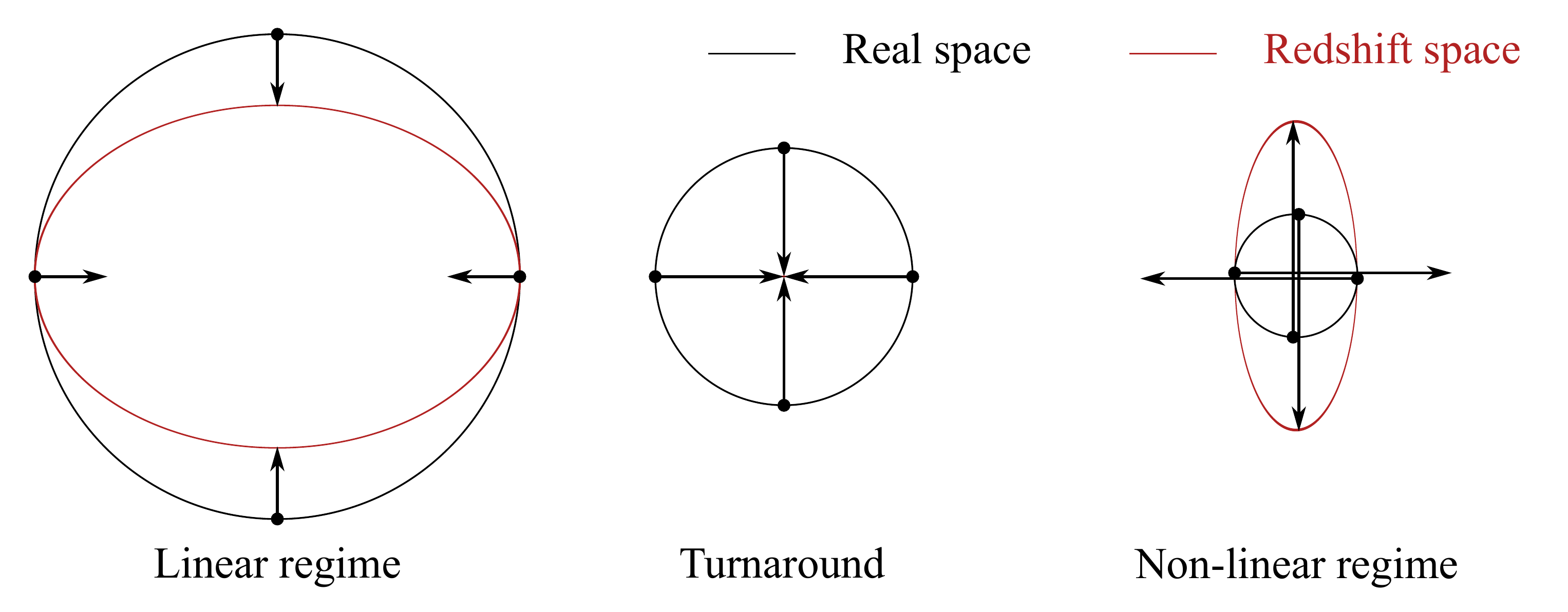}
        \caption{In the linear regime, a spherical density contrast appears squashed along the line of sight in redshift space. At smaller scales, velocities tend to be larger, originating the so-called fingers-of-god, appearing to be turned inside out. There is an intermediate point of turnaround, where the density contrast shell appears collapsed, when peculiar velocities in the line of sight exactly cancel the Hubble velocity.}
        \label{fig:RSDscheme}
    \end{figure}

    We show schematically in figure~\ref{fig:RSDscheme} how the \glspl{rsd} look
    like in redshift space for a radially symmetric distribution of galaxies.
    The galaxies, represented by dots, are falling towards the center with
    peculiar velocities represented by the arrows in the schematic image.
    All the galaxies on the same black circle (real space) have the same total
    peculiar velocity $\lvert \vecnotation{v} \rvert$.
    The observer is far away at some point below the undermost galaxy
    represented. 
    Then, due to the effect produced by peculiar velocities, galaxies will
    appear, in redshift space, at the positions represented by the red ellipses.
    Coherent infall velocities of galaxies between the center of the
    distribution and the observer will add to the Hubble expansion, while the
    velocities of those galaxies behind the center of the density contrast will
    subtract from the Hubble flow.

\begin{figure}[bt]
        \centering
        \includegraphics[width=\textwidth,trim=1.6cm 8.6cm 1.6cm 8.6cm,clip=true]{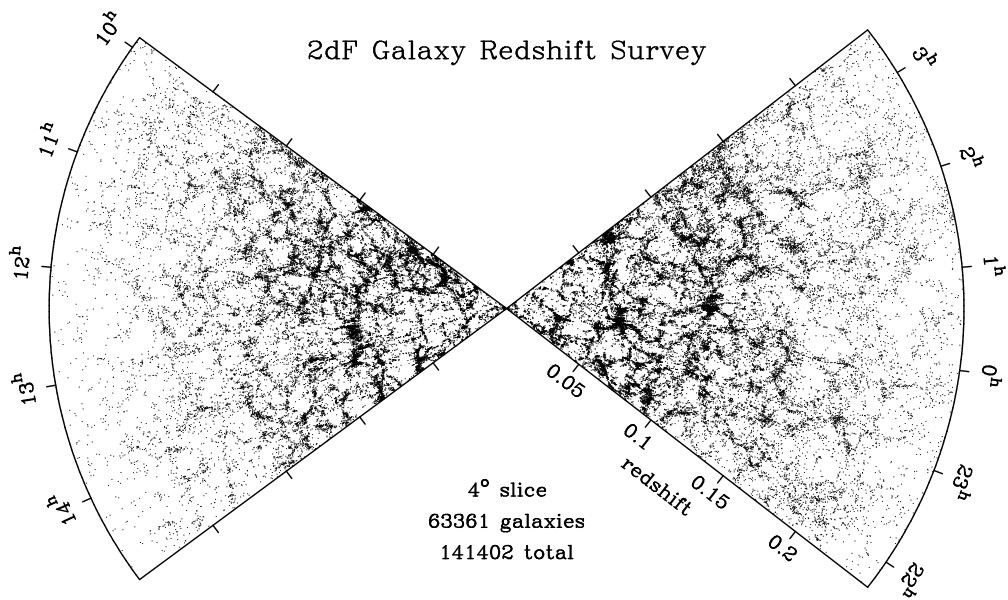}
        \caption{Distribution of galaxies in part of the 2dF galaxy survey, from \citet{Peacock2001Nature} (not all redshifts had been obtained at that time). The slices are \SI{4}{\degree} thick, centered at declination \SI{-2.5}{\degree} in the northern galactic pole and \SI{-27.5}{\degree} in the southern galactic pole.}
        \label{fig:slices}
    \end{figure}

    We can distinguish two regime scales.
    On large scales, for galaxies far from the center of the distribution, the
    distortion tends to be small because the gravitational pull is relatively
    weak.
    The distribution appears squashed along the line of sight.
    This is the so-called Kaiser effect.
    The situation changes at smaller scales, where virialized non-linear motions
    of galaxies closer to the center are composed of peculiar velocities that
    can even surpass the cosmic Hubble flow velocity, thus producing a smearing
    effect known as fingers-of-god.
    There is an intermediate point between the two regimes where the peculiar
    velocities exactly cancel out the Hubble flow velocity.
    In a real galaxy survey all these features are
    present, besides some other complicated characteristics about the coverage
    and the selection of galaxies.
    For example, the \gls{2df} \citep{tdfgrs2001,Peacock2001Nature}
    (figure~\ref{fig:slices}) is a survey limited by magnitude; most nearby
    galaxies are included in the catalog but, as the distance increases, only
    the brighter galaxies are selected because of the flux-limited window
    function.

    \section{The linear theory}
    \label{sec:lineartheoty}
    We have already deduced the continuity and Euler equations
    [eqs.~\eqref{eq:dynamicalequations}] for a (pressureless) fluid of matter in
    an expanding background in section~\ref{sec:evequations}.
    Before applying those results, a few important adjustments must be made.
    Those results give a good description of the clustering of the matter distribution
    field, which is composed mostly of dark matter. 
    Of course we do not observe dark matter directly, but rather discrete
    galaxies, which can at least be assumed to trace the underlying distribution
    of matter, so the statistical properties of the structure of the universe
    can be studied by observing the galaxy distribution.
    Thus the study of the statistical properties of the galaxy distribution is
    of great importance for uncovering the features of the large-scale structure
    of the universe.
    
    \subsection{The redshift-space distortion parameter}
    With the linear bias assumption $\delta_{\gal} = b(z) \delta_{\matter}$, the continuity equation \eqref{eq:flincont}
    for galaxies reads
    \begin{align}
        \label{eq:vdelta}
        \mathcal{H} a f \frac{\delta_{\gal}}{b} + \theta_{\gal} =
        \mathcal{H} a \, \beta \, \delta_{\gal} + \theta_{\gal} = 0,
    \end{align}
    where have we introduced the so-called \acrlong{rsd} parameter
    \begin{align}
        \label{eq:betarsd}
        \beta(z) \equiv \frac{f(z)}{b(z)}.
    \end{align}
    This definition is more general than the simple constant bias.
    Nevertheless, it is usual to extract measurements of $b$, $f$ and $\beta$ assumed
    constant within redshift bins.

    \subsection{The correlation function and the power spectrum in redshift space}
    With the redshift-space galaxy density contrast
    $\delta_{\gal}^{\redspace}(\vecnotation{s})$ adequately defined in all
    redshift space, inside and outside the region of the survey
    \citep{Hamilton_1998}, the redshift-space correlation function
    $\xi_{\gal}^{\redspace}$ is defined by
    \begin{align}
        \xi_{\gal}^{\redspace}(s_{12}, s_1, s_2) \equiv \bigl\langle
        \delta_{\gal}^{\redspace}(\vecnotation{s}_1) \delta_{\gal}^{\redspace}(\vecnotation{s}_2) \bigr\rangle.
    \end{align}
    Unlike in the real-space case, the redshift-space correlation function
    depends not only on the separation $s_{12} \equiv \lvert \vecnotation{s}_1 -
    \vecnotation{s}_2 \rvert$ of the galaxies in redshift space, but also in the
    redshift distances $s_1$ and $s_2$.
    Part of the symmetry present in $\xi_{\gal}(r_{12})$ is broken in redshift
    space due to the redshift distortions caused by the peculiar velocities of
    galaxies and possibly due to the heterogeneity of the selection function
    over different regions.
    However, the rotational symmetry about the observer at $s = 0$ is preserved,
    since the selection function does not depend on the direction of the vectors
    $\vecnotation{s}_1$ and $\vecnotation{s}_2$.

    If the angle between $\vecnotation{s}_1$ and $\vecnotation{s}_2$ is small
    enough, then $\xi_{\gal}^{\redspace}$ depends only on the parallel and
    perpendicular to the line of sight $\vecnotation{z}$ components
    $s_{\parallel}$ and $s_{\perp}$ of the separation $s_{12}$.
    This is the plane-parallel (or distant-observer) approximation.

    We also define the redshift Fourier modes $\hat
    \delta^{\redspace}(\vecnotation{k})$ in the same way of
    eqs.~\eqref{eq:Fmode} and \eqref{eq:invFmode}:
    \begin{align}
        \hat \delta_{\gal}^{\redspace}(\vecnotation{k}) &= \int \ud
        \vecnotation{s} \, e^{-i \vecnotation{k} \cdot \vecnotation{s}}
        \delta_{\gal}^{\redspace}(\vecnotation{s}), \\
        \delta_{\gal}^{\redspace}(\vecnotation{s}) &= (2\uppi)^{-3} \int \ud
        \vecnotation{k} \, e^{i \vecnotation{k} \cdot \vecnotation{s}} \hat
        \delta_{\gal}^{\redspace}(\vecnotation{k}).
    \end{align}
    The redshift-space power spectrum, similarly to the real-space case, is the Fourier transform of the redshift-space correlation function. In the plane-parallel approximation,
    \begin{align}
        \mathcal{P}_{\gal}^{\redspace}( k_{\parallel}, k_{\perp} ) \equiv \int \ud
        \vecnotation{s} \, e^{-i \vecnotation{k} \cdot \vecnotation{s}}
        \xi_{\gal}^{\redspace}(s_{\parallel}, s_{\perp} ),
    \end{align}
    where $k_{\parallel}$ and $k_{\perp}$ are the parallel and perpendicular to the line-of-sight components of the wavevector $\vecnotation{k}$.

\begin{figure}[bt]
        \centering
        \includegraphics[width=0.6\textwidth,trim=0cm 0cm 7.3cm 13.7cm,clip=true]{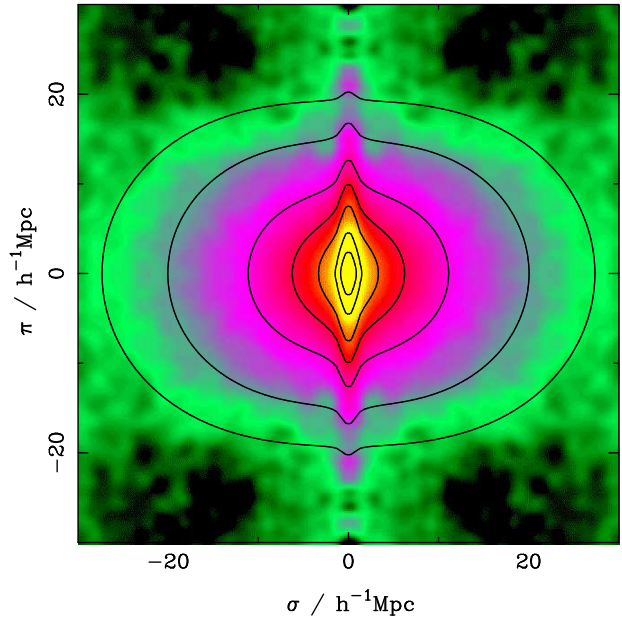}
        \caption{The redshift-space correlation function
            $\xi_{\gal}^{\redspace}(\sigma,\pi)$ for the \gls{2df} as a function of
            the transverse ($\sigma$) and radial ($\pi$) pair separations, from
            \citet{Peacock2001Nature}. The data are measured only in the first
            quadrant and mirrored in both axes for the purpose of
            illustrating deviations from circular symmetry. Redshift distortions
            are clearly identified, both the fingers-of-god elongations at small
            scales and the Kaiser effect at large distances. Superimposed are
            the contours $\xi_{\gal}^{\redspace} = \numlist[list-final-separator =
            {\text{ and }}]{10;5;2;1;0.5;0.2;0.1}$ showing model predictions
            with $\beta = \num{0.4}$. A pairwise velocity dispersion of
            \SI{400}{\km\per\s} was used for modelling the small-scale effects.}
        \label{fig:corr2df}
    \end{figure}

    As we saw in section~\ref{sec:distortionsinRS}, and now with the knowledge
    of the correlation function, we expect the contours of
    $\xi_{\gal}^{\redspace}$ to be compressed along the line of sight by
    galaxies falling into overdense regions. 
    This distortion of $\xi_{\gal}^{\redspace}$ on large scales is known as the
    Kaiser effect. 
    This effect offers a method for measuring the \acrlong{rsd} parameter
    $\beta$, which will then be detectable through the correlation function or
    the power spectrum.
    Figure~\ref{fig:corr2df} illustrates the distorted redshift-space
    correlation function computed for the \gls{2df}.

    \subsection{From real space to redshift space}
    Starting from the conservation of the number of galaxies in a survey,
    $n^{\redspace}(\vecnotation{s}) \, \ud \vecnotation{s} = n(\vecnotation{r})
    \, \ud \vecnotation{r}$, where $\vecnotation{r}$ and $\vecnotation{s} =
    \vecnotation{r} + \frac{1}{H_0} \vecnotation{v} \cdot \vers{r}$ are
    the positions in real and redshift space, respectively, and
    $\vecnotation{v}$ is the peculiar velocity field, it is possible to show
    \citep{Kaiser_1987,Hamilton_1998} that the observed Fourier modes of the galaxy density
    contrast in redshift space are related to the Fourier modes in real space by
    \begin{align}
        \hat \delta^{\redspace}_{\gal}(\vecnotation{k}) = \left( 1 + \beta \mu^2_{kz} \right)
        \hat \delta_{\gal}(\vecnotation{k})
    \end{align}
    in the plane-parallel (distant observer) approximation, where $\mu_{kz}
    \equiv \frac{1}{k} \vecnotation{k} \cdot \vers{z}$ is the cosine of the
    angle between the wavevector $\vecnotation{k}$ and the line-of-sight
    direction $\vers{z}$.
    It follows immediately that the power spectrum in redshift space is
    amplified by the square of that factor over the power spectrum in real
    space,
    \begin{align}
        \label{eq:powerspectrum}
        \mathcal{P}^{\redspace}_{\gal} (\vecnotation{k}) = \left( 1 + \beta \mu_{kz}^2
        \right)^2 \mathcal{P}_{\gal} (k).
    \end{align}
    Galaxies moving perpendicularly to the observer have $\mu_{kz} = 0$ and are
    thus not affected.
    In the next section we show how the \gls{rsd} parameter $\beta$ can be
    measured from the redshift-space power spectrum.

    \section{Measuring the redshift-space distortion parameter}
    \label{sec:measuringRSDbeta}
    Here we will present some of the methods that have been used, for example, by \citet{Ratcliffe_1998}, to obtain a measurement of the redshift-space distortion parameter.

    \subsection{The ratio of quadrupole-to-monopole moments of the red\-shift-space power spectrum}
\begin{figure}[tb]
        \centering
        \includegraphics[width=0.7\textwidth,trim=2cm 7.1cm 3.3cm 6.5cm,clip=true]{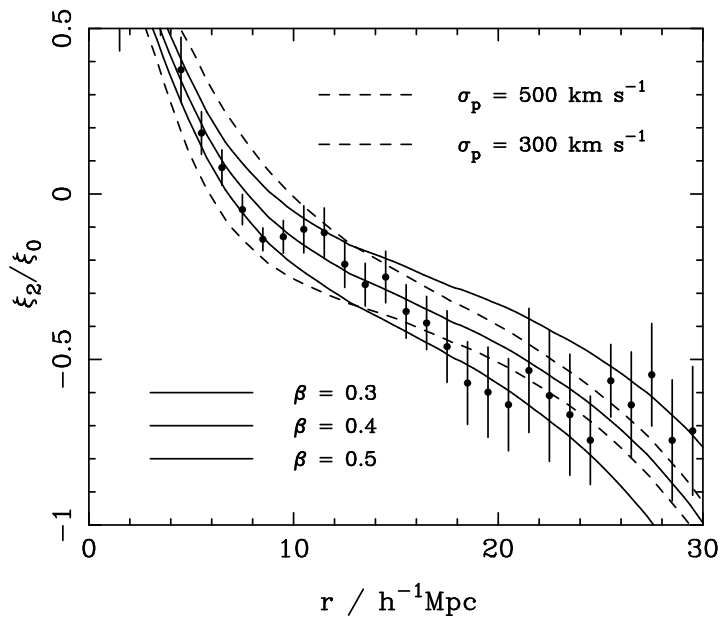}
        \caption{The quadrupole-to-monopole ratio $\xi^{\redspace}_2/\xi^{\redspace}_0$ as a function of the radii $r$, from \citet{Peacock2001Nature}. The quantity is positive at small scales, where the fingers-of-god effect dominates, and negative at large scales, dominated by coherent infall velocities. Solid lines show model predictions for $\beta = 0.3$, 0.4 and 0.5, with pairwise velocity dispersion $\sigma_{\text{p}} = \SI{400}{\km\per\s}$ for modelling the small-scale effects. The dashed lines are predictions for $\beta = 0.4$, with $\sigma_{\text{p}} = 300$ and \SI{500}{\km\per\s}. The ratio becomes more negative as $\beta$ increases and as $\sigma_{\text{p}}$ decreases.}
        \label{fig:qtmratio2df}
    \end{figure}

    In the plane-parallel approximation, one can write the redshift-space power
    spectrum as a sum of even harmonics $\mathcal{P}_{\ell}^{\redspace}(k)$:
    \begin{align}
        \mathcal{P}_{\gal}^{\redspace}(\vecnotation{k}) = \sum_{\text{$\ell$ even}} \mathscr{P}_{\ell} (\mu_{kz}) \mathcal{P}_{\ell}^{\redspace} (k),
    \end{align}
    where the harmonics are defined by
    \begin{align}
        \label{eq:harmonics}
        \mathcal{P}_{\ell}^{\redspace} \equiv \frac{2 \ell + 1 }{4 \uppi} \int
        \ud \Omega_{\vecnotation{k}} \, \mathscr{P}_{\ell}(\mu_{kz})
        \mathcal{P}_{\gal}^{\redspace}(\vecnotation{k}),
    \end{align}
    $\ud \Omega_{\vecnotation{k}}$ is the infinitesimal solid angle in Fourier
    space and $\mathscr{P}_{\ell}$ are the Legendre polynomials, given by
    \begin{align}
        \label{eq:Legendrepols}
        \mathscr{P}_{\ell}(\mu) = \frac{1}{2^{\ell} \ell!} \frac{\ud^{\ell}}{\ud
            \mu^{\ell}}
        \left[ \bigl( \mu^2 - 1 \bigr)^{\ell} \right].
    \end{align}
    The odd harmonics vanish by pair exchange symmetry and non-zero azimuthal
    harmonics ($Y_{\ell m}$ with $m \ne 0$) vanish by symmetry about the line of
    sight.
    In the linear regime, it can be shown that
    $\mathcal{P}_{\gal}^{\redspace}(\vecnotation{k})$ reduces to a sum of
    monopole, quadrupole and hexadecapole harmonics:
    \begin{align}
        \mathcal{P}_{\gal}^{\redspace}(\vecnotation{k}) = \mathscr{P}_0 (\mu_{kz}) \mathcal{P}_0^{\redspace}(k) + \mathscr{P}_2 (\mu_{kz}) \mathcal{P}_2^{\redspace}(k) + \mathscr{P}_4 (\mu_{kz}) \mathcal{P}_4^{\redspace}(k). 
    \end{align}
    Substituting \eq{eq:Legendrepols} into \eq{eq:harmonics}, it is easy to
    obtain each of these harmonics in terms of the true power spectrum.
    With $\mathscr{P}_0(\mu_{kz}) = 1$, $\mathscr{P}_2(\mu_{kz}) = \frac{1}{2}
    \bigl( 3 \mu_{kz}^2 - 1 \bigr)$ and $\mathscr{P}_4(\mu_{kz}) = \frac{1}{8}
    \bigl(3 - 30 \mu_{kz}^2 + 35 \mu_{kz}^4 \bigr)$, the monopole term is
    \begin{align}
        \mathcal{P}_{0}^{\redspace}(k) &= \frac{1}{4 \uppi} \int \ud
        \Omega_{\vecnotation{k}} \, \left(1 + \beta \mu_{kz}^2 \right)^2 \mathcal{P}_{\gal}(k) \nonumber \\
        &= \frac{1}{2} \int_{-1}^{1} \ud \mu_{kz} \, \left(1 + 2 \beta
            \mu_{kz}^2 + \beta^2 \mu_{kz}^4 \right) \mathcal{P}_{\gal}(k)
        \nonumber \\
        &= \left(1 + \frac{2}{3} \beta + \frac{1}{5} \beta^2 \right)
        \mathcal{P}_{\gal}(k),
    \end{align}
    the quadrupole is
    \begin{align}
        \mathcal{P}_2^{\redspace}(k) &= \frac{5}{4} \int_{-1}^{1} \ud \mu_{kz} \,
        \left(3 \mu_{kz}^2 - 1 \right) \left(1 + 2 \beta \mu_{kz}^2 + \beta^2
            \mu_{kz}^4 \right) \mathcal{P}_{\gal}(k) \nonumber \\
        &= \left( \frac{4}{3} \beta + \frac{4}{7} \beta^2 \right)
        \mathcal{P}_{\gal}(k)
    \end{align}
    and the hexadecapole is given by
    \begin{align}
        \mathcal{P}_4^{\redspace}(k) &= \frac{9}{16} \int_{-1}^1 \ud \mu_{kz} \,
        \left(3 - 30 \mu_{kz}^2 + 35 \mu_{kz}^4 \right) \left(1 - \beta \mu_{kz}^2
        \right)^2 \mathcal{P}_{\gal}(k) \nonumber \\
        &= \frac{8}{35} \beta^2 \mathcal{P}_{\gal}(k).
    \end{align}
    The hexadecapole harmonic is generally small and noisy, so it is of
    particular interest to compute the ratio of the quadrupole to the monopole
    harmonics of the redshift-space power spectrum:
    \begin{align}
        \frac{\mathcal{P}_2^{\redspace}(k)}{\mathcal{P}_0^{\redspace}(k)} = \frac{\frac{4}{3} \beta + \frac{4}{7} \beta^2}{1 + \frac{2}{3} \beta + \frac{1}{5} \beta^2}.
    \end{align}
    This same result applies for the quadrupole-to-monopole ratio of the
    redshift-space correlation function $\xi^{\redspace}_2/\xi^{\redspace}_0$.
    Thus measuring the ratio of quad\-ru\-pole-to-mon\-o\-pole moments of the
    redshift-space power spectrum (or correlation function) one can extract a
    measurement of the redshift-space distortion parameter $\beta$. The
    advantage of this method is that it uses quantities measured in redshift
    space only.
    Figure~\ref{fig:qtmratio2df} shows the results for the
    quadrupole-to-monopole ratio of the redshift-space correlation function for
    the \gls{2df} at different radii.

    \subsection{The ratio of redshift-space to real-space angle-averaged power spectra}
    The redshift-space power spectrum
    $\mathcal{P}_{\gal}^{\redspace}(\vecnotation{k})$ depends on the wavevector
    $\vecnotation{k}$ in Fourier space. 
    This can be explicitly put as a dependence on the modulus $k$ and on the
    angle of $\vecnotation{k}$ with the line-of-sight direction, that is,
    $\mathcal{P}_{\gal}^{\redspace}(\vecnotation{k}) =
    \mathcal{P}_{\gal}^{\redspace}(k, \mu_{kz})$.
    If we integrate over all $\mu_{kz}$ we get the angle-averaged redshift-space
    power spectrum $\mathcal{P}_{\gal}^{\redspace}(k)$ (the distinction with
    respect to the original redshift-space power spectrum
    $\mathcal{P}_{\gal}^{\redspace}(\vecnotation{k})$ is made clear by the
    explicit dependence on the modulus $k$ of the wavevector only):
    \begin{align}
        \mathcal{P}_{\gal}^{\redspace} (k) = \frac{\int_{-1}^{1} \ud \mu_{kz} \,
            \mathcal{P}_{\gal}^{\redspace} (k, \mu_{kz})}{\int_{-1}^{1} \ud \mu_{kz}}.
    \end{align}
    With $\mathcal{P}_{\gal}^{\redspace}(\vecnotation{k})$ from
    \eq{eq:powerspectrum}, this gives
    \begin{align}
        \mathcal{P}_{\gal}^{\redspace}(k) = \left[ 1 + \tfrac{2}{3} \beta +
            \tfrac{1}{5} \beta^2 \right] \mathcal{P}_{\gal}(k).
    \end{align}
    A similar expression also applies for the angle-averaged correlation
    function, by inverse Fourier transforming the equation above:
    \begin{align}
        \xi_{\gal}^{\redspace}(s) = \left[ 1 + \tfrac{2}{3} \beta + \tfrac{1}{5}
            \beta^2 \right] \xi_{\gal}(r).
    \end{align}
    Measurement of the real-space correlation function $\xi_{\gal}(r)$ is done
    by deprojecting the angular correlation function of a survey. 
    Since this process is known to be noisy, the method just discussed is more
    appropriate for large surveys.

    \chapter{The growth of structures in interacting dark energy models}
    \label{ch:coupling}
    We already obtained the evolution equations for fluid density and velocity
    perturbations in section~\ref{sec:evequations} in a general fashion,
    allowing the presence of an interaction term in the energy-momentum tensor
    conservation equation.
    Specially for the matter fluid, we combined those equations with the Poisson
    equation in sub-horizon scales to get a second order differential equation
    \eqref{eq:matterSOE} for $\delta_{\matter}$ in the non-interacting case. 
    Then, we have defined the growth rate of the matter perturbations as their
    logarithmic rate of change.
    We are now interested in seeing how the growth of structures is affected by
    the presence of interaction between dark energy and dark matter.
    The following study produced an interesting result from the theoretical
    point of view and has been submitted to the Journal of Cosmology and
    Astroparticle Physics for publication \citep{Marcondes2016}.

    The fact that the evolution of the growth rate can be solved
    approximately in an analytic form, 
    \begin{align}
        \label{eq:fOmegagamma}
        f(z) \approx \bigl[ \Omega_{\matter}(z) \bigr]^{\gamma(z)},
    \end{align}
    is remarkable.
    The approximation was first proposed by \citeauthor{Peebles1980}
    \citep{Peebles1980} for the matter dominated universe as $f(z=0) \approx
    \left( \Omega_{\matter,0}
    \right)^{0.6}$, followed by the more accurate approximation $\left(
        \Omega_{\matter,0} \right)^{4/7}$ by \citet{Lightman1990}.
    More generally, the approximation was also derived in dynamical \gls{de}
    models with zero curvature and slowly varying \gls{eos}
    \citep{WangSteinhardt1998} and in curved spaces \citep{Gong2009};
    in modified gravity models, the approximate solution was given in
    refs.~\citep{Linder2007,Gong2008}.
    This approximation has been shown very satisfactory until now for virtually
    any cosmological model without \gls{de}-\gls{dm} coupling, with $\gamma$
    varying accordingly (see, for example, \citet{Linder2005} and references therein).
    In the \gls{lcdm} model, the growth index is approximately \num{6/11}.
    Since growth of structures data spans a wide range of redshift and the
    growth index evolves with the redshift, it is worth exploring its
    parametrization as a function of the redshift. This can help distinguish
    between \gls{de} models and modified gravity models
    \citep{GannoujiPolarski2008,Dossett2010}.

    In order to investigate the influence of a \gls{de}-\gls{dm} interaction on
    the growth of structures, we now consider a simplified model of the
    universe composed exclusively of dark energy and dark matter. 
    Baryonic matter is not taken into account.
    The purpose of this chapter is to derive and present an analytical solution
    for the growth rate of (dark) matter perturbations as a function of the
    redshift in interacting \gls{de} models.
    A similar analysis is done by \citet{Simpson2011}, although with slightly
    different model and parametrization.
    Our approach generalizes the method employed by \citet{Tsujikawa2013}
    for the dynamical \gls{de} model without any interaction with \gls{dm}.
    The derivation is based on the expansion of the growth index and of the
    \gls{de} \gls{eos} parameter in terms of the \gls{de} density
    parameter $\Omega_{\de}(z)$.
    We also derive an expression for the \gls{rms} amplitude of
    perturbations $\sigma_8(z)$ and show that when the \gls{de} equation of
    state, the coupling, the \gls{de} energy density and the amplitude of
    perturbations at present are given, the evolution history of the growth of
    structures is fully determined analytically.
    This analytic solution of the growth can help us see clearly the influence
    of the interaction between dark sectors in the growth.
    The analytic form of $f\sigma_8(z)$ thus obtained enables us to test the
    interacting \gls{de} model by using \gls{lss} observations.

    We are going to analyze two different types of interacting \gls{de} models,
    one with an interacting term proportional to the \gls{de} density and other
    with a term proportional to the \gls{dm} density.
    In a general way, we can combine equations \eqref{eq:dynamicalequations}
    with the Poisson equation \eqref{eq:PoissonMatter} to get a second order
    differential equation for $\delta_{\dm}$ in the interacting case as
    well, without specifying the form of the interaction term $Q_0^{\dm}$.
    The resulting equation is
    \begin{align}
        \label{eq:fullSOE}
        \delta''_{\dm}
        - \left( \mathscr{Q} - \mathscr{H} \right) \delta'_{\dm} 
            - \left( \frac{3}{2} \mathcal{H}^2 \Omega_{\dm} + \mathscr{Q}' + \mathscr{H} \mathscr{Q} \right) \delta_{\dm} = 
            - \frac{i k^i \delta Q_i^{\dm}}{\bar\rho_{\dm}},
    \end{align}
    where we have defined
    \begin{align}
        \mathscr{Q} \equiv \frac{\bar Q_0^{\dm}}{\bar \rho_{\dm}} - \frac{\delta Q_0^{\dm}}{\bar\rho_{\dm} \delta_{\dm}} \qquad
        \text{and} \qquad
        \mathscr{H} \equiv \mathcal{H} - \frac{\bar Q_0^{\dm}}{\bar \rho_{\dm}}.
    \end{align}
    In the following sections we specify the interaction according to the two
    cases we want to study.

    \section{The interacting model \texorpdfstring{$Q^{\dm}_0 \propto
            \rho_{\de}$}{Q0(DM) proportional to the DE density}}
    \label{ss:CDEmodel}

    We start with a model with an interaction term in the \gls{dm}
    energy-momentum conservation equation that is proportional to the \gls{de}
    energy density,
    \begin{align}
        \text{\glsentryshort{cde}:} \qquad  \qquad Q_0^{\dm} &= {\bar
            Q}_0^{\dm} = -3 \mathcal{H} \zeta \bar\rho_{\de}, \label{eq:model2Q0}
    \end{align}
    where $\zeta$ is the coupling constant.
    The interaction in this \gls{cde} model has only an unperturbed part, since
    we are neglecting \gls{de} clustering.
    With \eq{eq:model2Q0}, the background evolution \eq{eq:fluidconsQ} reads
    \begin{align}
        \label{eq:BianchiDMmodel2}
        \bar{\rho}_{\dm}' + 3 \mathcal{H} \bar\rho_{\dm} = 3 \mathcal{H} \zeta \bar\rho_{\dm} \frac{1 - \Omega_{\dm}}{\Omega_{\dm}}.
    \end{align}
    Replacing $\bar Q_0^{\dm}/\bar\rho_{\dm} = - 3 \mathcal{H} \zeta \left(1 -
        \Omega_{\dm}\right)/\Omega_{\dm}$ and $\delta Q_0^{\dm}/\bar\rho_{\dm}\delta_{\dm} = 0$, 
    the evolution of the \gls{dm} perturbations \eqref{eq:fullSOE} reduces to
    \begin{align}
        \delta''_{\dm} 
        &+ \left(1 + 6 \zeta \tfrac{1-\Omega_{\dm}}{\Omega_{\dm}} \right) \mathcal{H} \delta'_{\dm} - {} \nonumber \\
       \label{eq:Model2eq}
       &-\tfrac{3}{2} \mathcal{H}^2 \delta_{\dm} \left[
           \Omega_{\dm}
           - 2 \zeta \tfrac{1 - \Omega_{\dm}}{\Omega_{\dm}} \left(
               1
               + \tfrac{\mathcal{H}'}{\mathcal{H}^2}
               + 3 \zeta \tfrac{1-\Omega_{\dm}}{\Omega_{\dm}} 
               - \tfrac{\Omega'_{\dm}}{\mathcal{H}\Omega_{\dm}} \tfrac{1}{1 - \Omega_{\dm}} 
   \right) \right]
       = 0.
    \end{align}
    The standard evolution $\delta''_{\dm} + \mathcal{H} \delta'_{\dm} - \tfrac{3}{2} \mathcal{H}^2 \Omega_{\dm} \delta_{\dm} = 0$ is recovered when $\zeta = 0$.
    Due to the presence of the interaction, the coefficient of $\delta_{\dm}$ in
    \eq{eq:Model2eq} can become positive as $\Omega_{\dm}$ decreases, leading
    to a decaying regime of the perturbation.
    It is evident that this negative growth rate cannot be described by the
    parametrization of $f$ with the growth index.
    This imposes a constraint on the values that the coupling can assume under
    this growth index parametrization, as we will se in
    section~\ref{ss:numcomparison}.

    \subsection{The growth of structure}
    \label{sss:CDEmodelpropde}
    To obtain the approximation $f \approx \Omega_{\dm}^{\gamma}$, we need to
    change the time derivatives $\partial/\partial \tau$ to $\partial/\partial
    a$ and write \eq{eq:Model2eq} in terms of $f$.
    We can carry out a power series expansion for the functions in terms of
    $\Omega_{\de}$ around zero, describing the time evolution in terms of the
    \gls{de} density parameter.
    In non-interacting models, a polynomial equation in $\Omega_{\de}$ can be
    obtained by equating coefficients in both sides, with its zero-th order
    coefficients vanishing identically and its coefficients for higher orders in
    $\Omega_{\de}$ giving the coefficients of $\gamma = \sum_{n=0}^{\infty}
    \gamma_n \left( \Omega_{\de}\right)^n$ in terms of the coefficients of
    $w_{\de} = \sum_{n=0}^{\infty} w_n \left( \Omega_{\de}\right)^n$ (see, for
    example, \citet{WangSteinhardt1998}).
    This form of parametrization has been shown useful in obtaining the analytic
    expression of the growth index in dynamical \gls{de} models and convenient
    for distinguishing the model from the \gls{lcdm} model
    \citep{WangSteinhardt1998,Tsujikawa2013}. 

    For the \gls{de}-\gls{dm} interaction model, we adopt the same strategy as
    that of the non-interacting cases \citep{Tsujikawa2013}.
    We do the expansion around $\Omega_{\de} = 0$ and assume that the ratio
    between the rate of change of the \gls{de} density parameter and the Hubble
    rate is negligible compared to the density parameter and to unity, at least
    in the regime of structure formation.
    Therefore, $\Omega'_{\de} \ll \mathcal{H} \Omega_{\de}$ in \eq{eq:Model2eq} and we are led to
    \begin{align}
        \delta''_{\dm} 
        &+ \left(1 + 6 \zeta \tfrac{1-\Omega_{\dm}}{\Omega_{\dm}} \right)
        \mathcal{H} \delta'_{\dm} - {} \nonumber \\
        \label{eq:ddotdeltaModel2}
        &-\tfrac{3}{2} \mathcal{H}^2 \delta_{\dm} \left\lbrace
           \Omega_{\dm}
           + 2 \zeta \tfrac{1 - \Omega_{\dm}}{\Omega_{\dm}} 
           \left[
               -\tfrac{1}{2} + 3 w_{\de} \left( 1 - \Omega_{\dm} \right) 
               - 3 \zeta \tfrac{1 - \Omega_{\dm}}{\Omega_{\dm}}
   \right] 
   \right\rbrace
       = 0.
    \end{align}
    After some manipulations, this is rewritten as
    \begin{align}
        \frac{\ud^2 \ln \delta_{\dm}}{\ud \ln a^2} &+ \left( \frac{\ud \ln \delta_{\dm}}{\ud \ln a} \right)^2 + \left[ \frac{1}{2} - \frac{3}{2} w_{\de} \left(1 - \Omega_{\dm} \right) + 6 \zeta \frac{1 - \Omega_{\de}}{\Omega_{\de}} \right] \frac{\ud \ln \delta_{\dm}}{\ud \ln a} - {} \nonumber \\
        {}& - \frac{3}{2} \Omega_{\dm} + 3 \zeta \frac{1 - \Omega_{\dm}}{\Omega_{\dm}} \left[\frac{1 - \Omega_{\dm}}{\Omega_{\dm}} \left( 3 \zeta - 3 w_{\de} \Omega_{\dm} \right) + \frac{1}{2} \right] = 0.
    \end{align}
    Substituting $f$, we have
    \begin{align}
        \label{eq:dfdlnaM2}
        \frac{\ud f}{\ud \ln a} + f^2 + f & \left[ \frac{1}{2} - \frac{3}{2} w_{\de} \left(1 - \Omega_{\dm} \right) + 6 \zeta \frac{1-\Omega_{\dm}}{\Omega_{\dm}} \right] - \frac{3}{2} \Omega_{\dm} + {} \nonumber \\
        {}& + 3 \zeta \frac{1 - \Omega_{\dm}}{\Omega_{\dm}} \left[\frac{1 - \Omega_{\dm}}{\Omega_{\dm}} \left( 3\zeta - 3 w_{\de} \Omega_{\dm} \right) + \frac{1}{2} \right] = 0,
    \end{align}
    which still has the first term parametrized by the scale factor.
    Next, we write 
    \begin{align}
        \label{eq:fchain}
        \frac{\ud f}{\ud \ln a} = \frac{\ud \Omega_{\dm}}{\ud \ln a} \frac{\ud f}{\ud \Omega_{\dm}}
    \end{align}
    and use the (background) energy conservation
    equations to substitute $\frac{\ud \Omega_{\dm}}{\ud \ln a}$.
    The total conservation equation gives
    \begin{gather}
        \ud \bar \rho + 3 \tfrac{\ud a}{a} \left(\bar \rho + \bar p \right) = 0, \nonumber \\
        %a^3 \ud \bar \rho + \ud (a^3) \bar \rho = - \ud (a^3) w_{\de} \bar \rho_{\de} \nonumber \\
        \ud (a^3 \bar \rho) = - \ud (a^3) w_{\de} \bar\rho_{\de}, \nonumber \\
        \label{eq:totcons}
        \ud \left( \frac{a^3 \bar\rho_{\dm}}{\Omega_{\dm}} \right) = - \ud (a^3) w_{\de} \bar \rho_{\dm} \frac{1-\Omega_{\dm}}{\Omega_{\dm}},
    \end{gather}
    where we have used $\bar\rho = \frac{\bar\rho_{\dm}}{\Omega_{\dm}}$ and $\bar\rho_{\de} = \bar\rho_{\dm} \frac{1-\Omega_{\dm}}{\Omega_{\dm}}$, while the \gls{dm} equation gives
    \begin{gather}
        \ud \bar \rho_{\dm} + 3 \tfrac{\ud a}{a} \bar \rho_{\dm} = 3 \tfrac{\ud a}{a} \zeta \bar\rho_{\dm} \tfrac{1 - \Omega_{\dm}}{\Omega_{\dm}}, \nonumber \\
        %a^3 \ud \bar \rho_{\dm} + \ud (a^3) \bar\rho_{\dm} = \ud(a^3) \zeta \bar\rho_{\dm} \tfrac{1 - \Omega_{\dm}}{\Omega_{\dm}} \nonumber \\
        \ud(a^3 \bar\rho_{\dm}) = \zeta \bar\rho_{\dm} \tfrac{1 - \Omega_{\dm}}{\Omega_{\dm}}  \ud(a^3),
    \end{gather}
    which can be inserted back in \eq{eq:totcons} to give
    \begin{gather}
        \zeta \bar \rho_{\dm} \frac{1 - \Omega_{\dm}}{\Omega_{\dm}} \frac{\ud (a^3)}{\Omega_{\dm}} - a^3 \bar\rho_{\dm} \frac{\ud \Omega_{\dm}}{\Omega_{\dm}^2} = - w_{\de} \bar \rho_{\dm} \frac{1 - \Omega_{\dm}}{\Omega_{\dm}} \ud (a^3), \nonumber \\
        3 \zeta \left(1 - \Omega_{\dm} \right) \ud \ln a - \ud \Omega_{\dm} = - 3 w_{\de} \Omega_{\dm} \left(1 - \Omega_{\dm} \right) \ud \ln a, \nonumber \\
        \label{eq:dOdlnaM2}
        \frac{\ud \Omega_{\dm}}{\ud \ln a} = 3 \left(1 - \Omega_{\dm} \right) \left( \zeta + w_{\de} \Omega_{\dm} \right).
    \end{gather}
    Substituting \eq{eq:dOdlnaM2} into \eq{eq:fchain} and dividing \eq{eq:dfdlnaM2} by $f$ we have
    \begin{align}
        3 &\left( \zeta + w_{\de} \Omega_{\dm} \right) \frac{1 - \Omega_{\dm}}{f}  \frac{\ud f}{\ud \Omega_{\dm}} + f + \frac{1}{2} - \frac{3}{2} w_{\de} \left(1 - \Omega_{\dm} \right) + 6 \zeta \frac{1 - \Omega_{\dm}}{\Omega_{\dm}} - {} \nonumber \\
        \label{eq:fullfModel2}
        {} &- \frac{3}{2} \frac{\Omega_{\dm}}{f} + 3 \zeta \frac{1 - \Omega_{\dm}}{f \Omega_{\dm}} \left[ \frac{1 - \Omega_{\dm}}{\Omega_{\dm}} \left( 3 \zeta - 3 w_{\de} \Omega_{\dm} \right) + \frac{1}{2} \right] = 0.
    \end{align}
    Finally, expanding eq.~\eqref{eq:fullfModel2} around $\Omega_{\de} = 0$ with $f = \left(\Omega_{\dm}\right)^{\gamma_0 + \gamma_1 \Omega_{\de} + \ldots}$, we arrive at the polynomial equation
    \begin{align}
        \label{eq:poleqCDE}
        \left[ 3 \left(1 - w_0 + 5 \zeta \right) - \gamma_0 \left( 5 - 6 w_0 - 6 \zeta \right) \right] \Omega_{\de}  
            + \frac{1}{2} \left[ 
            -\gamma_0^2
            + \gamma_0 \left(
                        1
                        +12 w_1
                        + 18 \zeta 
                    \right)
                    - {} \right. \nonumber \\
                    {} - \left. \vphantom{\gamma_0^2}  2 \gamma_1 \left(
                5 
                - 12 w_0
                - 12 \zeta
                    \right)
                - 6w_1
                + 6 \zeta \left( 
                    5
                    - 6 w_0
                    +6 \zeta
                \right)
            \right] \Omega_{\de}^2
        + \mathcal{O} \bigl( \Omega_{\de}^3 \bigr) = 0.
    \end{align}
    The zero-th order part is still identically zero even with non-zero $\zeta$.
    The equations of the higher order terms can be solved to give the modified growth index coefficients
    \begin{subequations}
        \label{eq:gammas}
        \begin{align}
            \gamma_0 &= \frac{3 \left(1 - w_0 + 5 \zeta \right)}{5 - 6 w_0 - 6 \zeta}, \\
            \gamma_1 &= \frac{- \gamma_0^2 + \gamma_0 \left(1 + 12 w_1 + 18 \zeta \right) - 6 w_1 + 6 \zeta \left(5 - 6 w_0 + 6 \zeta \right)}{2 \left( 5 - 12 w_0 -12 \zeta\right)}, \\
            \vdots \nonumber
        \end{align}
    \end{subequations}
    We note that positive $\zeta$ increases $\gamma_0$, the dominant part of the growth index.
    The well-known result $\gamma_0 = \frac{3 \left(1 - w_0 \right)}{5 - 6 w_0}$
    is recovered when $\zeta= 0$, giving $\gamma_0 = 6/11$ for \gls{lcdm}.
    With the standard values $w_0 = -1$ and $\zeta = 0$, the first-order
    coefficient is $\gamma_1 = \frac{3 \left(5 + 11 w_1 \right)}{2057}$, which
    may give a rather small contribution $\gamma_1 \Omega_{\de}$ to $\gamma$ for a
    slowly varying \gls{eos} parameter.

    Predictions made with $f = \left(\Omega_{\dm}\right)^{\gamma_0 + \gamma_1
        \Omega_{\de} + \ldots}$ can, in principle, be compared to growth rate measurements like
    those compiled in \citet{Dossett2010}.
    Those data, however, are generally obtained from measurements of the
    \gls{rsd} parameter $\beta = f/b$, where $b$ is the bias measuring how
    galaxies trace the matter density field, and thus can be bias-dependent.
    Usually, it is preferable to compare predictions with the bias-independent
    data of $f \sigma_{8}$ \citep{SongPercival2009}, the growth rate multiplied
    by the variance of the density field filtered at a scale $R =
    8\,h^{-1}\,\si{\mega\parsec}$, defined as
    \begin{align}
        \label{eq:s8def}
        \sigma_{R}^2 (z) \equiv \frac{1}{2 \uppi^2} \int_0^{\infty} \ud k \, k^2
        \mathcal{P}_{\dm}(k, z) \left| W( k R) \right|^2,
    \end{align}
    where $P(k, z)$ is the matter power spectrum and $W(kR)$ is the window function of the experiment in Fourier space.
    We derive $\sigma_8$ from $\delta_{\dm}$ starting with the definition of $f$,
    \begin{gather}
        \frac{\ud \Omega_{\dm}}{\ud \ln a} \frac{\ud \ln \delta_{\dm}}{\ud
            \Omega_{\dm}} = (\Omega_{\dm})^{\gamma} \quad \Rightarrow \nonumber \\
        \Rightarrow \quad 3 \left(1 - \Omega_{\dm} \right) \left(\zeta + w_{\de} \Omega_{\dm} \right) \frac{\ud \ln \delta_{\dm}}{\ud \Omega_{\dm}} = (\Omega_{\dm})^{\gamma} \quad \therefore \nonumber \\
        \therefore \quad \frac{\ud \ln \delta_{\dm}}{\ud \Omega_{\de}} = - \frac{\left( 1- \Omega_{\de} \right)^{\gamma}}{3 \Omega_{\de} \left[ \zeta + w_{\de} \left(1 - \Omega_{\de} \right) \right] }.
%        \ln \left( \frac{\delta_{\dm}}{\delta_{\dm,0}} \right) = \frac{\Delta^{(3)}_{\de}}{54 \omega_0} \left\lbrace 
%              11 \ell_0 
%              + 3 \beta_1 \left( \ell_1 - \alpha_1 \ell_0 \right)
%              -6 \tilde\alpha_1 \left(1 + \alpha_1^2 \right) 
%              + 3 \ell_0 \left[\ell_0 \left( \alpha_1-2 \right) - 2 \ell_1 \right] 
%            + \ell_0^3 \right\rbrace \nonumber \\
%            &\,\quad + \frac{ \Delta^{(2)}_{\de} }{12 \omega_0}  \left[2 \ell_1 + \ell_0 \beta_1 - \ell_0^2 -2 \left( \tilde\alpha_1 + \alpha_1^2\right) \right]  \nonumber \\
%            &\,\quad +\frac{\Delta^{(1)}_{\de}}{3 \omega_0} \left( \ell_0- \tilde\alpha_1 \right)
%            -\ln\left(\frac{\Omega_{\de}}{\Omega_{\de,0}}\right)^{1/3 \omega_0} ,
    \end{gather}
    We integrate backwards in $\Omega_{\de}$ from $\Omega_{\de,0}$ to $\Omega_{\de}(z)$ and expand it to obtain
    \begin{align}
        \ln &\frac{\delta_{\dm}}{\delta_{\dm,0}} = \ln \left( \frac{\Omega_{\de}}{\Omega_{\de,0}} \right)^{-1/3 \tilde w_0 }
        + \frac{\gamma_0 - \bar\omega_{01} }{3 \tilde w_0} \left(\Omega_{\de} - \Omega_{\de,0} \right) - {} \nonumber \\
        {} &- \frac{1}{6 \tilde w_0}\left[\frac{\gamma_0^2}{2}  - \gamma_0  \left( \frac{1}{2} + \bar\omega_{01} \right) - \gamma_1 + \frac{1}{\tilde w_0} \left( w_0 \bar\omega_{01} - w_2  + \frac{w_1 \tilde w_1}{\tilde w_0} \right) \right] \left( \Omega_{\de}^2 - \Omega_{\de,0}^2 \right) + {}  \nonumber \\
        \label{eq:sigma8}
        {} &+ \mathcal{O} \bigl(\Omega_{\de}^3\bigr) + \mathcal{O} \bigl( \Omega_{\de,0}^3 \bigr)
    \end{align}
    where we have introduced the definitions
    \begin{gather}
        \tilde w_n \equiv w_n + \zeta \qquad \text{and} \qquad
        \bar\omega_{01} \equiv \frac{w_0 - w_1}{\tilde w_0}.
    \end{gather}
    The time dependence of $\delta_{\dm}$ is parametrized by $\Omega_{\de}$. $\delta_{\dm,0}$ and $\Omega_{\de,0}$ represent their values today.
    Eq.~\eqref{eq:sigma8} then gives, up to the second order in $\Omega_{\de}$
    and $\Omega_{\de,0}$,
    \begin{align}
        \label{eq:sigma8m2}
        \delta_{\dm}(z) &= \delta_{\dm,0} \mathcal{D}_{\dm}(z;0),
    \end{align}
    with
    \begin{align}
        \mathcal{D}_{\dm}(z;0) &\equiv \left[ \frac{\Omega_{\de(z)}}{\Omega_{\de,0}}
        \right]^{-1/3 \tilde w_0} \exp \left[ \frac{\varepsilon_1
                \Delta_{\de}^{(1)} + \varepsilon_2 \Delta_{\de}^{(2)}}{3 \tilde
                w_0} \right]
    \end{align}
    the backward propagation function for the evolution of the \gls{dm}
    perturbation, as analogously defined in \eq{eq:backprop} for the matter
    perturbation, and 
    \begin{align}
        \varepsilon_1 &\equiv \gamma_0 - \bar\omega_{01}, \\
        \varepsilon_2 &\equiv -\frac{\gamma_0^{2}}{4} + \frac{\gamma_0}{2}
        \left(\frac{1}{2} + \bar \omega_{01} \right) + \frac{\gamma_1}{2} -
        \frac{1}{2 \tilde w_0} \left( w_0 \bar\omega_{01} - w_2 + w_1 \frac{\tilde w_1}{\tilde w_0} \right), \\
        \Delta_{\de}^{(n)} &\equiv \Omega_{\de}^n(z) - \Omega_{\de,0}^n.
    \end{align}
    Noting that $\mathcal{P}_{\dm}(k, z) = \left[ \mathcal{D}_{\dm}(z;0) \right]^2
    \mathcal{P}_{\dm,0}(k)$ and $\mathcal{D}_{\dm}$ is scale-independent, it
    follows directly from the definition \eqref{eq:s8def} that $\sigma_R^2 =
    \mathcal{D}_{\dm}^2 \sigma_{R,0}^2$, i.e., $\sigma_R$ satisfies the same equation \eqref{eq:sigma8m2} for $\delta_{\dm}$.
    Thus, at the scale $R = 8\,h^{-1}\,\si{\mega\parsec}$, we have
    \begin{align}
        \label{eq:normsigma8}
        \sigma_{8}(z) &= \sigma_{8,0} \left[ \frac{\Omega_{\de}(z)}{\Omega_{\de,0}} \right]^{-1/3 \tilde w_0} \exp \left[ \frac{\varepsilon_1 \Delta_{\de}^{(1)} + \varepsilon_2 \Delta_{\de}^{(2)}}{3 \tilde w_0} \right],
    \end{align}
    also up to the second order in $\Omega_{\de}$ and $\Omega_{\de,0}$.
    Note that there can be some inaccuracy in the computation of $\sigma_8(z)$ from eq.~\eqref{eq:normsigma8}, since we are integrating a function that has been expanded around $\Omega_{\de}=0$ from redshift zero, where $\Omega_{\de}$ is not so small, until $z$.
    This has the consequence of the errors of the expansion at low redshifts being accumulated for $\sigma_8$ at any redshift and constitutes a limitation of the method.
    We also note that if $|w_1|$ or $|w_2|$ is too large, it is possible that they can make the exponential in eq.~\eqref{eq:normsigma8} grow enormously.

    For the evaluation of $\Omega_{\de}(z)$, we have to use a recursive relation.
    The \gls{dm} and \gls{de} densities, in terms of the redshift, are 
    \begin{subequations}
        \begin{align}
            \bar\rho_{\dm}(z) &= \bar\rho_{\dm,0} \exp \left[ \int_0^z \frac{3}{1+\tilde z} \left(1 - \zeta \frac{\Omega_{\de}}{1 - \Omega_{\de}} \right) \ud \tilde z \right], \\
            \label{eq:consde}
            \bar \rho_{\de}(z) &= \bar\rho_{\de,0} \exp \left[ \int_0^z \frac{3}{1+\tilde z}\left[1 + w_{\de}(\tilde z) + \zeta \right] \ud \tilde z \right]. 
        \end{align}
    \end{subequations}
    The term $w^{\eff}_{\de} \equiv w_{\de} + \zeta$ in eq.~\eqref{eq:consde}
    can be seen as an effective \gls{de} equation of state within the
    alternative framework where the interaction term is absorbed into the
    equation of state and there is no net transfer of energy-momentum between
    \gls{de} and \gls{dm}.

    The zero-th order \gls{de} density parameter is obtained by setting $w_{\de} = w_0$ and neglecting the term $\zeta \frac{\Omega_{\de}}{1 - \Omega_{\de}} \approx \zeta \Omega_{\de} + \zeta \Omega_{\de}^2$,
    \begin{align}
        \Omega_{\de}^{(0)} = \frac{\bar\rho_{\de}^{(0)}}{\bar\rho_{\de}^{(0)} + \bar\rho_{\dm}^{(0)}} = \frac{\Omega_{\de,0} \left( 1 + z \right)^{3 \tilde w_0}}{1 - \Omega_{\de,0} + \Omega_{\de,0} \left(1 + z \right)^{3 \tilde w_0}}.
    \end{align}
    Now the density parameter up to the first order is calculated by using $w_{\de} = w_0 + w_1 \Omega_{\de}^{(0)}$ and $\zeta \frac{\Omega_{\de}}{1 - \Omega_{\de}} = \zeta \Omega_{\de}^{(0)}$,
    \begin{align}
        \label{eq:Omega1m2}
        \Omega_{\de}^{(1)}(z) = \frac{\Omega_{\de,0} \left(1 + z \right)^{3 \tilde w_0} \left[ 1 - \Omega_{\de,0} + \Omega_{\de,0} \left(1 + z \right)^{3 \tilde w_0} \right]^{\tilde w_1 / \tilde w_0}}{1 - \Omega_{\de,0} + \Omega_{\de,0} \left(1 + z \right)^{3 \tilde w_0} \left[1 - \Omega_{\de,0} + \Omega_{\de,0} \left(1 + z \right)^{3 \tilde w_0} \right]^{\tilde w_1 / \tilde w_0}}
    \end{align}
    With equations \eqref{eq:gammas}, \eqref{eq:normsigma8} and
    \eqref{eq:Omega1m2} we are now able to compute $f(z)$ and $\sigma_8(z)$ in
    this coupled model provided that we know the parameters $\zeta$, $w_n$,
    $\sigma_{8,0}$ and
    $\Omega_{\de,0}$.
    Once we know the coupling, \gls{de} \gls{eos} coefficients, \gls{de} density
    parameter and the mean perturbation amplitude at present we can determine
    analytically how structures have evolved and can compare these results with
    \gls{lss} observations.

    \subsection{Stability conditions}
    Interacting \gls{de} models with constant \gls{eos} have already been shown to suffer from instabilities with respect to curvature and dark energy perturbations \citep{Valiviita2008,He2009}.
    Depending on some combinations of the sign of the interaction and on the
    dark energy \gls{eos} being of the quintessence or phantom type,
    $\delta_{\de}$ and the potential $\phi$ (in the perturbation of the metric)
    can blow up.
    Table~\ref{tab:instabilities} summarizes the allowed regions for the
    interaction and the \gls{de} equation of state parameters in the \gls{cde}
    model as shown by \citet{Gavela2009}, which extends the model stability
    analysis of \citet{He2009} to negative values of $\zeta$.
    \begin{table}[t]
        \setlength\tabcolsep{36pt}
        \caption{Stability conditions of the \gls{cde} model.}
        \label{tab:instabilities}
        \centering
        \begin{tabular}{@{}lcc@{}}
            \toprule
            Constant \gls{eos}  &   Interaction sign  &   Condition  \\
            \midrule
            $w_{\de} < -1$    &   $\zeta < 0$   &   early-time instability   \\
            $w_{\de} < -1$    &   $\zeta > 0$   &   stable      \\
            $-1 < w_{\de} < 0$    &   $\zeta < 0$   &   stable  \\
            $-1 < w_{\de} < 0$    &   $\zeta > 0$   &   early-time instability \\
            \bottomrule
        \end{tabular}
    \end{table}

    These results strongly restrict the parameter space for interacting \gls{de}.
    As those references point out, such instabilities can be avoided by allowing the \gls{eos} to vary with time, which we do when we expand $w_{\de}$ in terms of $\Omega_{\de}$.
    However, before considering a time variable \gls{eos}, first we simplify our models by fixing $w_1$ so we have one less parameter to be constrained with the \gls{mcmc} method.
    We proceed in the next section to compare our results for the growth rate
    with numerical calculations provided by a modified version of the \gls{camb}
    \citep{CAMB}, in order to assess the reliability of our expressions and validate the method.

    \subsection{Comparison with full numerical computations in \glsentryshort{camb}}
    \label{ss:numcomparison}
    To test how effective our analytical result of the growth in the \gls{cde}
    model is, we compare it with the numerical $f(z)$ obtained in a modified
    version of \gls{camb}\footnote{\gls{camb} is a cosmological Boltzmann code
        commonly used for calculating theoretical radiation and matter power
        spectra, among other things, given the cosmological parameters of the
        standard \gls{lcdm} model or some of its derivatives.} for the interacting model.\footnote{The model implemented in \gls{camb} differed slightly from the analytical model by the presence of a baryonic component, accounting for \SI{4}{\percent} of the total energy density, which did not have a perceptible influence on the comparisons.}
    We are going to show that our analytic solution can be trusted and we can further use it
    to estimate the cosmological parameters with a \gls{mcmc} code, as a shortcut alternative to the full numerical computation to speed up the calculation.

\begin{figure}[tb]
        \centering
        \includegraphics[width=\textwidth]{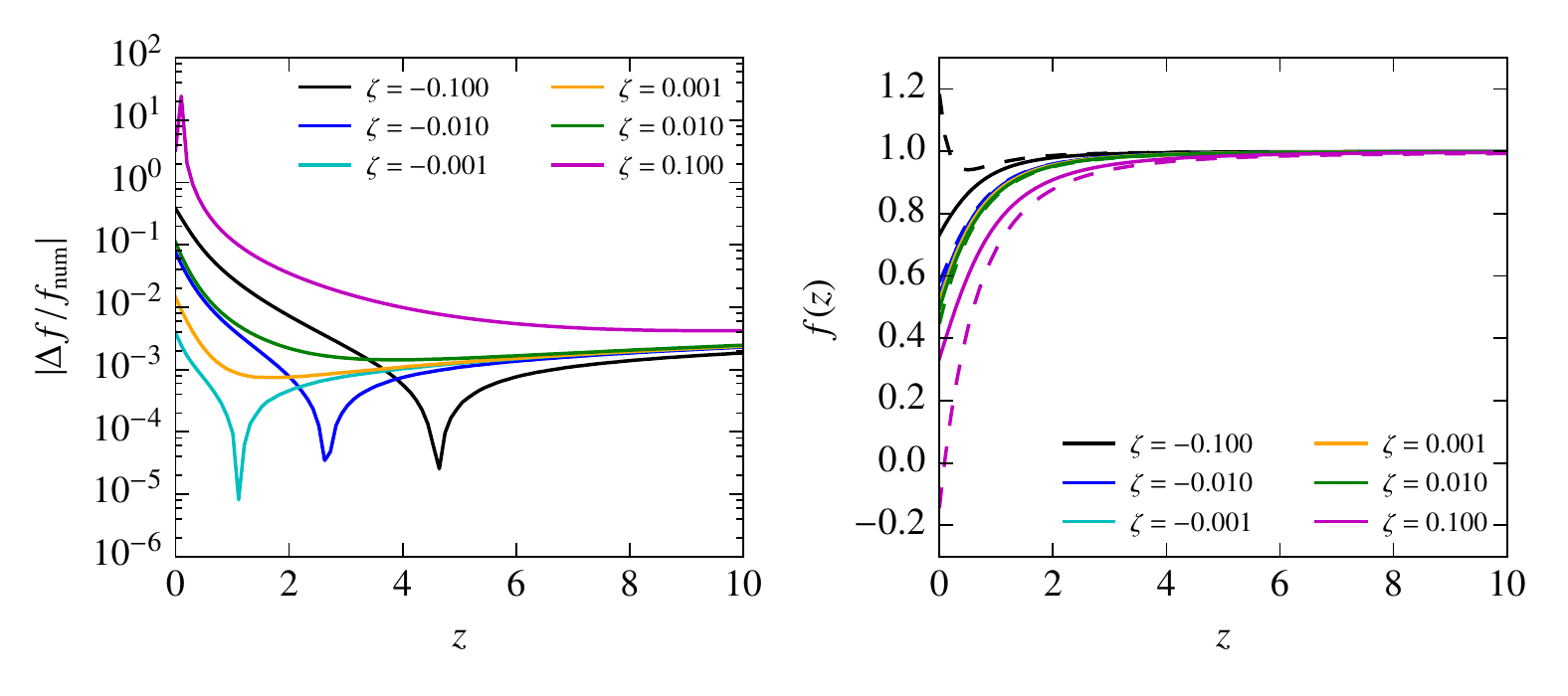}
        \caption{Comparison between analytical and numerical computations of $f(z)$ for the \gls{cde} model. In the left panel, the modulus of the relative differences, in logarithmic scale; in the right panel, the dashed lines represent the numerical results for $f(z)$, while the solid lines show our analytical results.}
        \label{fig:compModel2}
    \end{figure}

    We fix $w_1 = 0$ and $\Omega_{\de,0} = 0.7$ and calculate $f(z)$ with $z$ ranging from $0$ to $10$.
    According to the stability conditions given in the last section, the interaction constant in \gls{cde} can be negative, in which case the dark energy \gls{eos} must be of quintessence type, or the coupling be positive with phantom type \gls{de} \gls{eos}.
    We then fix $w_0 = -0.999$ and test the interaction constants $\zeta = -0.1, -0.01, -0.001$ and $w_0 = -1.001$ with $\zeta = 0.001, 0.01, 0.1$. 
    To distinguish these two tests, we name the models, respectively, \gls{cpde} and
    \gls{cqde}. 
    The comparisons are shown in figure~\ref{fig:compModel2} through the modulus of the difference $\Delta f \equiv f_{\anl} - f_{\tnum}$ divided by $f_{\tnum}$ (left panel), where ``anl'' and ``num'' stand for analytical and numerical computations.
    Over the range of the \gls{rsd} data (low redshift until $z \sim 1$), for a given $\zeta$, the discrepancy grows as we approach $z = 0$, which is expected from the fact that $\Omega_{\de,0}$ is as big as $0.7$.
    The discrepancy tends to decrease as $z$ increases, but only until a certain redshift, when it can start to grow, albeit slowly.
    In the $\zeta = 0.1$ case, $f_{\tnum}$ can become negative and the discrepancy is huge.
    This occurs because as $z$ decreases, $\frac{1-\Omega_{\dm}}{\Omega_{\dm}}$ increases and the second term inside the curly brackets in eq.~\eqref{eq:ddotdeltaModel2} dominates the coefficient of $\delta_{\dm}$ and changes its sign, leading to a negative growth.
    The analytical parametrization $(\Omega_{\dm})^{\gamma}$, on the other hand, obviously can never become negative.
    The numerical result for $\zeta = -0.1$ shows that $f$ grows very rapidly at small redshifts as $z$ goes to zero, a behavior that is opposite to the other cases.
    This is due to a change of sign in the coefficient of $\delta'_{\dm}$ (Hubble drag) in eq.~\eqref{eq:ddotdeltaModel2}.
    We discard the cases $\zeta = - 0.1$ and $\zeta = 0.1$ as they are not well described by eq.~\eqref{eq:fOmegagamma}
    and restrict $\zeta$ within the interval $\left[ -0.01, 0 \right]$ for the \gls{cqde} model and $\left[0, 0.01\right]$ for the \gls{cpde} model,
    which allow the difference between the numerical and the analytical results to be kept below about \SI{10}{\percent} 
    (with the other parameters fixed at reasonable values).
    The cusps observed in the curves of $|\Delta f/f_{\tnum}|$ in the two models mean a change of sign of $\Delta f$.
    The fact that the analytical and numerical curves cross themselves instead of converging to a common plateau, with $f_{\tnum}$ becoming smaller than $f_{\anl}$ as $z$ becomes larger, might indicate some contribution of a decaying mode of the perturbation, which is out of the scope of this work.

    The conclusion is that the \gls{mcmc} analysis can be made using the
    analytic approximations derived for $f$ in the interacting \gls{de} model,
    provided the parameters are restricted to the region where the discrepancy
    with respect to the numerical reference from \gls{camb} is reasonably
    small.
    In section~\ref{sec:data} we will present the \gls{rsd} data that we use to
    estimate the parameters of our interacting models via \gls{mcmc}.

    \section{The interacting model \texorpdfstring{$Q^{\dm}_0 \propto
            \rho_{\dm}$}{Q0(DM) proportional to the DM density}}
    \label{sec:Qrhodm}

    We have also analyzed the case of an interaction proportional do the
    \gls{dm} density,
    \begin{align}
        Q_0^{\dm} = - 3 \mathcal{H} \zeta \rho_{\dm}.
    \end{align}
    The perturbed spatial part is $\delta Q_i^{\dm}$ is set to zero.
    The background evolution is given by 
    \begin{align}
        \label{eq:rhodmcont_tao}
        {\bar\rho}'_{\dm} + 3 \mathcal{H} \bar\rho_{\dm} = 3 \mathcal{H} \zeta \bar \rho_{\dm}.
    \end{align}
    The evolution of the \gls{dm} perturbations comes from
    eq.~\eqref{eq:fullSOE} with $\frac{\bar{Q}_0^{\dm}}{\bar\rho_{\dm}} = \frac{\delta Q_0^{\dm}}{\bar\rho_{\dm} \delta_{\dm}} = - 3 \mathcal{H} \zeta$, giving
    \begin{align}
        \label{eq:dmSOE}
        \delta''_{\dm} + \left(1 + 3 \zeta \right) \mathcal{H} \delta'_{\dm} - \tfrac{3}{2} \mathcal{H}^2 \Omega_{\dm}  \delta_{\dm}   = 0.
    \end{align}
    This functional form of this equation is even simpler than the \gls{cde} case of section~\ref{ss:CDEmodel} with respect to the standard evolution, with only one extra term proportional to $\zeta$.

    \subsection{The growth of structure}
    In terms of $f$, with $\frac{\ud \Omega_{\dm}}{\ud \ln a} = 3 \Omega_{\dm} \left[ \zeta + w_{\de} \left(1 - \Omega_{\dm} \right) \right]$, the growth rate evolution equation is
    \begin{align}
        \label{eq:fgeneraleq}
        3 \left[ w_{\de} \left(1 - \Omega_{\dm} \right) + \zeta \right] \frac{\Omega_{\dm}}{f} \frac{\ud f}{\ud \Omega_{\dm}} + f + \frac{1}{2} - \frac{3}{2} w_{\de} \left(1 - \Omega_{\dm} \right) + 3 \zeta - \frac{3}{2} \frac{\Omega_{\dm}}{f} = 0,
    \end{align}
    which expanded in $\Omega_{\de}$ for $f = (\Omega_{\dm})^{\gamma_0 + \gamma_1 \Omega_{\de} + \ldots}$ gives the polynomial equation
    \begin{align}
        3 &\zeta \left(1 + \gamma_{0} \right) 
        + \frac{1}{2} \left[3\left(1 - w_0 \right) - \left(5 - 6 w_0 \right) \gamma_0 + 12 \zeta \gamma_1 \right] \Omega_{\de} + {} \nonumber \\
        {} &+ \frac{1}{4} \left[-{\gamma_0}^2 + 36 \zeta {\gamma_2} +2
            {\gamma_1} (12 {w_0}-5 - 3 \zeta)+ \left(1 + 12 w_1 \right) \gamma_0
            - 6 w_1 \right] \Omega_{\de}^2 + {} \nonumber \\
        \label{eq:M1poleq}
        {} &+ \mathcal{O} ( \Omega_{\de}^3 ) = 0.
    \end{align}
    Unlike eq.~\eqref{eq:poleqCDE}, this now has a zero-th order part that does not vanish identically regardless of the interaction or other parameters.
    In order for eq.~\eqref{eq:M1poleq} to hold, $3\zeta \left(1 + \gamma_0 \right) = 0$ must be satisfied.
    This implies $\zeta=0$, recovering the non-interacting results for $\gamma$ from the higher order terms, or $\gamma_0 = -1$ and $\gamma_1 = \frac{9 w_0 - 8}{12 \zeta}$ (with $\zeta \ne 0$) from the first-order coefficient, which does not seem to fit the observed growth unless perhaps with a fine tuning of the parameters.
    Also, note that this solution implies a non-smooth transition to zero interaction.
    Although numerically the growth rate in this model can still be, to some
    degree, well approximated by the power law $\Omega_{\dm}^{\gamma}$ form, as
    claimed by \citet{Linder2005}, analytically we can see that this form is not appropriate for a non-zero coupling in the interaction term proportional to $\rho_{\dm}$.

    \section{Observational constraints}
    \label{sec:data}
    In this section we present the dataset used to constrain the parameters of
    our \gls{cde} model.
    However, an adjustment to our growth rate $f$, calculated in a universe
    where \gls{dm} and \gls{de} interact, is needed before comparing with
    $f\sigma_8$ data obtained assuming a standard cosmology.
    We explain in detail how the comparison must be made, then describe the
    statistical method employed in the analysis and discuss our results.

    \subsection{The data}
    One way of measuring the growth of structures is through the effect of
    redshift-space distortions.
    We have seen in section~\ref{sec:lineartheoty} that the galaxy power spectrum
    $\mathcal{P}^{\redspace}$ observed in redshift space is expected to be amplified
    with respect to the real power spectrum $\mathcal{P}(k)$ by a factor that
    depends on the growth rate and on the cosine of the angle between the
    movement of the galaxies and the observation, according to
    \eq{eq:powerspectrum}.
    Multipole analysis of the anisotropy of the redshift-space power spectrum or
    correlation function in the redshift survey allows the observational
    determination of $\beta$, as we have also seen in
    section~\ref{sec:measuringRSDbeta}.
    Thus one gets the measurement $\beta \sigma_{8,\gal} = f
    \sigma_{8,\matter}$ of the growth of structure.
    The advantage of using $f \sigma_8$ rather than just $f$ to compare with
    model predictions is that the estimator $\beta \sigma_{8,\gal}$ is
    independent of the bias model.
    Also, the determination of $\beta$ is affected only weakly by changes in the
    cosmology (through the determination of distances)
    \citep{SongPercival2009,Basilakos2013}.
    
    In table~\ref{tab:data} we list measurements of growth rate with their
    errors for various redshifts from different surveys like the \gls{2df},
    the \gls{6df}, the \gls{sdss}, the \gls{boss} and the WiggleZ dark energy survey.
    \begin{table}[tb]
        \setlength\tabcolsep{11pt}
        \caption{Observed growth rate data and their respective references.}
        \label{tab:data}
        \centering
        \sisetup{table-format=1.3(1)}
        \begin{tabular}{@{}S[table-format=1.3]SccS[table-format=1.2]Sc@{}}
            \toprule
            {$z$} &   {$f \sigma_8(z)$} &   Ref.  & &  {$z$} &   {$f \sigma_8(z)$}  &   Ref.  \\
            \midrule
            0.02   &    0.360   \pm   0.040 &   \citep{HudsonTurnbull2012} & &   0.40   &    0.419   \pm   0.041 &   \citep{Tojeiroetal2012}\\
            0.067   &   0.423   \pm   0.055 &   \citep{Beutleretal2012} & &  0.41   &    0.450   \pm   0.040 &   \citep{Blakeetal2011}\\
            0.10   &    0.37   \pm    0.13  &   \citep{Feix2015} & &  0.50   &    0.427   \pm   0.043 &   \citep{Tojeiroetal2012}\\
            0.17   &    0.510   \pm   0.060 &   \citep{SongPercival2009,Percivaletal2004}  & &   0.57   &    0.427   \pm   0.066 &   \citep{Reidetal2012}\\
            0.22   &    0.420   \pm   0.070 &    \citep{Blakeetal2011} & &   0.60   &    0.430   \pm   0.040 &   \citep{Blakeetal2011}\\
            0.25   &    0.351   \pm   0.058 &   \citep{Samushia01032012} & &  0.60   &    0.433   \pm   0.067 &   \citep{Tojeiroetal2012}\\
            0.30   &    0.407   \pm   0.055 &   \citep{Tojeiroetal2012} & &  0.77   &    0.490   \pm   0.180 &   \citep{Guzzo:2008ac,SongPercival2009}\\
            0.35   &    0.440   \pm   0.050 &   \citep{SongPercival2009,Tegmark2006} &  &  0.78   &    0.380   \pm   0.040 &   \citep{Blakeetal2011}\\
            0.37   &    0.460   \pm   0.038 &   \citep{Samushia01032012} & &  0.80   &    0.47    \pm   0.08 &   \citep{delaTorreetal2013}\\
            \bottomrule
        \end{tabular}
    \end{table}
    Most of those data are measured using \gls{rsd} and others are based on
    direct measurements of peculiar velocities
    \cite{HudsonTurnbull2012,Turnbulletal2012,Davisetal2011} or galaxy
    luminosities \cite{Feix2015}.

    \subsubsection{Corrections to the growth rate due to the altered continuity
    equation}

    In a standard cosmology, the coherent motion of galaxies is connected to the
    growth rate through the galaxy continuity equation $\theta_{\gal} = -
    \mathcal{H} \beta \delta_{\gal}$, built upon the matter continuity equation
    $\theta_{\matter} = - \mathcal{H} f_{\matter} \delta_{\matter}$ with the
    density bias assumption $\delta_{\gal} = b \delta_{\matter}$ and without any
    bias for the velocities ($\theta_{\gal} = \theta_{\matter}$).
    Whether the \gls{rsd} parameter $\beta$ is measured from the power
    spectrum or from peculiar velocities, the $f \sigma_8$ data are based on the
    correspondence between $f/b$ and the velocity divergence as established by
    the continuity equation.
    When an interacting matter component is involved, these continuity equations
    do not hold anymore.
    We will now see what quantity corresponds to the velocity divergence
    $\theta_{\gal}$ in an interacting \gls{de} model.

    We need to start over from the baryons and \gls{dm} continuity equations in
    the interacting model to write a continuity equation for matter on which a
    continuity equation for galaxies can be based.
    The two matter fluids now behave differently, one coupled to the dark energy
    fluid and the other uncoupled.
    For baryons, we still have $\delta'_{\baryons} + \theta_{\baryons} = 0$.
    With $Q_0^{\dm} = - 3 \mathcal{H} \zeta \bar \rho_{\de}$, the \gls{dm}
    continuity equation was obtained in section~\ref{sec:evequations},
    \begin{align}
        \delta'_{\dm} + 3 \mathcal{H} \zeta \frac{\bar \rho_{\de}}{\bar \rho_{\dm}} \delta_{\dm} + \theta_{\dm} = 0.
    \end{align}
    Since the matter density $\rho_{\matter}$ is the sum of the densities
    $\rho_{\baryons}$ and $\rho_{\dm}$, the matter perturbation is
    $\delta_{\matter} = \left( \bar \rho_{\baryons} \delta_{\baryons} + \bar
        \rho_{\dm} \delta_{\dm} \right) / \bar \rho_{\matter}$ and its time
    derivative is
    \begin{align}
        \delta'_{\matter} &= - 3 \mathcal{H} \zeta \frac{\bar \rho_{\de}}{\bar
            \rho_{\matter}} \delta_{\matter} - \frac{\bar \rho_{\baryons}
            \theta_{\baryons} + \bar \rho_{\dm} \theta_{\dm}}{\bar \rho_{\matter}}
    \end{align}
    where we have also used the background evolution equations of each component
    from \eq{eq:fluidconsQ}.
    Substituting the time derivative,
    \begin{align}
        \mathcal{H} \left( \frac{\ud \ln \delta_{\matter}}{\ud \ln a} + 3 \zeta \frac{\bar \rho_{\de}}{\bar \rho_{\matter}} \right) \delta_{\matter}  + \frac{\bar \rho_{\baryons} \theta_{\baryons} + \bar \rho_{\dm} \theta_{\dm}}{\bar \rho_{\matter}} = 0.
    \end{align}
    Recognizing $\theta_{\matter}$ by the term $\left( \bar\rho_{\baryons}
        \theta_{\baryons} + \bar \rho_{\dm} \theta_{\dm} \right)/ \bar
    \rho_{\matter}$, as usual, gives the continuity equation altered by the
    interaction
    \begin{align}
        \mathcal{H} \tilde f_{\matter} \delta_{\matter} + \theta_{\matter} = 0,
    \end{align}
    where $\tilde f_{\matter} \equiv f_{\matter} + 3 \zeta \frac{\bar
        \rho_{\de}}{\bar \rho_{\matter}}$ is the modified growth rate, with the
    usual $f_{\matter} \equiv \frac{\ud \ln \delta_{\matter}}{\ud \ln a}$.

    We maintain the assumption that galaxies trace the matter field via
    $\delta_{\gal} = b \delta_{\matter}$ and $\theta_{\gal} = \theta_{\matter} =
    \theta$, so the galaxy continuity equation is now
    \begin{align}
        \label{eq:modbeta}
        \mathcal{H} \tilde \beta \delta_{\gal} + \theta = 0,
    \end{align}
    with $\tilde \beta \equiv \tilde f/ b$.
    Therefore, this modified growth rate function is the quantity that
    effectively corresponds to the coherent motion of galaxies if there is an
    interaction between \gls{dm} and \gls{de} according to the \gls{cde} model
    considered here.
    Also, the \gls{rsd} parameter that is effectively measured from the power
    spectrum is $\tilde \beta$, since the modeling of the Kaiser effect,
    including its nonlinear features, relies on a continuity equation like
    \eq{eq:modbeta}\footnote{Higher-order terms are generally neglected in the
        continuity equation.} to substitute the velocity divergence in favor of
    the density multiplied by the (thus modified) growth rate.
    The same argument applies about the treatment of nonlinear effects like the
    \gls{fog} (see, for example, refs.~\citep{OhSeon2016,Taruya2010}).
    We then just need to add the term $3 \zeta \frac{\bar \rho_{\de}}{\bar
        \rho_{\matter}}$ to the growth rate $f_{\dm} = \Omega_{\dm}^{\gamma}$
    obtained in section~\ref{sss:CDEmodelpropde} before comparing those
    predictions to the $f \sigma_8$ data.
    In our simplified model with the matter sector composed of dark matter only,
    without baryonic matter, the modified growth rate is $\tilde f_{\dm} =
    f_{\dm} + 3 \zeta \frac{1 - \Omega_{\dm}}{\Omega_{\dm}}$.

    \subsection{The statistical method}
    We perform a posterior likelihood analysis with flat priors for the
    parameters. 
    In order to do that, we employ our analytic formula in computing the
    theoretical growth, implement it in a \gls{mcmc} program in python and carry
    out the data fitting by using a simple Metropolis algorithm
    \citep{Geyer_2011,emcee,Heavens2010}.
    The proposal function in the algorithm is a multivariate normal distribution
    centered at the current state of the Markov chain.
    Its covariance matrix is a diagonal matrix where each diagonal element is
    equal to the square of a fraction of the prior interval of its corresponding
    parameter, adjusted by hand to give an acceptance ratio roughly between
    $0.2$ and $0.5$ in the Metropolis algorithm \citep{emcee}. 
    The likelihoods are computed as $\log \mathcal{L} = - \sum_{i=1}^{N}
    \log\left(\sigma_i \sqrt{2\uppi}\right) - \chi^2/2$, with
    \begin{align}
        \chi^2 = \sum_{i=1}^{N} \frac{\left[f\sigma_8^{\text{\scriptsize
                        (obs)}}(z_i) - \tilde f \sigma_8^{\text{\scriptsize (th)}}(z_i)\right]^2}{\sigma_i^2}.
    \end{align}
    $N$ is the number of points in the dataset, $\sigma_i$ the errors in the
    measurements, ``obs'' stands for the observed data and ``th'' is our
    theoretical prediction by using the analytic formula on the growth. We then
    compute the unnormalized posterior $P(X|D) \propto P(D|X) \, \pi(X)$ for the
    parameter-space point $X$ given the dataset $D$, according to the Bayesian
    theorem, where $P(D|X)$ is the likelihood $\mathcal{L}$ and $\pi(X)$ is the
    prior.
    Our \gls{mcmc} code evolves the chains checking for convergence after each
    $N_{\text{\scriptsize steps}}$ and keeps running until they match the
    convergence criteria.
    The starting points are chosen randomly with uniform probability within the
    prior ranges for each parameter.

    \subsubsection{Convergence criterion}
    For monitoring the convergence of the chains, we implemented the
    multivariate extension by \citet{Brooks_1998} of the method proposed by 
    \citet{Gelman_1992}.
    Having the starting points of eight chains chosen randomly, we run the
    chains through $2 N$ iterations discarding the first $N$ to avoid the burn-in
    period.
    The choice of discarding half of chain sample is an attempt to maximize the
    efficiency in the convergence diagnosis.
    Keeping more iterations would make the samples vary too little, causing the
    detection to be attained later than necessary.

    Let us consider a $p$-dimensional vector parameter $\theta$, a random variable with mean
    vector $\mu$ and variance $\sigma^2$ which we want to estimate through the
    \gls{mcmc} method with $m$ independent chains. 
    We write $\theta^{(i)}_{jt}$ denoting its $i$-th element in chain $j$ at
    time (step) $t$.
    We take the estimate $\hat{\mu} = \bar{\theta}_{\cdot\cdot}$, with the bar
    denoting an average over the set corresponding to the index that has been
    replaced with a dot.
    In this case, the average is over the sequence of $N$ elements of the sample
    chain and then over the $m$ chains.
    To estimate the variance, we calculate the $p$-dimensional between-chain
    covariance matrix $B/N$ defined by
    \begin{align}
        \label{eq:betweenseqvar}
        \frac{B}{N} = \frac{1}{m - 1} \sum_{j=1}^{m} \bigl( \bar \theta_{j \cdot} -
            \bar \theta_{\cdot\cdot} \bigr) \bigl( \bar \theta_{j \cdot} - \bar
            \theta_{\cdot \cdot} \bigr)^{\dagger}
    \end{align}
    and the within-chain covariance matrix
    \begin{align}
        W = \frac{1}{m\left(N -1\right)} \sum_{j=1}^m \sum_{t=1}^N
        \bigl(\theta_{jt} - \bar \theta_{j \cdot} \bigr) \bigl( \theta_{jt} -
        \bar \theta_{j \cdot} \bigr)^{\dagger}.
    \end{align}
    The covariance matrix $\sigma^2$ could be estimated by a weighted average of
    $B$ and $W$,
    \begin{align}
        \hat\sigma_{\tplus}^2 = \frac{N-1}{N} W + \frac{B}{N}.
    \end{align}
    However, this would be an unbiased estimate of the true covariance
    $\sigma^2$ only if the starting points of the chains were drawn from the
    target distribution.
    Instead, this overestimates $\sigma^2$ if the distribution of the starting
    points is appropriately overdispersed, as it needs to be in order for the
    \gls{mcmc} algorithm to sample a large volume of the parameter space thus
    enabling the algorithm to be more likely to detect the eventual existence of
    multiple modes in the distribution.
    We account for the sampling variability of $\hat \mu$ using the covariance estimate
    \begin{align}
        \hat V = \hat \sigma_{\tplus}^2 + \frac{B}{mN} = \frac{N-1}{N} W + \left(1 +
            \frac{1}{m} \right) \frac{B}{N}.
    \end{align}
    We can then monitor $\hat V$ and $W$, determining convergence when a certain
    distance measure between the two covariance matrices indicates that they are
    satisfactorily close.
    This distance must be a rotationally invariant measure.
    In the one-dimensional case, we would use the square root of the variance
    ratio $R = \bigl(\hat V/\sigma^2 \bigr)^{1/2}$, which is called the \gls{srf}, or
    better yet its overestimate $\hat R = \bigl( \hat V / W \bigr)^{1/2}$, underestimating
    $\sigma^2$, unknown, by $W$.
    This is the \gls{psrf}.
    In the multidimensional case, the \gls{mpsrf} is a scalar measure that
    approaches \num{1} (from above) as convergence is achieved.
    It is given by the maximum \gls{srf} of the linear projection of $\theta$
    along any arbitrary vector $\alpha$,
    \begin{align}
        \hat R^p = \left( \max_{\alpha} \frac{\alpha^{\dagger} \hat V
                \alpha}{\alpha^{\dagger} W \alpha} \right)^{1/2}.
    \end{align}
    It is possible to show \citep{Mardia1980}, however, that $\bigl(\hat R^p
    \bigr)^2$ is equal to the largest eigenvalue of the positive definite matrix $W^{-1}
    \hat V$, as long as both $\hat V$ and $W$ are nonsingular, positive definite
    and symmetric.

    Convergence is achieved when $\hat R^p - 1$ is smaller than some required
    precision $\epsilon$. 
    In our runs, which we describe next, we sought for convergence within
    $\epsilon \sim \num{e-6}$, the \gls{lcdm} case going even further at
    $\epsilon \sim \num{e-7}$, a higher precision achieved
    without extra computational cost due to the reduced number of parameters in
    the \gls{lcdm} model.

    \subsection{The results}
    \label{ss:results}
\begin{figure}[t]
        \centering
        \includegraphics[width=0.6\textwidth]{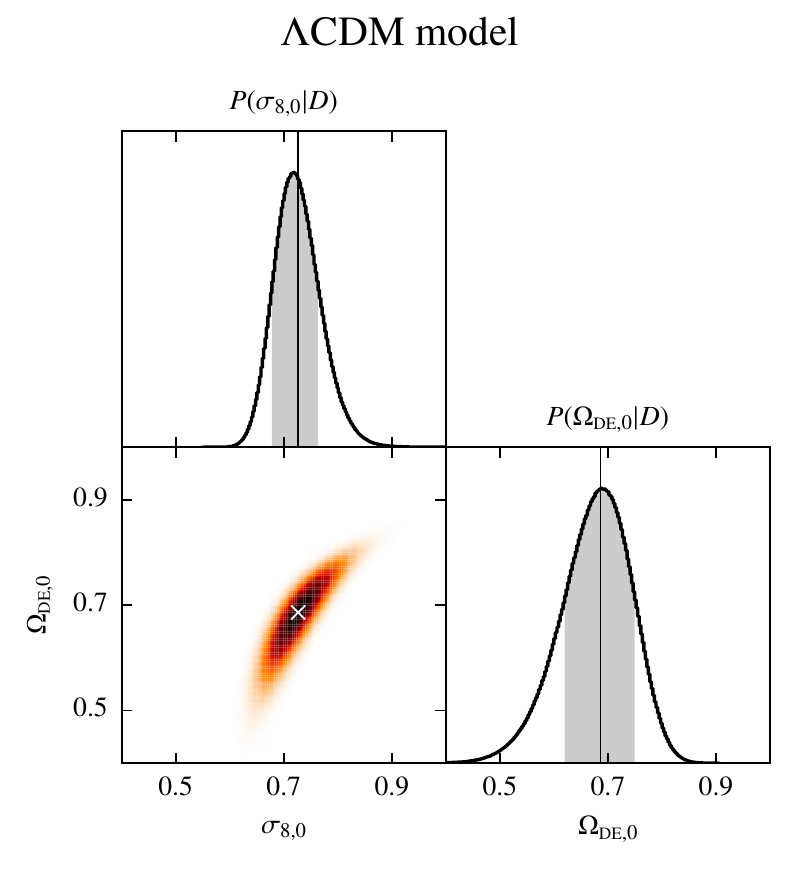}
        \caption{Histograms for the values of the \gls{de} density parameter and
            dark matter \gls{rms} fluctuation today at the scale of
            $8\,h^{-1}\,\si{\mega\parsec}$ in the \gls{lcdm} model.  The
            vertical thin lines mark the best-fit values, and the grey area
            under the histograms show the $1\sigma$ \gls{cl}. In the 2D
            histogram, the colors map the parameter space points to their
            unnormalized posterior values, from white (lowest values) to black
            (highest values), with shades of orange representing intermediate
            values. The white cross marks the best-fit point.}
        \label{fig:LCDMgrid}
    \end{figure}

    For comparison purposes, we first constrain a simple \gls{lcdm} model with
    the two free parameters $\sigma_{8,0}$ and $\Omega_{\de,0}$. Their best-fit
    values are used in the subsequent analysis when we compare the fitting of
    our models to the \gls{lcdm}'s fitting with the same data in the end of this section.
    The $f\sigma_8$ data from table~\ref{tab:data} provide the following $1\sigma$ \gls{cl} for the parameters: $\sigma_{8,0} = 0.7195^{+0.0440}_{-0.0415}$, $\Omega_{\de,0} = 0.6889^{+0.0606}_{-0.0691}$, with the best-fit values $\sigma_{8,0} = 0.7266$ and $\Omega_{\de,0} = 0.6864$ (see figure~\ref{fig:LCDMgrid}). 
    The priors used, always flat in this and in all subsequent analyses, were $\left[ 0.4, 1.0 \right]$ for both parameters and we summarize the results in table~\ref{tab:allmodels}.
    The growth rate determined by the \gls{eos} parameters is
    \begin{align}
        \text{\gls{lcdm}:}    & \qquad f(\Omega_{\dm}) = \left( \Omega_{\dm} \right)^{0.5455 + 0.0073 \left(1 - \Omega_{\dm} \right)}
    \end{align}
    regardless of the resulting best-fit $\sigma_{8,0}$ and $\Omega_{\de,0}$. 
    The growth index today is $\gamma = 0.5505$, up to first order in $\Omega_{\de}$.
    In the following, we present the results for the interacting \gls{de} models.

    \subsubsection{The coupled \gls{de} models}
    \label{sss:model1}
    \begin{figure}[!t]
        \centering
        \subfloat[\label{fig:fdiffxi}Effect of $\zeta$ on $\tilde f(z)$ (top) and
        $\sigma_8(z)$ (bottom
        panel).]{\includegraphics[width=0.47\textwidth]{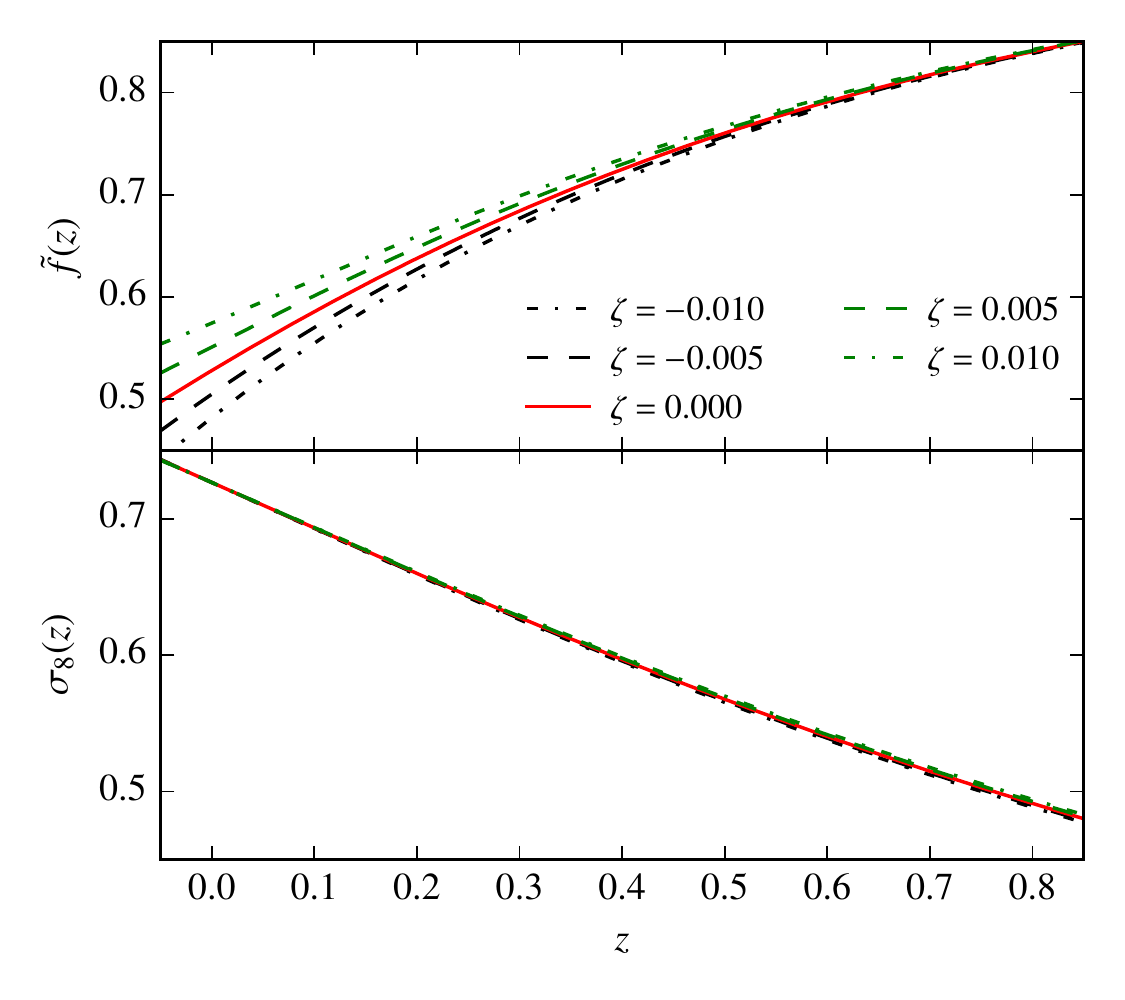}}
        \hspace{16pt}
        \subfloat[\label{fig:fs8diffxi}Influence of different values of $\zeta$
        on the product $\tilde f \sigma_{8}$.]{\includegraphics[width=0.47\textwidth]{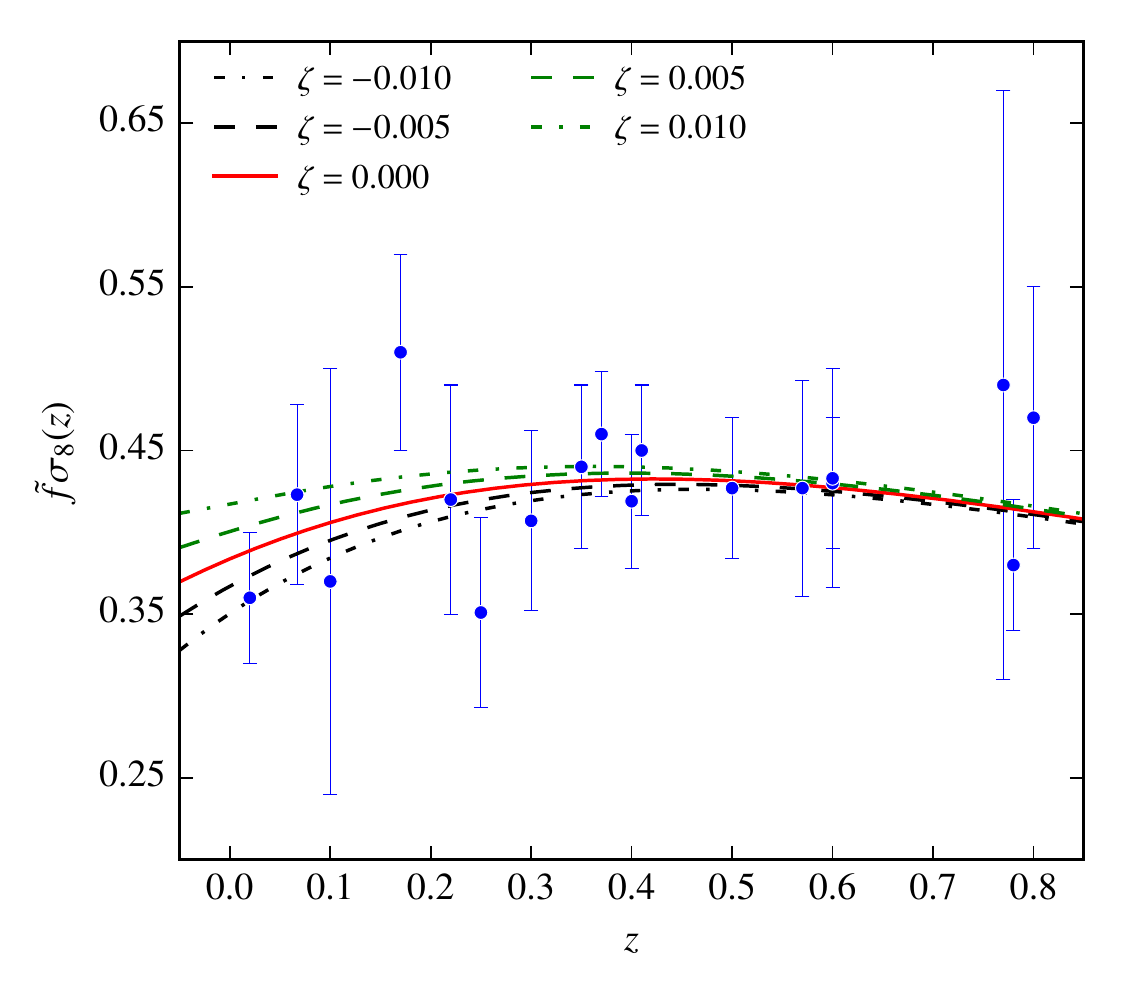}}
        \caption{Evolution of the growth of structures in the coupled \gls{de} model for varying values of the coupling $\zeta$. The negative values (black lines) correspond to the \gls{cqde} model and the positive values (green lines) to the \gls{cpde} model. In both cases we use $w_0 = -1$ for simplification, since we are interested in seeing the effect of the coupling only. The red line is the \gls{lcdm} result. The data from table~\ref{tab:data} are also plotted in (b).}
        \label{fig:differentxi}
    \end{figure}
    Besides $\Omega_{\de,0}$ and $\sigma_{8,0}$, \gls{cde} has other free parameters: $w_0$, $w_1$ and the coupling constant $\zeta$.
    However, before trying to constrain all these parameters together, we first fix $w_1 = 0$ and see if we can have a good indication of $w_0 \ne -1$.
    Not being able to constrain $w_0$ alone in the equation of state means that we will certainly not be able to constrain $w_0$ and $w_1$ together.
    We show in figure~\ref{fig:differentxi} the effect of the interaction on
    $\tilde f(z)$, $\sigma_8(z)$ and on the product $\tilde f \sigma_8(z)$ with $\Omega_{\de,0}$ and $\sigma_{8,0}$ fixed at their \gls{lcdm} best-fit values and with $w_0 \rightarrow -1$.
    In \ref{fig:fdiffxi} (top panel) we can clearly see influence of the interaction on the growth rate.
    The coupling $\zeta$ causes a shift of opposite sign to the growth rate $f$
    (not shown), but a larger shift of equal sign to the modified rate $\tilde
    f$, the shift getting larger as $z$ gets closer to zero.
    The impact of the interaction on $\sigma_8$ (bottom panel is barely
    perceptible.

    We choose the priors based on our comparison with the numerical result for
    $f(z)$, given in section~\ref{ss:numcomparison}.
    As discussed in section~\ref{ss:CDEmodel}, in order to avoid changing the
    sign of the coefficient of $\delta_{\dm}$ and to keep discrepancies with
    respect to the numerical solutions small, values of $\zeta$ should be small,
    of the order \num{e-2}, so we use the prior $\left[ 0 , 0.01 \right]$ for
    $\zeta$ in the phantom case and  $\left[ -0.01, 0 \right]$ in the
    quintessence case.
    $\Omega_{\de,0}$ can be assumed any value in the interval $\left(0.0, 1.0 \right]$.
    \begin{table}[tb]
        \centering
        \setlength{\tabcolsep}{11.6pt}
        \caption{Priors, best-fit values and $1\sigma$ \gls{cl} ranges for the parameters of all models. Central values are shown only for reasonably well constrained parameters.}
        \label{tab:allmodels}
        \sisetup{table-format=1.4}
        \begin{tabular}{@{}lccSc@{}}
            \toprule
            Model   &   Parameter   &   Prior   &   {Best-fit}    &   $1\sigma$ \gls{cl}  \\
            \midrule
            \multirow{2}{*}{\gls{lcdm}}   &   $\sigma_{8,0}$	 & 	$\left[0.4, 1.0\right]$	 & 	0.7266	 & 	$0.7195^{+0.0440}_{-0.0415}$	\\
                                    &   $\Omega_{\de,0}$	 & 	$\left[0.4, 1.0\right]$	 & 	0.6864	 & 	$0.6889^{+0.0606}_{-0.0691}$	\\            
            \midrule
            \multirow{4}{*}{\gls{cpde}}   &   $\zeta$	 & 	$\left[0.00,
                0.01\right]$	 &  \num{7.8e-5}	 & 	$\left[0.0034, 0.0100\right]$	\\
                    &   $\sigma_{8,0}$	 & 	$\left[0.2, 1.4\right]$	 & 	0.6750
                    & 	$0.6322^{+0.0473}_{-0.0293}$	\\
                    &   $\Omega_{\de,0}$	 & 	$\left(0.0, 1.0\right]$	 & 0.6712	 & 	$0.6939^{+0.0652}_{-0.0731}$	\\
                    &   $w_0$	 & 	$\left[-3.0, -1.0\right)$	 & 	-1.4173	 & $\left[-2.1042, -1.0000\right]$	\\            
            \midrule
            \multirow{4}{*}{\gls{cqde}}   &   $\zeta$	 & 	$\left[-0.01,
                0.00\right]$	 & 	-0.0100	 & 	$\left[-0.0069, 0.0000\right]$	\\
                    &   $\sigma_{8,0}$	 & 	$\left[0.2, 1.4\right]$	 & 	0.7230  & $0.7513^{+0.1262}_{-0.0598}$	\\
                    &   $\Omega_{\de,0}$	 & 	$\left(0.0, 1.0\right]$	 & 	0.6533	 & $0.7032^{+0.0667}_{-0.0705}$	\\
                    &   $w_0$	 & 	$\left(-1.0, -0.3\right]$	 & 	-0.9977	 & $\left[-1.0000, -0.5552\right]$	\\            
            \midrule
            \multirow{3}{*}{$w$CQDE}    &   $\zeta$	 & 	$\left[-0.01,
                0.00\right]$	 & 	-0.0100	 & 	$\left[-0.0100, -0.0031\right]$	\\
                &   $\sigma_{8,0}$	 & 	$\left[0.2, 1.4\right]$	 & 	0.7240	 &
                $0.7166^{+0.0412}_{-0.0386}$	\\
                &   $\Omega_{\de,0}$	 & 	$\left(0.0, 1.0\right]$	 & 	0.6546 & 	$0.6737^{+0.0512}_{-0.0702}$	\\
            \bottomrule
        \end{tabular}
    \end{table}    
    \begin{figure}[bt]
        \centering
        \includegraphics[width=0.75\textwidth]{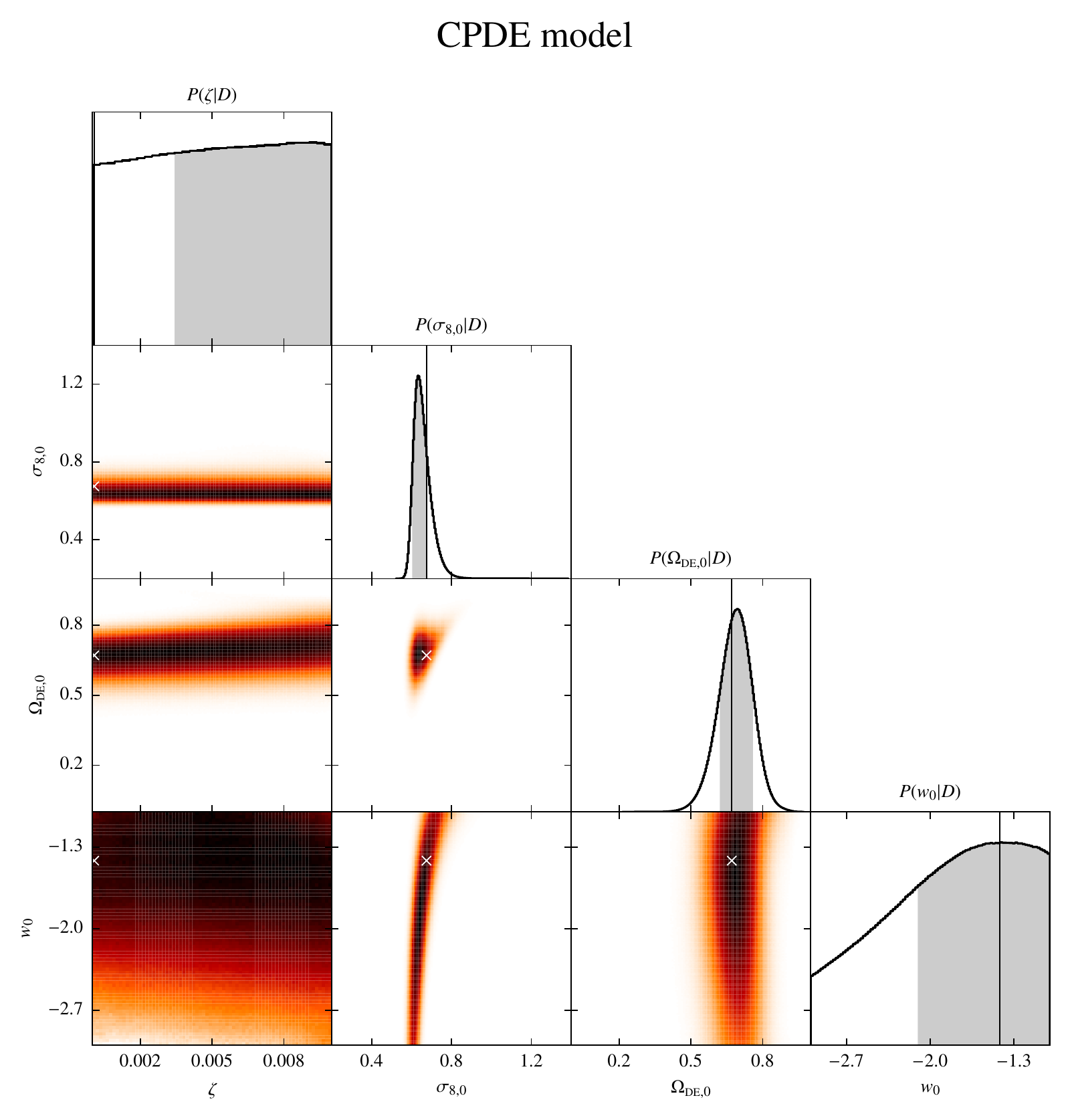}
        \caption{Histograms for the free parameters of \gls{cpde}. The vertical
            thin lines mark the best-fit values, and the grey area under the
            histograms show the $1\sigma$ \gls{cl}. In the 2D histograms, the
            colors map the parameter space points to their unnormalized
            posterior values, from white (lowest values) to black (highest
            values), with shades of orange representing intermediate values. The
            white crosses mark the best-fit point. Due to the large
            uncertainties in the measurements of $f\sigma_8(z)$, the data could
            not constrain the interaction and the \gls{eos} parameter.}
        \label{fig:phantomgrid}
    \end{figure}
    \begin{figure}[bt]
        \captionsetup[subfigure]{position=top}
        \centering
        \subfloat[\gls{cqde} model]{\label{fig:model2Qgrid}\includegraphics[width=0.56\textwidth]{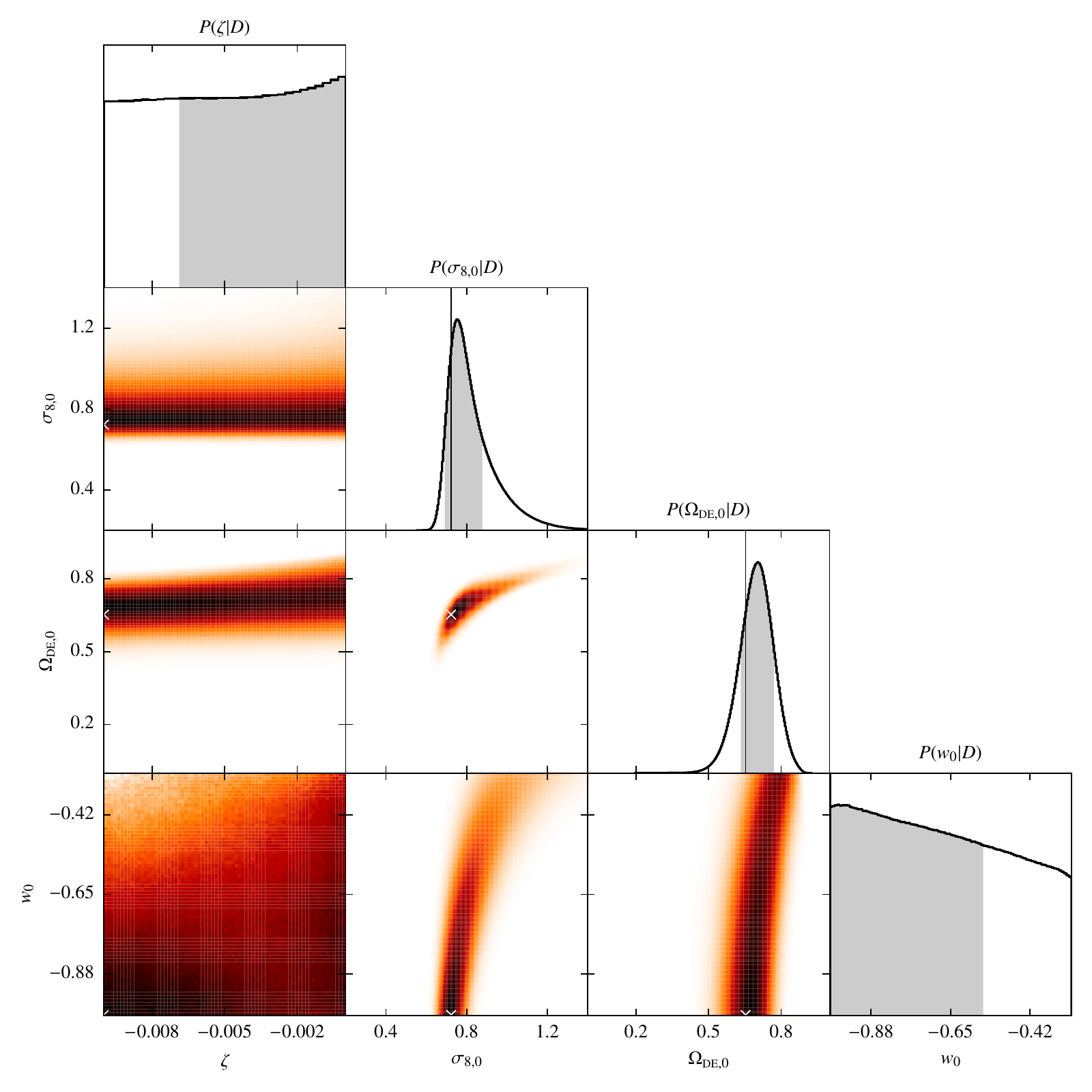}} 
        \subfloat[$w$CQDE model]{\label{fig:model2Lgrid}\includegraphics[width=0.43\textwidth]{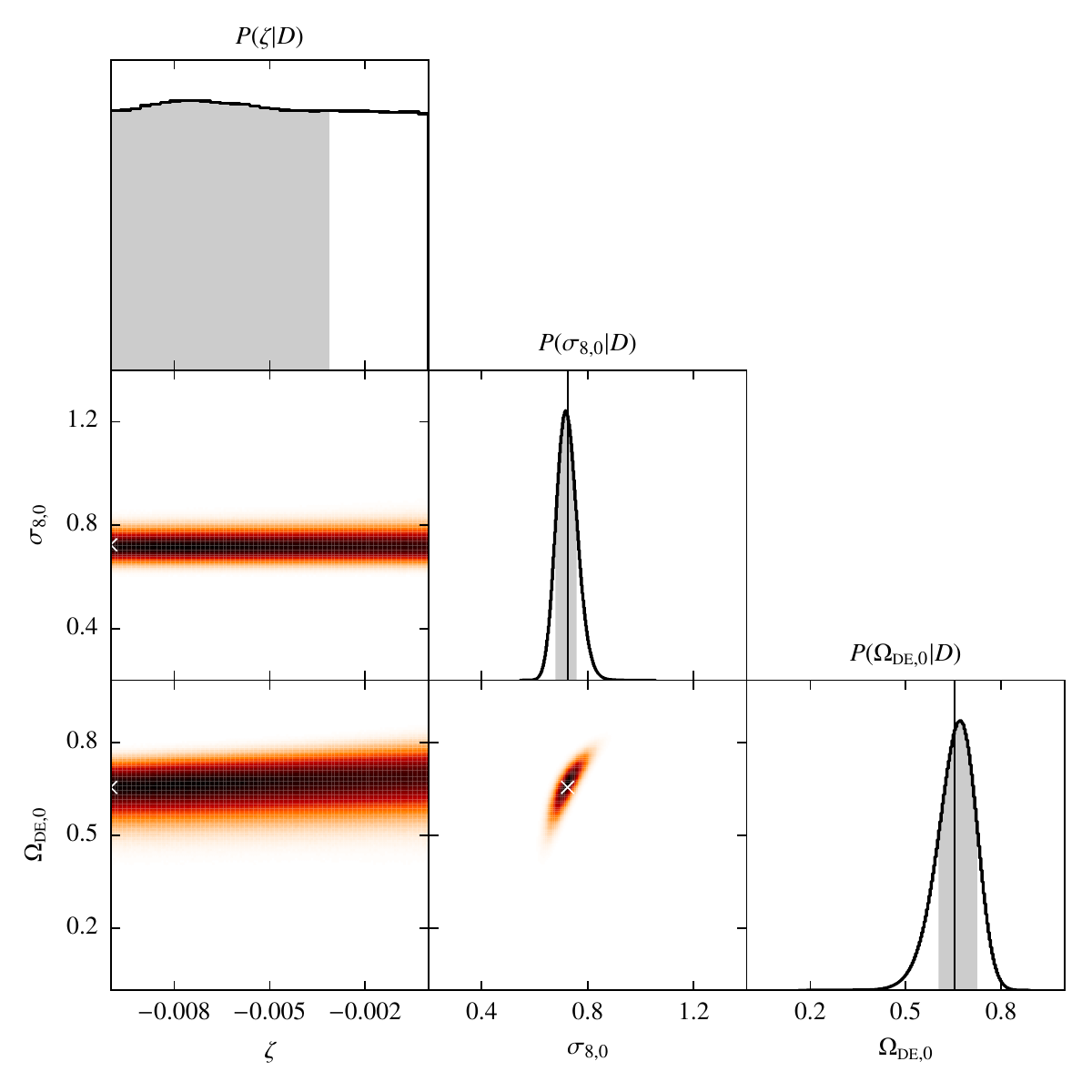}}
        \caption{Marginalized posterior distributions for (a) \gls{cqde} and (b)
            $w$CQDE models. The vertical thin lines mark the best-fit values,
            while the grey areas under the histograms in the diagonal show the
            $1\sigma$ \gls{cl}.  In the 2D histograms, the colors map the
            parameter space points to their unnormalized posterior values, from
            white (lowest values) to black (highest values), with shades of
            orange representing intermediate values. The white crosses mark the
            best-fit point. As we can see from the results of $w$CQDE, fixing
            the \gls{eos} parameter is not sufficient to constrain the
            interaction coupling in the already too tight prior.}
        \label{fig:quintessencegrid}
    \end{figure}

    Table~\ref{tab:allmodels} summarizes the priors and the fitting results and we show in figures~\ref{fig:phantomgrid} and \ref{fig:model2Qgrid} the marginalized distributions for \gls{cpde} and \gls{cqde}, respectively.
    We prefer to express the $1\sigma$ \gls{cl} intervals of the unconstrained parameters without reporting a central value.
    Because of the large uncertainties of the data, the method was not able to
    constrain $w_0$ and $\zeta$ with $f \sigma_8$ data alone, as can be seen from the histograms of the marginalized distributions.
    This hints the fact that such set of parameters can only be better
    constrained if we combine the $f\sigma_8$ data with other kinds of observations, e.g.~the \gls{cmb}.
    The best-fit values encountered lead to the growth rates
    \begin{align}
        \text{\gls{cpde}:}     & \qquad f(\Omega_{\dm}) = \left( \Omega_{\dm}
        \right)^{0.5371 + 0.0058 \left( 1-  \Omega_{\dm} \right)}, \\
        \text{\gls{cqde}:}     & \qquad f(\Omega_{\dm}) = \left( \Omega_{\dm}
        \right)^{0.5290 - 0.0147 \left( 1-  \Omega_{\dm} \right)},
    \end{align}
    for the two models as functions of $\Omega_{\dm}$.
    The best-fit $\Omega_{\de,0}$ gives, for each model, the growth index today
    $\gamma = 0.5410$ and $\gamma = 0.5194$ respectively, up to first order in the density parameter.

    \subsubsection{On the unconstrained parameters}
    The models considered in our work cannot have all their parameters
    satisfactorily constrained due to the large uncertainties in the
    measurements of the large-scale structure.
    This difficulty motivated us to try to obtain a more conclusive
    determination of the interaction constant by fixing one more parameter,
    $w_0$ in the equation of state.
    We analyze the case of \gls{cqde} with the \gls{eos} fixed in its best-fit value
    $w_0 = -0.997728$.
    The choice of \gls{cqde} over \gls{cpde} is because this class of models gives,
    according to \citet{Gavela2009}, the best fit to \gls{lss}
    data.\footnote{Which model gives the best fit to the data that we used here
        could be evaluated by comparing their Bayesian evidences. However, this
        analysis is out of the scope of this work.}
    We then run this \gls{cqde} model with the \gls{eos} parameters fixed at $w_0 =
    -0.997728$ and $w_1 = 0$, which we call \gls{wcqde}. 
    The results are shown in figure~\ref{fig:model2Lgrid} and in
    table~\ref{tab:allmodels}.
    We have obtained the growth rate
    \begin{align}
        \text{\gls{wcqde}:}   &  \qquad f(\Omega_{\dm}) =
        \left(\Omega_{\dm}\right)^{0.5290 - 0.0147 \left(1 - \Omega_{\dm}\right)},
    \end{align}
    with today's value of the growth index $\gamma = 0.5194$.
    This pretty much coincides with the \gls{cqde} result, since the best-fit
    values of all parameters are practically identical.

    We see that even when we keep the equation of state fixed, although the region of
    $1\sigma$ \gls{cl} has been considerably reduced for $\sigma_{8,0}$ and
    $\Omega_{\de,0}$, the growth of structure data cannot constrain very well
    all the parameters either because the measurements are not very precise or
    the prior is too tight.
    Relaxing this prior for $\zeta$ would compromise the analysis, as the results
    for $f$ would not be so reliable, as discussed in
    section~\ref{ss:numcomparison}.
    This last result reinforces the need of additional observables in order to
    get fully satisfactory constraints and make assertive conclusions about a
    possible detection of a \gls{de}-\gls{dm} interaction.

    Indeed, \citet{YangXu2014apr} used \gls{cmb}, \gls{bao} and
    \gls{sneIa} in addition to $f\sigma_8$ data to constrain an interacting $w$CDM
    model (IwCDM) which is equivalent to our \gls{cqde} model.
    \citet{Murgia2016} also combined \gls{cmb}
    temperature and polarization, gravitational lensing and supernovae data with
    \gls{bao}/\gls{rsd} data to constrain their models MOD1 and MOD2, identical
    to our models \gls{cqde} and \gls{cpde}, respectively.
    \citet{AndreXiaodong2016} have combined the latest Planck
    \gls{cmb} data, \gls{bao}, \gls{sneIa}, $H_0$ and \gls{rsd} data to
    constrain several parameters of their models, which also include our models
    \gls{cqde} and \gls{cpde} (models I and II in ref.~\citep{AndreXiaodong2016}).
    In all these works, the authors 
    obtained the growth by numerically computing the perturbation equations and
    compared with observational datasets.
    Their results are consistent with our treatment by employing the analytic
    formula on computing the growth.
    All these results converge that $f\sigma_8$ data alone cannot help to
    constrain well the model parameters due to the large uncertainty of the
    current data.

    \subsubsection{Comparing the growth in different models}
\begin{figure}[ptb]
        \centering
        \includegraphics[width=0.82\textwidth]{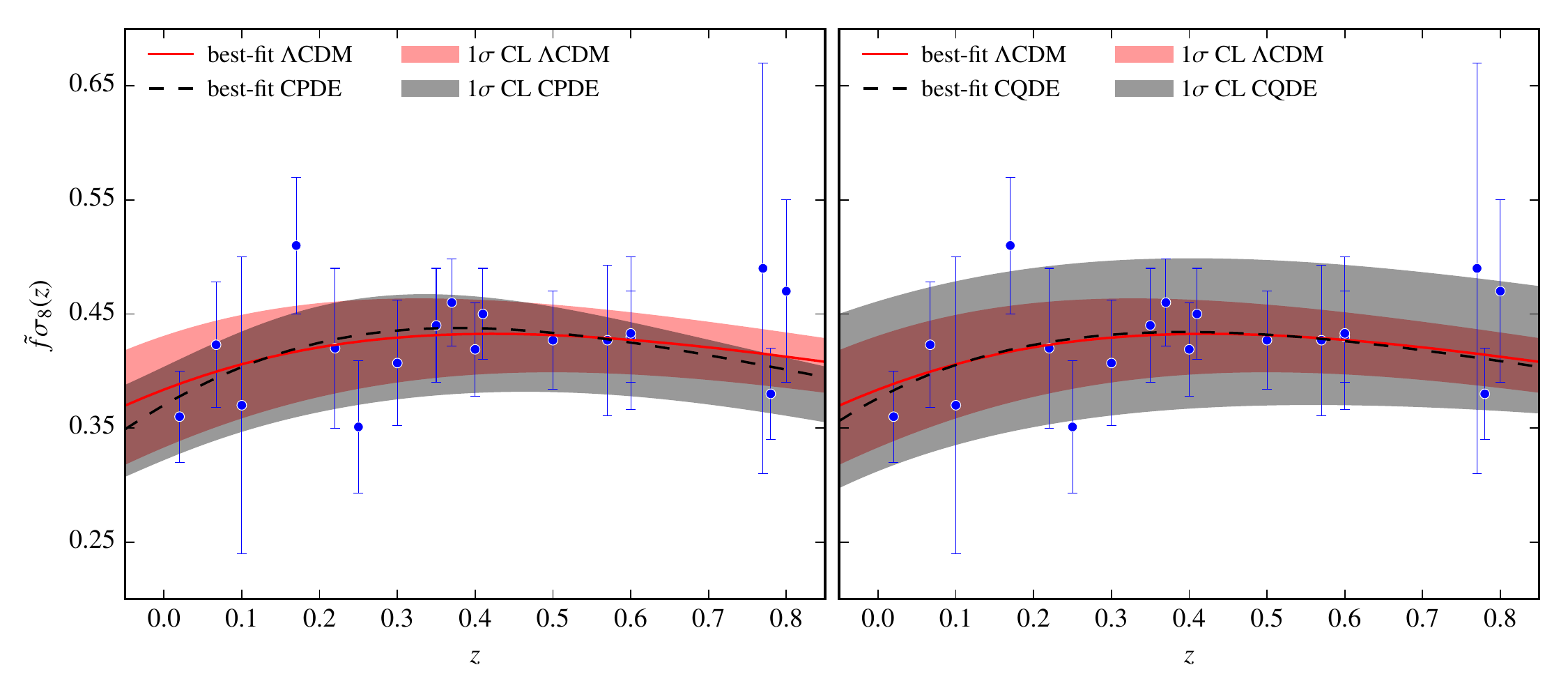}
        \caption{Comparisons of best-fit and $1\sigma$-range $\tilde f \sigma_8(z)$ between \gls{cpde} and \gls{lcdm} (left panel) and between \gls{cqde} and \gls{lcdm} (right panel). The blue data points are listed in table~\ref{tab:data}.}
        \label{fig:fs8onesigma}
    \end{figure}

    In figure~\ref{fig:fs8onesigma} we plot separately each of the interacting
    models' best-fit $\tilde f \sigma_8(z)$, together with the \gls{lcdm}'s best-fit
    over the redshift range of the data.
    \begin{figure}[ptb]
        \centering
        \includegraphics[width=.82\textwidth]{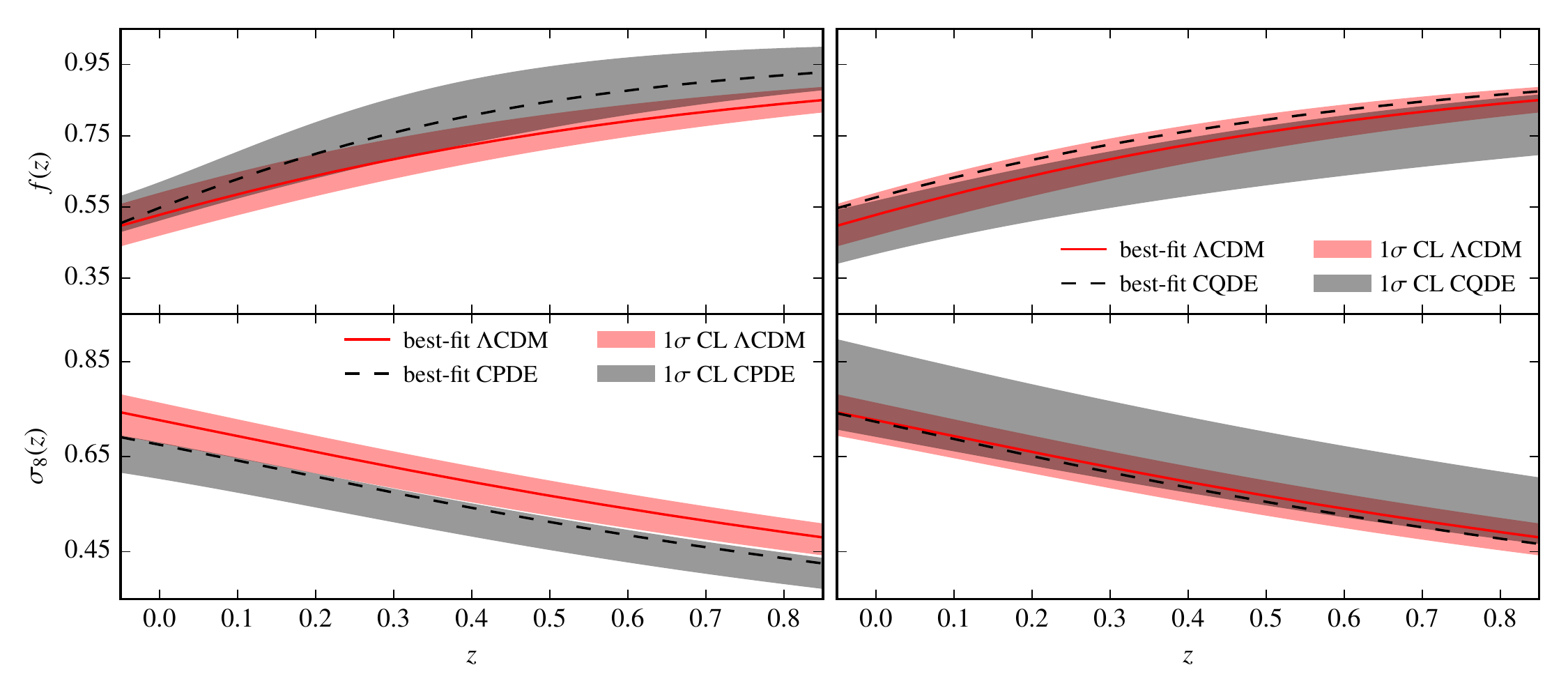}
        \caption{Comparisons of best-fit and $1\sigma$-range $f(z)$ (upper panels) and $\sigma_8(z)$ (lower panels) between \gls{cpde} and \gls{lcdm} (left panels) and between \gls{cqde} and \gls{lcdm} (right panels).}
        \label{fig:fonesigmas8onesigma}
    \end{figure}
    \begin{figure}[pbt]
        \centering
        \includegraphics[width=.82\textwidth]{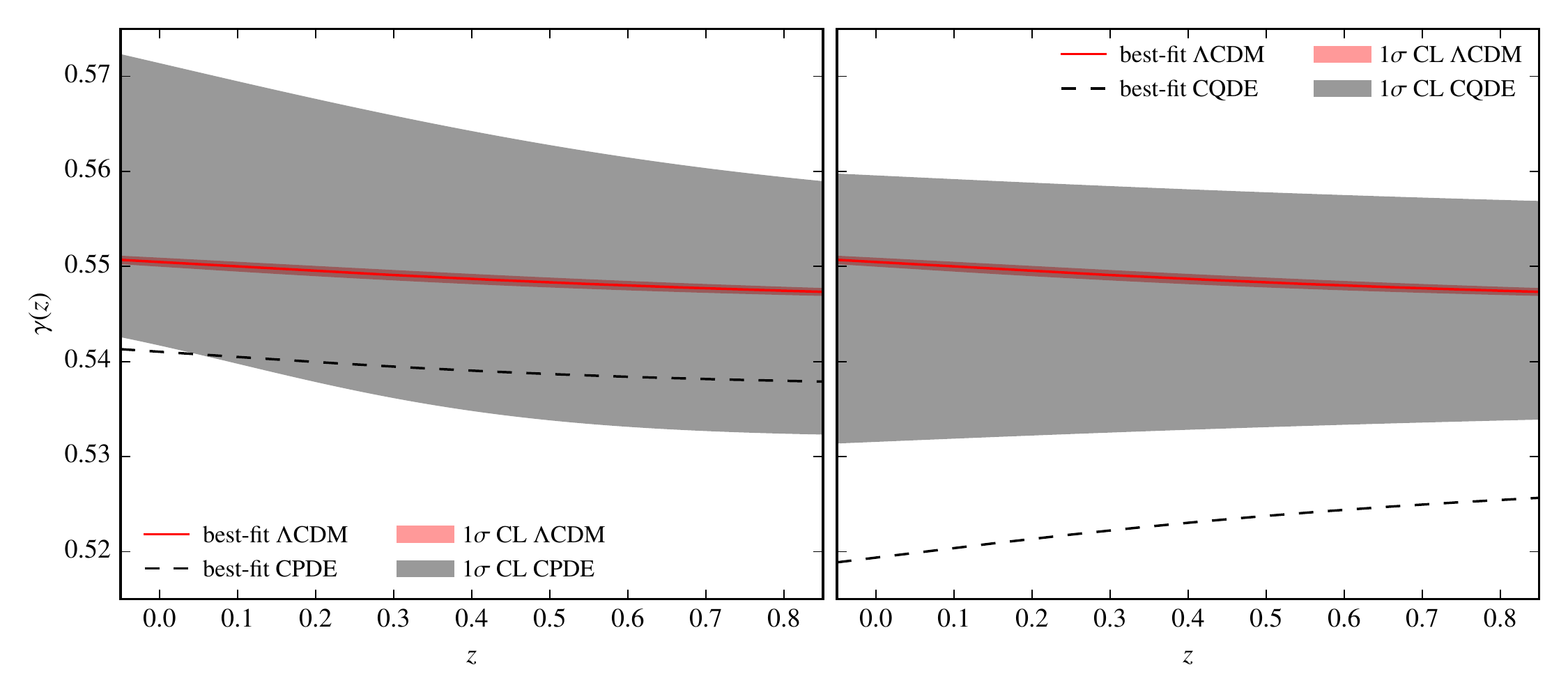}
        \caption{Comparisons of best-fit and $1\sigma$-range $\gamma(z)$ between \gls{cpde} and \gls{lcdm} (left panel) and between \gls{cqde} and \gls{lcdm} (right panel).}
        \label{fig:gammaonesigma}
    \end{figure}
    We note that the best-fit $\tilde f \sigma_8(z)$ in the \gls{cde} models is
    generally lower than that in \gls{lcdm}, but as the redshift decreases, it
    surpasses \gls{lcdm} around redshift $z = 0.5$ and becomes smaller again
    around $z = 0.1$, the difference being slightly larger in the \gls{cpde}
    case (left panel) due to the best-fit point more distant from the \gls{lcdm}
    best-fit.

    The discrepancies between the models become more apparent when we look at
    the $1\sigma$ ranges and at the functions $\tilde f(z)$, $\sigma_8(z)$ and
    $\gamma(z)$ separately.
    In order to do that, we perform linear error propagation on the fitted
    parameters.
    We simplify the task by centralizing the $1\sigma$ \gls{cl} intervals,
    getting the values listed in table~\ref{tab:1CLparams}, then propagate the
    errors through eqs.~\eqref{eq:gammas}, \eqref{eq:fOmegagamma},
    \eqref{eq:normsigma8} and \eqref{eq:Omega1m2}.
    \begin{table}[bt]
        \setlength\tabcolsep{13.2pt}
        \caption{Centralized $1\sigma$ \gls{cl} intervals of the free parameters in \gls{lcdm} and interacting models for the linear error propagation.}
        \label{tab:1CLparams}
        \centering
        \sisetup{table-format=1.4(1)}
        \begin{tabular}{@{}cSSS@{}}
            \toprule
            {Parameter}    &   {\gls{lcdm}}    &   {\gls{cpde}}    &   {\gls{cqde}}    \\
            \midrule
            $\zeta \pm \Delta \zeta$     & 0.0   &  0.0067 \pm 0.0033     &     -0.0034 \pm 0.0034  \\
            $\sigma_{8,0} \pm \Delta \sigma_{8,0}$     &  0.7209 \pm 0.0426   & 0.6412 \pm 0.0383     &     0.7845 \pm 0.0930 \\
            $\Omega_{\de,0} \pm \Delta \Omega_{\de,0}$    &   0.6846 \pm 0.0649 &   0.6900 \pm 0.0692 &   0.7013 \pm 0.0686 \\
            $w_0 \pm \Delta w_0$    &  -1.0 &   -1.5521 \pm 0.5521  &   -0.7776 \pm 0.2224    \\
            \bottomrule
        \end{tabular}
    \end{table}

    Although the \gls{cqde}'s best-fit is closer to \gls{lcdm} than \gls{cpde}'s
    best-fit, \gls{cqde} presents a wider $1\sigma$ range, encompassing the
    entire \gls{lcdm} $1\sigma$ range (see figure~\ref{fig:fs8onesigma}).
    \gls{cpde}'s $1\sigma$ range is about as wide as \gls{lcdm}'s.
    The three models are overall consistent within $1\sigma$ \gls{cl}.

    In figure~\ref{fig:fonesigmas8onesigma} we analyze the unmodified $f(z)$ and
    $\sigma_8(z)$ separately. 
    Faster growth rate means less dark matter in the past and explains the
    corresponding lower  amplitudes $\sigma_8$ for \gls{cpde}, which presents
    higher $f(z)$ compared to \gls{lcdm}.
    The opposite happens in \gls{cqde}.
    The differences between the interacting models and \gls{lcdm} appear to
    enhance as $z$ increases.
    The interacting models' $1\sigma$ ranges are consistent with \gls{lcdm}
    except for \gls{cpde}'s $1\sigma$-range $\sigma_8$, which is only marginally
    consistent with \gls{lcdm} at low redshifts.

    The $1\sigma$ range interval of $\gamma(z)$ in \gls{lcdm} (see
    figure~\ref{fig:gammaonesigma}) is very tight because the only uncertainty
    involved is in the $\Omega_{\de,0}$ parameter, which is well constrained.
    The best-fit growth index is lower than \gls{lcdm}'s best-fit in the two \gls{cde} models, falling closer to \gls{lcdm} in the \gls{cpde} case and outside its own $1\sigma$ range in the \gls{cqde} case. 
    However, their $1\sigma$ ranges are still consistent with \gls{lcdm} in the redshift interval we are considering.

    %%%%%%%%%%%%%% SEPARATION

    \chapter{A tentative detection with non-virialized galaxy clusters}
    \label{ch:nvclusters}

    Cosmological simulations in a \gls{lcdm} universe and observations of galaxy
    and gas distributions suggest that clusters of galaxies are still
    accreting mass and thus are not expected to have achieved equilibrium.
    In this chapter we investigate the possibility to evaluate the departure
    from virial equilibrium in order to detect effects from a \gls{de}-\gls{dm}
    interaction in that balance.
    The \LI\ model, a simple model for the interacting sector, has been
    considered previously
    \citep{Bertolami2007,Bertolami2009,Bertolami2012,Abdalla2009,Abdalla2010,He2010}.
    We now employ optical observations in order to get mass profiles and
    intracluster gas temperatures of a set of galaxy clusters through weak
    lensing and X-ray data.
    We then perform evaluations of observed virial ratios, interaction strength,
    rest virial ratio and departure from equilibrium factors with a Monte Carlo
    method for error estimation.
    This work resulted in a paper published in the Monthly Notices of the Royal
    Astronomical Society \citep{LeDelliouetal2015}.

    \section{The \LI\ model with interaction}
    \label{sss:intmodel}
    Also known as the cosmic energy equation, the \LI\ equation
    \citep{Peebles1993,Avelino2013} generalizes the energy conservation equation
    of a system of non-relativistic particles interacting gravitationally in an
    expanding cosmological background.
    Here we include the dark energy--dark matter interaction in this balance.
    The interaction is described by a phenomenological model similar to that of
    section~\ref{sec:Qrhodm}, with interaction proportional to the \gls{dm}
    density.
    In an \gls{flrw} background,
    \begin{subequations}
        \begin{gather}
            \label{eq:rhodmcont_t}
            \dot \rho_{\dm} + 3 H \rho_{\dm} = 3 H \zeta \rho_{\dm}, \\ 
            \dot \rho_{\de} + 3 H \rho_{\de} (1 + w_{\de}) = -3 H \zeta \rho_{\dm}.
        \end{gather}
    \end{subequations}
    \Eq{eq:rhodmcont_t} is the same as \eq{eq:rhodmcont_tao} but in terms of the
    cosmic time.
    We abandon the notation that used bars to indicate unperturbed quantities
    since we are not using perturbation theory in this chapter.
    Denoting the kinetic and the potential energy densities of the dark
    matter component by $\kin$ and $\pot$, with $\kin  + \pot = \rho_{\dm}$,
    the resulting \LI\ equation for dark matter \citep{He2010} reads
    \begin{align}
        \label{eq:LIdm}
        \dot \rho_{\dm} + H \bigl[ \left( 2 + 3 \zeta \right) \kin 
            + \left(1 + 3\zeta \right) \pot \bigr] = 0.
    \end{align}
    The time derivative vanishes when in equilibrium, yielding the interacting
    virial balance as
    \begin{align}
        \label{eq:virbalance}
        \frac{\kin}{\pot} = - \frac{1 + 3\zeta}{2 + 3\zeta}.
    \end{align}
    However, we want to take into account departures from equilibrium.
    Note that certainty of convergence of the energy density towards
    equilibrium, together with other magnitude restriction considerations
    \citep[e.g.][]{HeWangAbdalla2011}, prescribes from \eq{eq:LIdm} to exclude
    values of $\zeta$ lower than \num{-1/3}.

    \subsection{The non-virialized model}
    \label{sss:nvmodel}
    In order to simplify the calculation, we approximate the departure of $\kin$
    from the equilibrium as proportional to the departure of $\pot$.
    That is, from \eq{eq:virbalance},
    \begin{align}
        \label{eq:approxdept}
        \dkin \approx - \frac{1 + 3\zeta}{2 + 3\zeta} \dpot,
    \end{align}
    so that \eq{eq:LIdm} becomes
    \begin{align}
        \label{eq:forint}
        \left[ 1 - \tfrac{1 + 3\zeta}{2 +3\zeta} \right] \dpot = - H \bigl[ \left(2 + 3\zeta \right) \kin + \left( 1+ 3 \zeta \right) \pot \bigr],
    \end{align}
    then the virial ratio is given by
    \begin{align}
        \label{eq:solveforxi}
        \frac{\kin}{\pot} =  - \frac{1 + 3 \zeta}{2 + 3 \zeta} - \frac{1}{(2 + 3 \zeta)^2} \frac{\dpot}{H \pot}.
    \end{align}
    The virial balance \eq{eq:virbalance} is corrected by a term we call \gls{dfe}.
    For the approximation \eq{eq:approxdept} to remain valid, the \gls{dm} halo
    has to be close to the virial equilibrium.
    We thus need to check that
    \begin{align}
        \label{eq:cond7}
        \left\lvert  \frac{\dpot}{H \pot} \right\rvert \ll \left(2 + 3\zeta \right) 
            \left( 1 + 3\zeta\right).
    \end{align}
    A simple solution for $\zeta$ can be obtained in that case as long as $\kin$, $\pot$, $\dpot$ and $H$ can be observed.
    Dividing \eq{eq:forint} by $H\pot$, we get
    \begin{align}
        \left[ 1 - \frac{1 + 3 \zeta}{2 + 3\zeta} \right] \frac{\dpot}{H \pot} = 
            - \left(2 + 3\zeta \right) \frac{\kin}{\pot}
            - \left(1 + 3\zeta \right);
    \end{align}
    then, in terms of the theoretical virial ratio $\kin/\pot$ and the term $\dpot/H \pot$, we have the quadratic equation
    \begin{align}
        9 \zeta^2 \left( 1 + \tfrac{\kin}{\pot} \right) + 3 \zeta \left(4 \tfrac{\kin}{\pot}  + 3 \right) + \left( 4 \tfrac{\kin}{\pot} + 2 + \tfrac{\dpot}{H \pot} \right) = 0
    \end{align}
    with the solution
    \begin{align}
        \label{eq:quadraticsolution}
        \zeta = \frac{- \left(3 + 4 \tfrac{\kin}{\pot} \right) + \sqrt{1 - 4 \tfrac{\dpot}{H\pot} \left( 1 + \tfrac{\kin}{\pot} \right)}}{6 \left( 1 + \tfrac{\kin}{\pot} \right)}
    \end{align}
    verifying the classical, non-interacting and virialized result $\zeta= 0 $ for $\dpot  = 0$ and $\kin/\pot = - 1/2$.
    Note that \eq{eq:quadraticsolution} is singular at $\kin/\pot = -1$.
    This singularity has its origin in the approximation \eq{eq:approxdept} and
    corresponds to $\zeta \rightarrow - \infty$.

    We now need to evaluate $\rho_{K}$, $\rho_{W}$, $\dot{\rho}_{W}$
    and $H$ from cluster observations.
    However, as we will see in section~\ref{sss:VandDeval}, the factor
    $\dpot/H\pot$ in the \gls{dfe} is not a pure observable and depends on
    $\zeta$.

    \section{The evaluation from clusters}
    \label{sec:evaluation}
    For each cluster, we have access to the total mass distribution through
    weak-lensing observations and to the cluster's kinetic state.
    The former is given through \gls{nfw} profile parameter
    fits, from which we can derive the potential energy, and the latter is
    evaluated through the cluster's X-ray temperature.
    In the following we provide the framework to make contact between such
    observables and the theoretical scheme we have presented above.
    
    \subsection{The \glsentryshort{nfw} density profile and weak-lensing mass}
    The \gls{nfw} density profile \citep{Navarro1996} found in $N$-body
    simulations is widely used to fit observed clusters in order to parametrize
    their mass distribution.
    The profile has two parameters and can be expressed in
    different ways, being the classical form in terms of a scale radius $\rs$
    and the corresponding density $\rhos$:
    \begin{align}
        \rho_{\mathrm{NFW}}(r) = \frac{\rhos}{\tfrac{r}{\rs} \left(1 + \tfrac{r}{\rs} \right)^2}.
    \end{align}
    From spherical collapse considerations, an ``edge'' can be defined at the
    virial radius of a cluster, assumed to extend to a distance $r_{200}$ within
    which the mean density is about \num{200} times the background density.    
    This suggests defining the \gls{nfw} concentration parameter $\nfwc \equiv
    r_{200} / \rs$.
    Integrating the profile yields the mass profile
    \begin{align}
        \label{eq:massprofile}
        M_{\mathrm{NFW}}(r) = \frac{M_{200}}{\ln(1+\nfwc) - \tfrac{\nfwc}{1 + \nfwc}} \left[ \ln \left( 1 + \nfwc \tfrac{r}{r_{200}} \right) - \frac{ \nfwc \tfrac{r}{r_{200}}}{1 + \nfwc \tfrac{r}{r_{200}}} \right],
    \end{align}
    where $M_{200}$ is the mass inside $r_{200}$.
    The density profile in terms of $\nfwc$, $r_{200}$ and $M_{200}$ reads
    \begin{align}
        \label{eq:NFWprofile}
        \rho_{\mathrm{NFW}}(r) = \frac{M_{200}}{4\uppi (r_{200})^3 \left[ \ln (1 + \nfwc) - \tfrac{\nfwc}{1 + \nfwc} \right] } \frac{\nfwc^2}{\tfrac{r}{r_{200}} \left( 1 + \nfwc \tfrac{r}{r_{200}} \right)^2}.
    \end{align}

    \subsection{The potential and kinetic energy density evaluations}
    With the profile \eqref{eq:NFWprofile}, the potential energy density is simply
    \begin{align}
        \label{eq:potstate}
        \pot \equiv - \frac{4 \uppi}{\tfrac{4}{3}\uppi(r_{200})^3} \int_0^{r_{200}} \frac{\rho(r) G M(r)}{r} r^2 \, \ud r = - \frac{3 G (M_{200})^2}{4 \uppi (r_{200})^4 f_{\nfwc}},
    \end{align}
    with
    \begin{align}
        f_{\nfwc} \equiv \frac{\left(1 + \nfwc\right) \bigl[ \ln(1 + \nfwc ) - \nfwc \left( 1 + \nfwc\right)^{-1} \bigr]^2}{\nfwc \left\lbrace \tfrac{1}{2} \bigl[ \left(1 + \nfwc\right) - \left(1 + \nfwc\right)^{-1} \bigr] - \ln(1 + \nfwc) \right\rbrace}.
    \end{align}
    On the other hand, in order to evaluate the kinetic state of the cluster we
    use published X-ray observations, where we just need to obtain the X-ray
    temperature to get the equipartition formula
    \begin{align}
        \label{eq:kinstate}
        \kin = \frac{3}{2} N \frac{\kB \tx}{V} = \frac{9}{8\uppi} \frac{M_{200}}{r_{200}^3} \frac{\kB \tx}{\icgmu\mH},
    \end{align}
    where the equivalent number of particles $N$ is computed from the total mass
    given by weak-lensing as $M_{200}/\icgmu\mH$, $\icgmu$ is the mean molecular
    mass in the intracluster gas, $\mH$ is the proton mass, $V$ is the
    volume and $\tx$ is the observable X-ray temperature.
    An advantage of this method over using a scale relation $\sigx$-$\tx$
    between the galaxy velocity dispersion and the X-ray temperature, as in
    previous works
    \citep{Bertolami2007,Bertolami2009,Bertolami2012}.
    is that the error from the scatter is avoided.
    We use a single compounded temperature $\tx$ extracted from the X-ray flux
    of the central region of radius $r_{500}$, i.e., the radius within which the
    density is \num{500} times the background density, which implies $r_{500} <
    r_{200}$.
    This is justified since this temperature already largely encompasses the
    turnaround of the temperature profile
    \citep{vikhlinin2005,pratt2007,moretti2011}, therefore representing well the
    total density averaged temperature (the so-called virial temperature).
    
    At this point the virial ratio can be evaluated, combining
    eqs.~\eqref{eq:potstate} and \eqref{eq:kinstate}, as
    \begin{align}
        \label{eq:virratio14}
        \frac{\kin}{\pot} = - \frac{3}{2} \frac{r_{200}}{GM_{200}} \frac{\kB \tx}{\icgmu\mH} f_{\nfwc}.
    \end{align}

    \subsection{The departure from equilibrium evaluation}
    \label{sss:VandDeval}
    The \gls{dfe} factor can be rewritten as
    \begin{align}
        - \frac{1}{(2 + 3\zeta)^2} \frac{\dpot}{H \pot} = - \frac{1}{(2 + 3 \zeta)^2} \frac{\ppot}{H\pot} \dot r_{200},
    \end{align}
    with the prime indicating differentiation with respect to $r_{200}$.
    From \eq{eq:potstate} we get
    \begin{align}
        \frac{\ppot}{\pot}  = \frac{\nfwc g_{\nfwc} - 3}{r_{200}}, \quad \text{with} \quad
        g_{\nfwc} \equiv \frac{\ln(1 + \nfwc) - \nfwc\left(1 +
                \nfwc\right)^{-1}}{\tfrac{1}{2}\left(\nfwc+2\right) - \left(1 +
                \nfwc\right)\ln(1+\nfwc)}.
    \end{align}
    $\dot r_{200}$ still remains to be evaluated.
    We write the kinetic density as
    \begin{align}
        \kin = \frac{3}{2} \frac{M_{200}}{V} \sigx^2,
    \end{align}
    in terms of a one-dimensional velocity dispersion $\sigx$, thus defined as
    $\sigx^2 = \frac{\kB\tx}{\icgmu\mH}$.
    Now we define a theoretical average velocity dispersion $v_{\mathrm{th}}$ the
    cluster would have if it were at virial equilibrium, adiabatically evolving
    from the current state.
    The theoretical virial ratio is given by \eq{eq:virbalance}, with a theoretical kinetic density
    \begin{align}
        \rho_{\mathrm{kin,th}} = \frac{3}{2} \frac{M_{200}}{V} (v_{\mathrm{th}})^2.
    \end{align}
    This definition combined with eqs.~\eqref{eq:virbalance} and \eqref{eq:potstate} leads to
    \begin{align}
        (v_{\mathrm{th}})^2 = \frac{2}{3} \frac{1 + 3\zeta}{2 + 3\zeta} \frac{GM_{200}}{r_{200} f_{\nfwc}}.
    \end{align}
    Finally, we evaluate the time evolution of $r_{200}$ by taking its difference with the velocity dispersion
    \begin{align}
        \dot r_{200} = \sigx - v_{\mathrm{th}}.
    \end{align}
    We obtain the \gls{dfe} factor
    \begin{align}
        \label{eq:Out(xi)}
        - \frac{\dpot/H\pot}{(2 + 3\zeta)^2} = - \frac{1}{(2 + 3 \zeta)^2}
        \frac{\nfwc g_{\nfwc} - 3}{Hr_{200}} \left[ \left( \frac{\kB
                    \tx}{\icgmu\mH} \right)^{1/2} - \left(\frac{2}{3}\frac{1+
                    3\zeta}{2+3\zeta}\frac{GM_{200}}{r_{200}f_{\nfwc}}
            \right)^{1/2} \right].
    \end{align}
    With this equation we estimate the \gls{dfe} due to ``standard'' dynamical
    sources (e.g.~cluster collisions, AGN and supernova feedback, dynamical
    friction) combining observations and the dark energy model, leaving no room
    for degeneracy in the determination of $\zeta$.
    The \gls{dfe} presented here appears model dependent in its explicit
    reference to the interaction strength;
    however, the method can use any model we want that gives a definite
    shift to the virial balance.

    \section{Computations for a set of non-virialized clusters}
    Cosmologically interesting observations of clusters are produced in many
    surveys and studies such as \citet{Okabe2010} and \citet{Aghanim2012}.
    In order to maximize our sample, while being able to
    separately evaluate from observations the kinetic and potential energy
    states of each cluster, we have restricted inputs to weak-lensing \gls{nfw} fit
    parameters, X-ray derived $\nfwc_{500}$ \gls{nfw} fits, and X-ray temperatures.

    \subsection{The sample}
    In order to try to minimize any systematics due to observational
    uncertainties, the clusters in our sample should present well determined
    X-ray gas temperature, as well as \gls{nfw} profile fitted to the mass
    distribution obtained with weak lensing observations.

    Most of the 22 clusters in our sample come from \citet{Okabe2010}. 
    Their \gls{nfw} profiles are described by best-fit virial masses $M_{\rmn{vir}}$,
    concentration parameters $\cvir$ and masses $M_{200}$ estimated from
    this three-dimensional model fitting. 
    Those are the Abell clusters A68, A115, A209, A267, A383, A521, A586, A611,
    A697, A1835, A2219, A2261, A2390, A2631 and also RX J1720.1+2638, RX
    J2129.6+0005, ZwCl 1454.8+2233 and ZwCl
    1459.4+4240.
    Their weak lensing data are shown in table~\ref{tbl:subset2}. 
    \renewcommand{\arraystretch}{1.3}
    \begin{table}
        \setlength\tabcolsep{33.4pt}
        \caption{Weak lensing masses $M_{200}$, $M_{\rmn{vir}}$ and
            concentration $\cvir$ for the Okabe's clusters. Masses are in units of $h^{-1}10^{14}M_{\odot}$. }
        \label{tbl:subset2}
        \centering
        \begin{tabular}{@{}lccc@{}}
            \toprule
            Cluster & $M_{200} $ & $\mvir$ & $\cvir$ \\
            \midrule
            A68 & $4.45^{+1.75}_{-1.35}$ & $5.49^{+2.56}_{-1.81}$ & $4.02^{+3.36}_{-1.82}$ \\
            A115 & $4.24^{+2.60}_{-1.79}$ & $5.36^{+4.08}_{-2.45}$ & $3.69^{+5.03}_{-2.04}$ \\
            A209 & $10.62^{+2.17}_{-1.81}$ & $14.00^{+3.31}_{-2.60}$ & $2.71^{+0.69}_{-0.60}$ \\
            A267 & $3.23^{+0.82}_{-0.69}$ & $3.85^{+1.08}_{-0.88}$ & $6.00^{+2.11}_{-1.58}$ \\
            A383 & $3.11^{+0.88}_{-0.69}$ & $3.62^{+1.15}_{-0.86}$ & $8.87^{+5.22}_{-3.05}$ \\
            A521 & $4.58^{+1.00}_{-0.88}$ & $5.85^{+1.45}_{-1.22}$ & $3.06^{+1.01}_{-0.79}$ \\
            A586 & $6.29^{+2.26}_{-1.69}$ & $7.37^{+2.89}_{-2.08}$ & $8.38^{+3.52}_{-2.52}$ \\
            A611 & $5.47^{+1.31}_{-1.11}$ & $6.65^{+1.75}_{-1.42}$ & $4.23^{+1.77}_{-1.23}$ \\
            A697 & $9.73^{+1.86}_{-1.61}$ & $12.36^{+2.68}_{-2.21}$ & $2.97^{+0.85}_{-0.69}$ \\        
            A1835 & $10.86^{+2.53}_{-2.08}$ & $13.69^{+3.65}_{-2.86}$ & $3.35^{+0.99}_{-0.79}$ \\
            A2219 & $7.75^{+1.89}_{-1.60}$ & $9.11^{+2.54}_{-2.06}$ & $6.88^{+3.42}_{-2.16}$ \\
            A2261 & $7.97^{+1.51}_{-1.31}$ & $9.49^{+2.01}_{-1.69}$ & $6.04^{+1.71}_{-1.31}$ \\
            A2390 & $6.92^{+1.50}_{-1.29}$ & $8.20^{+1.93}_{-1.63}$ & $6.20^{+1.53}_{-1.28}$ \\
            A2631 & $4.54^{+0.89}_{-0.78}$ & $5.24^{+1.15}_{-0.98}$ & $7.84^{+3.54}_{-2.28}$ \\
            RX J1720 & $3.48^{+1.28}_{-0.99}$ & $4.07^{+1.65}_{-1.22}$ & $8.73^{+5.60}_{-3.08}$ \\
            RX J2129 & $5.29^{+1.76}_{-1.38}$ & $6.71^{+2.73}_{-1.96}$ & $3.32^{+2.16}_{-1.34}$ \\
            ZwCl 1454 & $2.80^{+1.39}_{-1.03}$ & $3.45^{+2.02}_{-1.36}$ & $4.01^{+3.44}_{-1.96}$ \\
            ZwCl 1459 & $3.77^{+1.17}_{-0.98}$ & $4.40^{+1.50}_{-1.20}$ & $6.55^{+3.34}_{-2.18}$ \\
            \bottomrule
        \end{tabular}
    \end{table}
    We also include four more clusters from \citet{Aghanim2012}: A520, A963, A1914 and A2034 (data in table~\ref{tbl:subset1}).
    \begin{table}
        \setlength\tabcolsep{70pt}
        \caption{Weak lensing masses $M_{500}$ and concentration $\nfwc_{500}$
            for the Planck Collaboration's clusters. Masses are in units of
            $\num{e14}M_{\odot}$. Data from \citet{Aghanim2012}.}
        \label{tbl:subset1}
        \centering
        \begin{tabular}{@{}lcc@{}}
            \toprule
            Cluster & $M_{500}$ & $\nfwc_{500}$ \\
            \midrule
            A520 & $4.1^{+1.1}_{-1.2}$ & $1.4 \pm 0.6$ \\
            A963 & $4.2^{+0.9}_{-0.7}$ & $1.2 \pm 0.2$ \\
            A1914 & $4.7^{+1.6}_{-1.9}$ & $2.0 \pm 0.2$ \\
            A2034 & $5.1^{+2.1}_{-2.4}$ & $1.8 \pm 0.3$ \\
            \bottomrule
        \end{tabular}
    \end{table}
    Weak lensing masses $M_{500}$ and best fitting NFW concentration parameter
    $\nfwc_{500}$ are given instead of $\mvir$, $\cvir$ and $M_{200}$ for these clusters.
    However, the error bars for $\nfwc_{500}$ were estimated from the X-ray data,
    since they are not given by ref.~\citep{Aghanim2012}.
    The spectroscopically determined temperatures $\tx$ are measured within
    $r_{500}$ and are all given in ref.~\citep{Aghanim2012}, with the exceptions of A115 and A697 from \citet{Landry2013} and A611 from \citet{Kenneth2008}. 
    Uncertainties correspond to $1\sigma$ \gls{cl}. 
    Errors in redshifts (see table~\ref{tbl:obsdata}) are not specified but can
    be safely neglected compared to the errors in other quantities (the typical
    spectroscopic redshift error is around \SI{1}{\percent}).

    \subsubsection{The uniformization of the NFW profiles}

    It would be interesting if we had all clusters described by the same parameters, in a uniform way, so we can apply the same method for all of them, in a single code.
    In what follows we describe how we proceed to convert the \gls{nfw} profile parameters for those clusters with given $M_{500}$ and $\nfwc_{500}$ to $M_{200}$ and $\nfwc$.

    In general, within a radius $r_{\Delta}$ we have
    \begin{align}
        \Delta = \frac{M_{\Delta}}{\tfrac{4}{3} \uppi (r_{\Delta})^3 \bar \rho (z) },
    \end{align}
    which results in the radius
    \begin{align}
        \label{eq:rDelta}
        r_{\Delta} = \frac{1}{H(z)} \sqrt[3]{\frac{2 G M_{\Delta} H(z)}{\Delta}}
    \end{align}
    given the mass $M_{\Delta}$.
    For the latter set of clusters, with \gls{nfw} profiles specified by $M_{500}$ and $\nfwc_{500} = r_{500}/\rs$ (rather than $\nfwc_{200}$, which we called just $\nfwc$), the parameter $\rs = r_{200}/\nfwc = r_{500}/\nfwc_{500}$ comes immediately by using \eq{eq:rDelta} for $\Delta=500$.
    Using SymPy \citep{sympy} in python, we get $M_{200}$ and $r_{200}$ by solving simultaneously an equation similar to \eq{eq:massprofile}---but parametrized by $M_{500}$ and $\nfwc_{500}$ and evaluated at $r_{200}$ to give $M_{200}$---and \eq{eq:rDelta} with $\Delta=200$:
    \begin{subequations}
        \label{eq:systemM200r200}
        \begin{align}
            M_{200} &= \frac{M_{500}}{\ln(1 + \nfwc_{500}) - \tfrac{\nfwc_{500}}{1 + \nfwc_{500}}} \left[ \ln \left( 1 + \nfwc_{500} \tfrac{r_{200}}{r_{500}} \right) - \frac{\nfwc_{500} \tfrac{r_{200}}{r_{500}}}{1 + \nfwc_{500} \tfrac{r_{200}}{r_{500}}} \right], \\
            r_{200} &= \frac{1}{H(z)} \sqrt[3]{\frac{GM_{200}H(z)}{100}}.
        \end{align}
    \end{subequations}
    Then we can finally compute $\nfwc = r_{200} /\rs$.

    For the former set, with \gls{nfw} profiles specified by $M_{200}$, $\mvir$ and $\cvir$, ``vir'' would correspond to some $\Delta_{\text{vir}}$ around \num{200} but this value can vary with the redshift.
    Then we proceed as follows.
    We compute $r_{200}$ with \eq{eq:rDelta} and solve
    \begin{align}
        M_{200} = \frac{\mvir}{\ln(1 + \cvir) - \frac{\cvir}{1 + \cvir}} \left[ \ln \left( 1 + \cvir \tfrac{r_{200}}{\rvir} \right) - \frac{\cvir \tfrac{r_{200}}{\rvir}}{1 + \cvir \tfrac{r_{200}}{\rvir}} \right]
    \end{align}
    for $\rvir$. $\Delta_{\text{vir}}$ can also be determined now with $\rvir$ and $\mvir$, inverting \eq{eq:rDelta}.
    Finally, we have $\rs = \rvir/\cvir$ and $\nfwc = r_{200}/\rs$.
    The errors are estimated using a Monte Carlo method that we describe in the
    next section.
    With $\nfwc$ and $M_{200}$, we can now proceed to the computation of the virial ratios.

    \subsection{The Monte Carlo estimation of errors}
    \label{sec:MC}

    We apply a Monte Carlo method to propagate uncertainties through the numerical solutions.
    Multiple random realizations of each cluster are performed, with the observables assuming values drawn from a distribution that reflects the $1\sigma$ confidence intervals from the original asymmetrical uncertainties.
    Then, we carry the computations for all realizations of each cluster and analyze the final distribution of the quantities of interest in order to get their error intervals.

    \subsubsection{Uncertainties in $M_{200}$ and $\nfwc_{200}$}
    These two \gls{nfw} parameters are always positive.
    In order to guarantee that their uncertainties will not lead to negative
    values in any of the random realizations, we choose the log-logistic
    distribution, for it being a non-negative probability distribution whose
    \gls{pdf} and \gls{cdf} have simple analytical forms.
    If $X$ is a random variable following a log-logistic distribution, its \gls{pdf} with parameters $(\alpha, \beta)$ is
    \begin{align}
        f_{X}(x; \alpha, \beta) = \frac{ \left(\beta/ \alpha \right)\left( x/\alpha
            \right)^{\beta-1}}{\left[ 1 + \left( x/\alpha \right)^{\beta} \right]^2}
    \end{align}
    and the \gls{cdf} is given by
    \begin{align}
        F_{X}(x; \alpha, \beta) = \frac{1}{ 1 + \left( x/\alpha \right)^{-\beta}}.
    \end{align}

    For a given observable $X$ with measured value $x^{+\Delta x_1}_{-\Delta x_2}$, we would like the Monte Carlo generating distribution to match the following criteria:
    \begin{enumerate}[(i)]
        \item \label{i:mode} The maximum probability coincides with the nominal measure;
        \item \label{ii:CDF} The probability of $X$ lying between $x-\Delta x_2$ and $x+\Delta x_1$ is \sepercent;
        \item \label{iii:PDF} The \gls{pdf} has the same value at the two points $x-\Delta x_2$ and $x+\Delta x_1$, so the interval in condition \eqref{ii:CDF} corresponds to the \sepercent\ most probable values, i.e., $1\sigma$ \gls{cl}
    \end{enumerate}
    For the log-logistic distribution, these conditions are translated as
    \begin{enumerate}[(i)]
        \item $\alpha \left( \frac{\beta - 1}{\beta + 1} \right)^{1/\beta} = x$ (for $\beta > 1$);
        \item $F_X(x+\Delta x_1; \alpha, \beta) - F_X(x-\Delta x_2; \alpha, \beta) = 0.68$;
        \item $f_X(x-\Delta x_2;\alpha, \beta) = f_X(x+\Delta x_1;\alpha, \beta)$.
    \end{enumerate}
    However, these are too many conditions for a distribution which has only two parameters.
    We choose to relax condition \eqref{i:mode} and solve \eqref{ii:CDF} and \eqref{iii:PDF} for $\alpha$ and $\beta$.
    In practice, our resulting maximum probabilities usually happen to lie very close to $x$.

    When extracting the $1\sigma$ \gls{cl}, we take the opposite direction and
    get a best-fit log-logistic \gls{pdf} for the distributions of $M_{200}$ and
    $\nfwc$,
    now solving (\ref{ii:CDF}) and (\ref{iii:PDF}) for $x-\Delta x_2$ and $x+\Delta
    x_1$. 
    The maximum probability of the distribution is assigned to the nominal value $x$.

    We have also used the log-normal distribution to check whether our choice of
    distribution could be biasing our results.
    The log-normal \gls{pdf} and \gls{cdf} are given by
    \begin{align}
        f_X(x; \mu, \sigma) = \frac{\exp\left\lbrace-\frac{\left( \ln x - \mu
                \right)^2}{2\sigma^2}\right\rbrace}{x \sigma \sqrt{2 \uppi}}, \quad F_X(x; \mu,\sigma) = \frac{1}{2} \erfc \left[ - \frac{\ln x - \mu}{\sigma \sqrt{2}} \right],
    \end{align}
    where $\erfc(x)$ is the complementary error function and $\mu$ and $\sigma$
    are the Gaussian parameters of the distribution of $\ln X$.

    We applied this Monte Carlo procedure for the clusters in our sample. 
    However, the log-normal distribution could not satisfy our requirements
    (\ref{ii:CDF}) and (\ref{iii:PDF}) for all clusters in the first group.
    Nevertheless, we were able to verify in the other cases, where the
    log-normal distribution works, that the confidence intervals obtained with
    the two distributions are very similar, within a few percent of displacement
    between their extremities.
    The maximum probability can vary a little more between the two distributions
    because (\ref{i:mode}) is not being satisfied, but we are more concerned with
    the confidence intervals, since we use uniform distributions for $\nfwc$,
    $M_{200}$ and $\tx$ in the evaluation of the virial ratios, interaction
    strength and departure from equilibrium.
    We believe, then, that the use of the log-logistic distribution with the
    requirements that we propose for the estimation of errors for $M_{200}$ and
    $\nfwc$ is a reasonable choice, as it works for all clusters and the results
    seem not to be biased.

    \subsubsection{Virial ratios and interaction strength fittings}
    \label{sss:generalfit}
    Table~\ref{tbl:obsdata} summarizes the data used for computation of the
    virial ratios and interaction strengths according to the steps described in
    section~\ref{sec:evaluation}.
    \begin{table}[!tb]
        \setlength\tabcolsep{20.5pt}
        \caption{Redshift, temperature and compiled \gls{nfw} parameters of the
            22 galaxy clusters. Temperatures are given in \si{\keV} and masses
            in units of $h^{-1} \num{e14} M_{\odot}$. X-ray data from
            \citet{Aghanim2012,Landry2013,Kenneth2008}.}
        \label{tbl:obsdata} %
        \centering
        \begin{tabular}{@{}lcccc@{}}
            \toprule  % 150414-2342,2352,2358 2600
            Cluster  & $z$  &  $\kB \tx$ &  $M_{200}$  & $\nfwc$\\
            \midrule
            A68 & $0.255$ & $8.3 \pm 0.3$ & $4.45^{+1.75}_{-1.35}$ & $2.49^{+3.12}_{-1.65}$ \\
            A115 & $0.197$ & $8.9^{+0.6}_{-0.7}$ & $4.24^{+2.60}_{-1.79}$ & $1.86^{+3.52}_{-1.48}$ \\
            A209 & $0.206$ & $6.6 \pm 0.2$ & $10.62^{+2.17}_{-1.81}$ & $1.90^{+0.81}_{-0.63}$ \\
            A267 & $0.230$ & $5.6 \pm 0.1$ & $3.23^{+0.82}_{-0.69}$ & $3.95^{+3.00}_{-1.96}$ \\
            A383 & $0.188$ & $4.1 \pm 0.1$ & $3.11^{+0.88}_{-0.69}$ & $5.59^{+6.31}_{-3.51}$ \\
            A520 & $0.203$ & $7.9 \pm 0.2$ & $4.20^{+2.30}_{-1.70}$ & $2.20^{+1.10}_{-0.80}$ \\
            A521 & $0.248$ & $6.1 \pm 0.1$ & $4.58^{+1.00}_{-0.88}$ & $2.18^{+1.13}_{-0.83}$ \\
            A586 & $0.171$ & $7.8^{+1.0}_{-0.8}$ & $6.29^{+2.26}_{-1.69}$ & $4.90^{+5.82}_{-3.16}$ \\
            A611 & $0.288$ & $7.1^{+0.6}_{-0.5}$ & $5.47^{+1.31}_{-1.11}$ & $3.01^{+1.99}_{-1.37}$ \\
            A697 & $0.282$ & $8.8^{+0.7}_{-0.6}$ & $9.73^{+1.86}_{-1.61}$ & $2.17^{+0.94}_{-0.73}$ \\
            A963 & $0.206$ & $5.6 \pm 0.1$ & $5.10^{+1.70}_{-1.40}$ & $2.10 \pm 0.40$ \\
            A1835 & $0.253$ & $8.4 \pm 0.1$ & $10.86^{+2.53}_{-2.08}$ & $2.35^{+1.27}_{-0.93}$ \\
            A1914 & $0.171$ & $8.5 \pm 0.2$ & $4.20^{+2.90}_{-2.00}$ & $3.20^{+0.70}_{-0.60}$ \\
            A2034 & $0.113$ & $6.4 \pm 0.2$ & $4.30^{+4.00}_{-2.40}$ & $2.90^{+0.90}_{-0.70}$ \\
            A2219 & $0.228$ & $9.6^{+0.3}_{-0.2}$ & $7.75^{+1.89}_{-1.60}$ & $4.59^{+4.11}_{-2.53}$ \\
            A2261 & $0.224$ & $6.1^{+0.6}_{-0.5}$ & $7.97^{+1.51}_{-1.31}$ & $4.31^{+2.37}_{-1.72}$ \\
            A2390 & $0.231$ & $9.1 \pm 0.2$ & $6.92^{+1.50}_{-1.29}$ & $4.40^{+2.63}_{-1.86}$ \\
            A2631 & $0.278$ & $7.5^{+0.4}_{-0.2}$ & $4.54^{+0.89}_{-0.78}$ & $5.57^{+4.24}_{-2.78}$ \\
            RX J1720 & $0.164$ & $5.9 \pm 0.1$ & $3.48^{+1.28}_{-0.99}$ & $4.87^{+6.98}_{-3.45}$ \\
            RX J2129 & $0.235$ & $5.6 \pm 0.1$ & $5.29^{+1.76}_{-1.38}$ & $4.67^{+6.21}_{-3.19}$ \\
            ZwCl 1454 & $0.258$ & $4.6 \pm 0.1$ & $2.80^{+1.39}_{-1.03}$ & $2.18^{+3.25}_{-1.57}$ \\
            ZwCl 1459 & $0.290$ & $6.4 \pm 0.2$ & $3.77^{+1.17}_{-0.98}$ & $4.10^{+4.56}_{-2.56}$ \\        
            \bottomrule
        \end{tabular}
    \end{table}    
    We assume flat distributions within the range $[x-\Delta x_2, x+\Delta x_1]$
    for the inputs in the form $x^{+\Delta x_1}_{-\Delta x_2}$ in the generation
    of random realizations for our Monte Carlo method.

    Inspection of the final distributions of virial ratios and interaction strengths suggests the use of log-normal distributions to fit (the negative of) the data.
    However, due to the nature of these quantities and their domains, we include a location parameter to allow the distribution to be shifted from the origin.
    The log-normal \gls{pdf} is then
    \begin{align}
        f_X(x; \mu,\sigma,\loc) = \frac{\exp\left\lbrace-
            \left[\ln(x-\loc)-\mu\right]^2/2\sigma^2 \right\rbrace}{\left(
                x-\loc \right) \sigma\sqrt{2\uppi}}
    \end{align}
    with $\loc$ being the location parameter.
    We take the 68\% most probable values and the maximum probability of this
    log-normal PDF to yield the resulting value $x^{+\Delta x_1}_{-\Delta x_2}$
    of the quantity $X$.
    The fits obtained are especially good for the interaction strengths, as we
    show in section~\ref{sec:interaction}.

    In addition, we introduce two selection criteria which we apply to the
    values of the interaction strength obtained with this method to conserve
    realizations: one physical, discussed in section~\ref{sss:intmodel}, selects
    $\zeta \ge -1/3$, while the other avoids numerical problems, discussed in
    section~\ref{sss:nvmodel}, by keeping only $9\zeta \le \num{200}$ (see discussion in
    section~\ref{sec:interaction}).

    \subsubsection{Reliability of the results}
    Our analysis considers samples of 8600 random realizations of each cluster.
    However, some of them happen to have no solution for $\zeta$, or to have a
    solution outside the domain established by \eq{eq:LIdm}.
    We remove these cases from the analysis, which makes the samples
    considerably smaller for some clusters.
    That is the case for A68, A115, A520 and A1914, for which we are left with
    only about 650 (on average) realizations.
    A possible explanation for such a large fraction of these samples not having
    a physical solution for $\zeta$ could be linked with the dynamical activity
    of those clusters \citep[e.g.][]{Markevitch2005,Barrena2013}, so their
    virial states are not as close to equilibrium for our method to be
    applicable.

    The gas distribution in clusters can be used as a probe of the recent past 
    dynamical activity of a cluster, since the gas responds quickly to 
    perturbations in the gravitational potential, for instance, due to cluster 
    merger and/or collision \citep{AndradeSantos2012}.
    Visual inspections of Chandra X-ray images show that all clusters except
    A115, A520 and A1914 have rather undisturbed and symmetric gas distribution,
    suggesting that they are not dynamically active. 

    For comparison purposes, we define a ``success rate'' (SR) as the fraction
    of realizations satisfying our selection criteria in the total generated for
    each cluster.
    Clusters like A1835, A209 and A2261 present this fraction very close or
    equal to \num{1}.
    The success rates for all clusters are presented in table~\ref{tab:successrate}.
    \begin{table}
        \setlength\tabcolsep{11.2pt}
        \caption{Fraction of Monte Carlo-produced realisations for each cluster.
        The SR here presented were computed for our Monte Carlo samples of size
        \num{8600}. Tests have shown that there is no significant variation of the SR with
        the size of the sample.}
        \label{tab:successrate}
        \begin{tabular}{@{}lcccccccccccccccccccccccccccccc@{}}
            \toprule
            Cluster  & \mcs{A115} & \mcs{A1835} & \mcs{A1914} & \mcs{A2034} & \mcs{A209} & \mcs{A2219} \\
            SR & \mcs{$0.06$} & \mcs{$1.00$} & \mcs{$0.09$} & \mcs{$0.58$} & \mcs{$1.00$} & \mcs{$0.79$} \\
            \midrule
            Cluster  & \mcs{A2261} & \mcs{A2390} & \mcs{A2631} & \mcs{A267} & \mcs{A383} & \mcs{A520} \\
            SR & \mcs{$1.00$} & \mcs{$0.75$} & \mcs{$0.68$} & \mcs{$0.61$} & \mcs{$1.00$} & \mcs{$0.05$} \\
            \midrule
            Cluster & \mcf{A521} & \mcf{A586} & \mcf{A611} & \mcf{A68} & \mcf{A697} \\
            SR & \mcf{$0.67$} & \mcf{$0.86$} & \mcf{$0.80$} & \mcf{$0.11$} & \mcf{$0.99$} \\
            \midrule
            Cluster & \mcf{A963} & \mcf{RX J1720} & \mcf{RX J2129} & \mcf{ZwCl 1454} & \mcf{ZwCl 1459}\\
            SR & \mcf{$0.91$} & \mcf{$0.71$} & \mcf{$1.00$} & \mcf{$0.64$} & \mcf{$0.63$} \\  
            \bottomrule 
        \end{tabular}
    \end{table}
    In the histograms, we use a number of bins proportional to the size of the samples.

    In order to evaluate the consistency of our method, we consider a cluster
    from $N$-body simulation, similar to those of \citet{Machado2013}, in a
    cosmology with $\Omega_{\matter} = 0.3$, $\Omega_{\de} = 0.7$, $h = 0.72$ and no
    interaction in the dark sector, so the virial ratio should be very
    close\footnote{Some variations can be introduced by the effects of
        projection translating from simulation to observables.} to $-0.5$ and
    interaction compatible with zero. 
    The data for this simulated cluster are $M_{200} = 18.0 \, h^{-1} \num{e14}
    M_{\odot}$, $z = 0.0$, $\nfwc = 3.0$ and $\tx = \SI{7.3 \pm 0.8}{\keV}$. 
    The uncertainty in the temperature comes from the $\sigx$--$\tx$ scatter
    relation \citep{Xue2000}
    \begin{align}
        \sigx = 10^{2.49 \pm 0.02} \tx^{0.65 \pm 0.03},
    \end{align}
    from which $\tx$ was computed for a one-dimensional velocity dispersion of
    $\sigx = \SI{1125}{\km\per\second}$.
    Because the observed virial ratio is linear with the temperature, the only
    source of errors in this case, its histogram for all random realizations
    produced in our code reflects clearly the uniform distribution assigned to
    the input temperature.
    Fitting that uniform distribution we find a virial ratio of $-0.47 \pm 0.04$
    from the central \sepercent\ most probable values.

    With the analysis of section~\ref{sss:generalfit}, for the interaction
    strength we get $9\zeta = 0.05_{-0.54}^{+0.69}$, therefore compatible with the
    simulation. 
    The theoretical virial ratio is also in accordance with the classic value,
    $(\kin/\pot)_{\mathrm{th}} = -0.51 \pm 0.05$, while the \gls{dfe}
    being $-0.08 \pm 0.06$ satisfies our condition \eqref{eq:cond7}.

    \section{Analysis of the results}
    \label{sec:results}

    In this section we present and discuss the outcome of our analysis starting
    with the \glspl{ovr}, the \glspl{tvr} estimates from
    combining their \gls{dfe} factors and interaction strengths $\zeta$.
    Throughout this section, we present the constraints on the virial ratios,
    interaction and \gls{dfe} in figures~\ref{fig:MCVR_all}, \ref{fig:allxi6895},
    \ref{fig:allvrout6895} and \ref{fig:OB}, all obtained for each cluster according to
    the method described in the previous section. 
    These results are summarized in table~\ref{tab:results}.
    \begin{table}
        \setlength\tabcolsep{9.5pt}
        \caption{Virial ratios, interactions, theoretical virial ratios and departure from equilibrium from the log-normal fits.}
        \label{tab:results}
        \centering
        \begin{tabular}{@{}lcccc@{}}
            \toprule
            Cluster &   OVR &   $\zeta$  &  TVR & DfE  \\
            \midrule
            A520 & $-0.96 \pm 0.02$ & $-53.28^{+19.87}_{-44.20}$ & $-0.96 \pm 0.02$ & $-0.003 \pm 0.003$ \\ 
            A1914 & $-0.95 \pm 0.02$ & $-48.60^{+18.59}_{-43.09}$ & $-0.96 \pm 0.02$ & $-0.004 \pm 0.003$ \\ 
            A115 & $-0.94 \pm 0.03$ & $-35.72^{+17.85}_{-43.34}$ & $-0.95 \pm 0.03$ & $-0.01 \pm 0.01$ \\ 
            A68 & $-0.93 \pm 0.03$ & $-31.62^{+13.90}_{-37.11}$ & $-0.94 \pm 0.03$ & $-0.01 \pm 0.01$ \\ 
            A521 & $-0.89 \pm 0.05$ & $-13.14^{+5.67}_{-24.80}$ & $-0.90 \pm 0.05$ & $-0.01 \pm 0.03$ \\ 
            A267 & $-0.88 \pm 0.06$ & $-10.50^{+5.45}_{-24.20}$ & $-0.89 \pm 0.06$ & $-0.01 \pm 0.04$ \\ 
            A2390 & $-0.87 \pm 0.06$ & $-13.08^{+5.90}_{-22.60}$ & $-0.89 \pm 0.05$ & $-0.05 \pm 0.04$ \\ 
            A2631 & $-0.87 \pm 0.07$ & $-11.56^{+5.75}_{-22.91}$ & $-0.89 \pm 0.06$ & $-0.04 \pm 0.05$ \\ 
            A611 & $-0.85^{+0.07}_{-0.08}$ & $-8.76^{+4.58}_{-20.08}$ & $-0.87 \pm 0.07$ & $-0.03 \pm 0.05$ \\ 
            ZwCl 1459 & $-0.85 \pm 0.08$ & $-7.41^{+4.43}_{-21.56}$ & $-0.86^{+0.07}_{-0.08}$ & $-0.02 \pm 0.06$ \\ 
            A2219 & $-0.84 \pm 0.08$ & $-9.28^{+4.48}_{-18.55}$ & $-0.87 \pm 0.06$ & $-0.07 \pm 0.05$ \\ 
            ZwCl 1454 & $-0.82 \pm 0.09$ & $-3.95^{+2.42}_{-15.49}$ & $-0.82^{+0.09}_{-0.10}$ & $0.02 \pm 0.06$ \\ 
            RX J1720 & $-0.81 \pm 0.10$ & $-4.47^{+2.94}_{-16.34}$ & $-0.83 \pm 0.09$ & $-0.02 \pm 0.07$ \\ 
            A697 & $-0.81 \pm 0.07$ & $-8.17^{+3.62}_{-10.75}$ & $-0.84 \pm 0.06$ & $-0.04 \pm 0.06$ \\ 
            A963 & $-0.78^{+0.07}_{-0.10}$ & $-4.59^{+1.98}_{-13.39}$ & $-0.80^{+0.08}_{-0.09}$ & $-0.01 \pm 0.05$ \\ 
            A2034 & $-0.77^{+0.08}_{-0.10}$ & $-4.37^{+2.23}_{-13.35}$ & $-0.80^{+0.08}_{-0.09}$ & $-0.02 \pm 0.06$ \\ 
            A586 & $-0.77 \pm 0.11$ & $-3.83^{+2.74}_{-12.51}$ & $-0.80^{+0.09}_{-0.10}$ & $-0.07 \pm 0.08$ \\ 
            A1835 & $-0.70^{+0.06}_{-0.07}$ & $-4.84^{+2.10}_{-3.78}$ & $-0.75 \pm 0.06$ & $-0.07 \pm 0.06$ \\ 
            A383 & $-0.61^{+0.08}_{-0.11}$ & $-0.41^{+1.04}_{-3.44}$ & $-0.60^{+0.09}_{-0.12}$ & $0.05 \pm 0.10$ \\ 
            A209 & $-0.60 \pm 0.05$ & $-1.92^{+1.01}_{-1.41}$ & $-0.64 \pm 0.05$ & $-0.05 \pm 0.08$ \\ 
            RX J2129 & $-0.60^{+0.09}_{-0.12}$ & $-0.68^{+1.23}_{-4.04}$ & $-0.63^{+0.10}_{-0.12}$ & $-0.02 \pm 0.10$ \\ 
            A2261 & $-0.56^{+0.06}_{-0.07}$ & $-0.99^{+1.00}_{-1.55}$ & $-0.60 \pm 0.07$ & $-0.03 \pm 0.09$ \\ 
            TOTAL & --- & $-1.99^{+2.56}_{-16.00}$ & $-0.79 \pm 0.13$ & --- \\  
            \bottomrule
        \end{tabular}
    \end{table}

    \subsection{The observed virial ratios}
    \label{ss:MCVR}
\begin{figure}[bt]
        \centering
        \subfloat[\gls{ovr} for the cluster A2261 from a sample of \num{8600}
        random realizations. $\kin/\pot = \num{-0.56 \pm 0.06}$.]{\label{fig:MCVR_A2261}\includegraphics[width=0.48\textwidth]{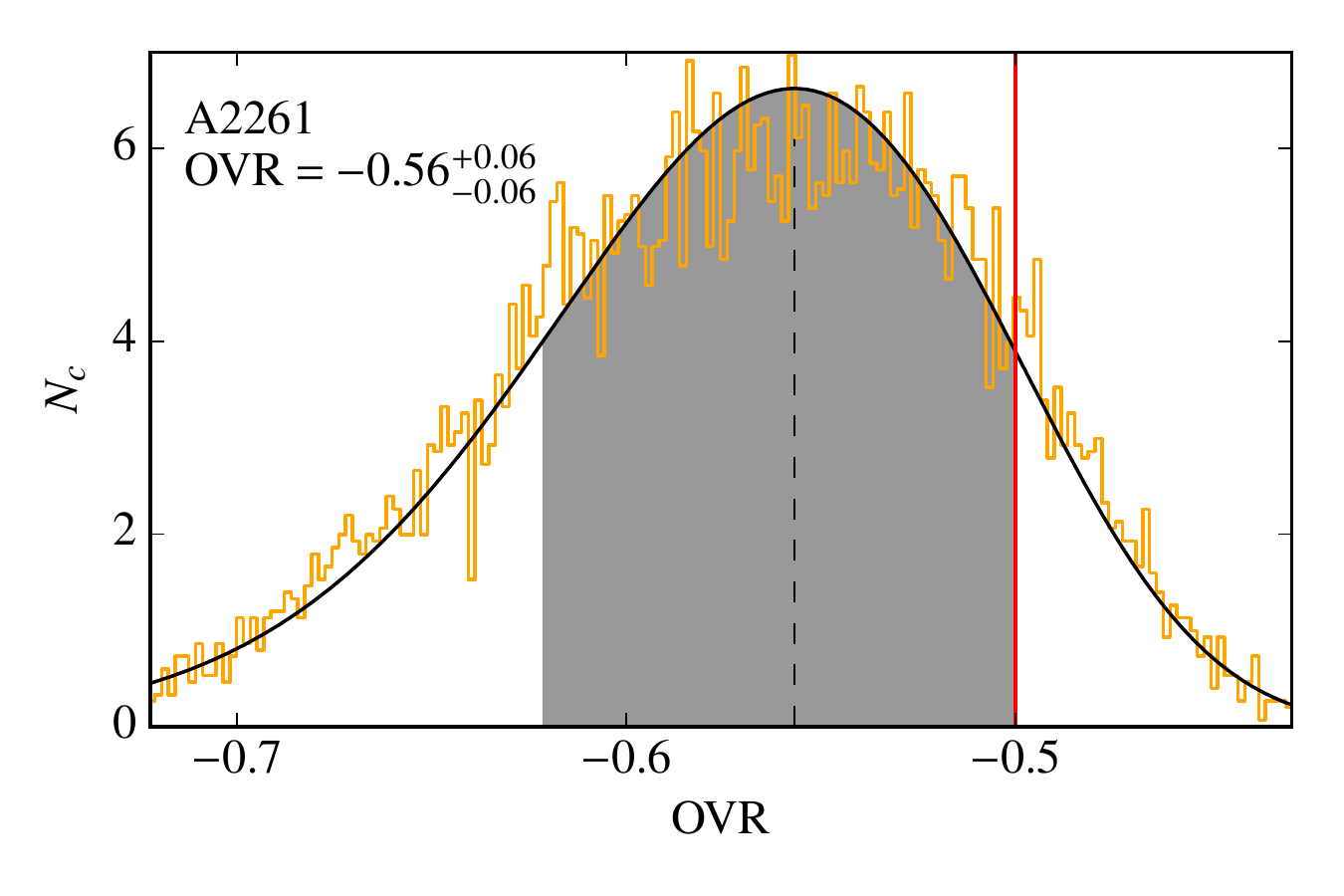}} \hspace{10pt}
        \subfloat[\glspl{ovr} with error bars indicating \nfpercent\ and \sepercent\ \glspl{cl} for all clusters.]{\label{fig:MCVR_all}\includegraphics[width=0.48\textwidth]{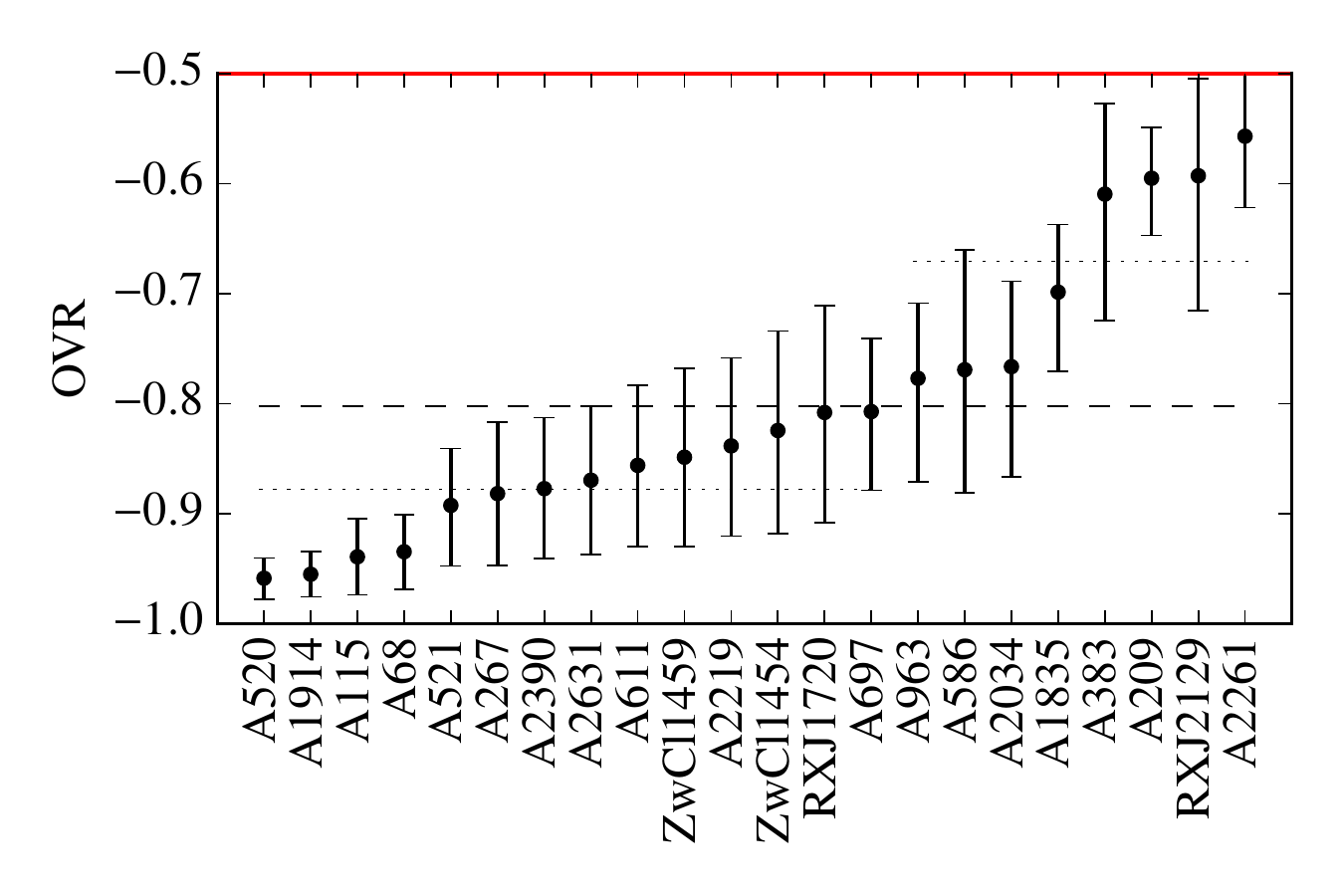}}
        \caption{Panel (a) shows in detail the distribution of the \gls{ovr} for
            the cluster A2261 when we apply the Monte Carlo method
            (section~\ref{sec:MC}). $N_c$ is the normalized count of Monte Carlo
            clusters per bin of \gls{ovr}. The shade area
            correspond to the \sepercent\ \gls{cl}. The dashed
            vertical line indicates the most probable value while the red solid
            line denotes the classic value. Panel (b) presents the most probable
            values and confidence levels for each cluster. The 
            error bars give the \sepercent\ \gls{cl}. The mean of the most
            probable values is signaled by the dashed black line. The two dotted
            lines show the means for the two groups of clusters defined in the
            text and the solid red line marks the classic value.}
        \label{fig:MCVR}
    \end{figure}

    The \gls{ovr} is obtained from applying the method described in
    section~\ref{sec:MC} to \eq{eq:virratio14}. 
    The histograms of \gls{ovr} produced for each cluster are very similar to
    the ones obtained for their theoretical counterparts
    (section~\ref{ss:theoreticalVR}).
    As an example, we present in figure~\ref{fig:MCVR_A2261} the distribution
    obtained for the cluster A2261.
    It represents the histogram distribution of the \gls{ovr}
    obtained from our Monte Carlo sampling of mass, temperature and
    concentration ranges.
    Superimposed is the log-normal fit, with shaded area corresponding to
    the \sepercent\ \gls{cl}.
    The red vertical line marks the theoretical non-interacting value, while the
    dashed line gives the most likely value.

    We summarize the results of the \gls{ovr} with their corresponding
    asymmetrical errors in figure~\ref{fig:MCVR_all}, where we have shown the
    theoretical non-interacting virial ratio as a horizontal red line.
    We have ordered the clusters by increasing \gls{ovr} and keep
    this order for the rest of the work.

    We have represented the mean value for the whole sample with a dashed line.
    We identify two groups of similar virial ratios separated by the global mean
    and for each group we represent their means by the dotted lines.
    The dispersion of the ratios may reflect the diversity of the equilibrium
    conditions.
    However, the first group seems to have less scatter than the second one. 

    With this robust non-linear treatment of error propagation, all clusters
    exclude \num{-0.5} at $1\sigma$, with the only exception of A2261, which includes
    that value marginally.

    \subsection{The interaction strength}
    \label{sec:interaction}
    As previously mentioned, we solve \eq{eq:solveforxi} for $\zeta$ with the
    \gls{dfe} term \eqref{eq:Out(xi)} using the Monte Carlo method of section~\ref{sec:MC}.
    \Eq{eq:quadraticsolution} is singular at $\kin/\pot = \num{-1}$, which
    corresponds to $\zeta$ infinite.
    This is a limitation of our method that we deal with by restricting the
    interaction strength to $\lvert 9\zeta \rvert \le \num{200}$ such that the
    histograms would be legible. 
    Note that this introduces a cut in the histograms of the \glspl{tvr}
    (section~\ref{ss:theoreticalVR}), which reflects the limitation of
    our method.

\begin{figure}[!t]
        \centering
        \includegraphics[width=\textwidth]{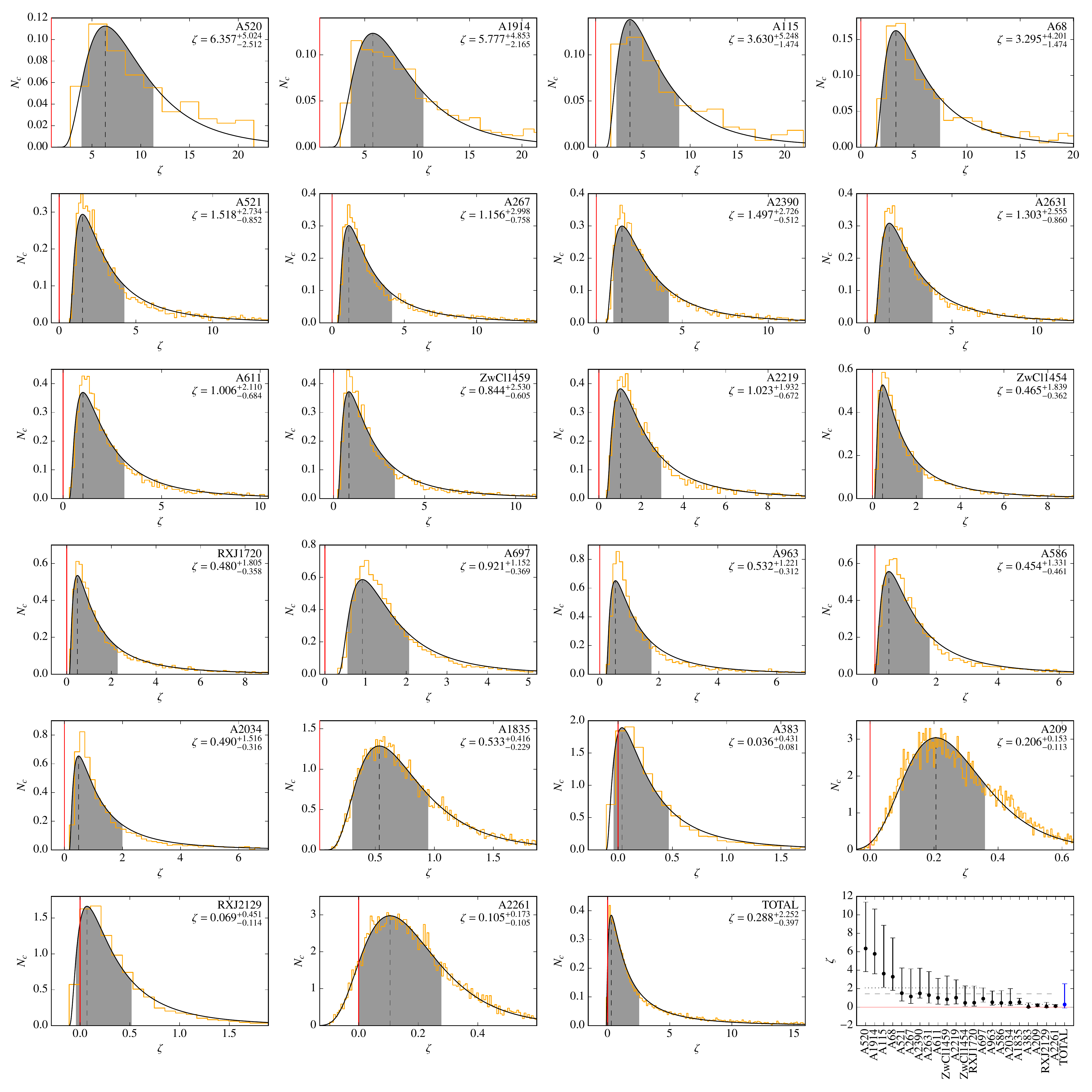}
        \caption{Histograms for the distributions of interaction strength
            $\zeta$ with their log-normal fits for each cluster. The two
            lowest rightmost panels present the cumulated histogram of all the
            clusters and the most probable values and error bars for each
            cluster and compounded distribution in blue. The shaded areas under
            the histograms fits mark the \sepercent\ \gls{cl}. The vertical
            dashed lines point the most probable values, while the solid red
            line, when shown, marks the $\zeta=0$ position. The horizontal
            dotted lines represent the means for each group of clusters, while
            the horizontal dashed line marks the overall mean.}
        \label{fig:allxi6895}
    \end{figure}

    Figure~\ref{fig:allxi6895} represents the histograms of interaction
    strengths for all clusters, with the two lowest rightmost panels displaying
    the compounded distribution for all the studied clusters and the most
    probable values of $\zeta$ for all clusters together with their error bars
    for the \sepercent\ \gls{cl}.
    As previously done, the vertical dashed line gives the most probable value,
    and the shaded areas correspond to the \sepercent\ \gls{cl}.
    The red solid lines indicate the $\zeta=0$ absence of interaction.
    The whole sample mean of the most probable values is given by the horizontal
    dashed line.
    Each group previously singled out also present their group mean as
    horizontal dotted lines.
    Finally, we add the most probable value and error bars of the compounded
    distribution in blue.

    Of the 22 clusters, all of them except A586, A383, RX J2129 and A2261 (marginally)
    display a $1\sigma$ detection favoring positive $\zeta$.

    In this model, the interaction strength should be the same for all clusters.
    The global mean $\bar \zeta = \num{1.44}$ is compatible with 13 of the 22 clusters: A521, A267, A2390,
    A2631, A611, ZwCl 1459, A2219, ZwCl 1454, RX J1720, A697, A963, A586 and A2034.
    However, three of the clusters, as well as the compounded distribution that
    yields $\zeta = 0.288^{+2.252}_{-0.397}$, display compatibility with no interaction.
    This points to problems in our method, namely when it assumed small
    deviation from equilibrium while the results have important variations.
    Actually, the highest boundary is limited by the results of
    \citet{Salvatelli2013} to\footnote{\citeauthor{Salvatelli2013} have found
        $0.07 \le \zeta \le 0.30$ at \nfpercent\ \gls{cl}.} \num{0.30}.
    Nevertheless, we concentrate on the present scheme and leave the solutions
    to a forthcoming work.

    \subsection{The theoretical virial ratios}
    \label{ss:theoreticalVR}
\begin{figure}[!tbp]
        \centering
        \includegraphics[width=\textwidth]{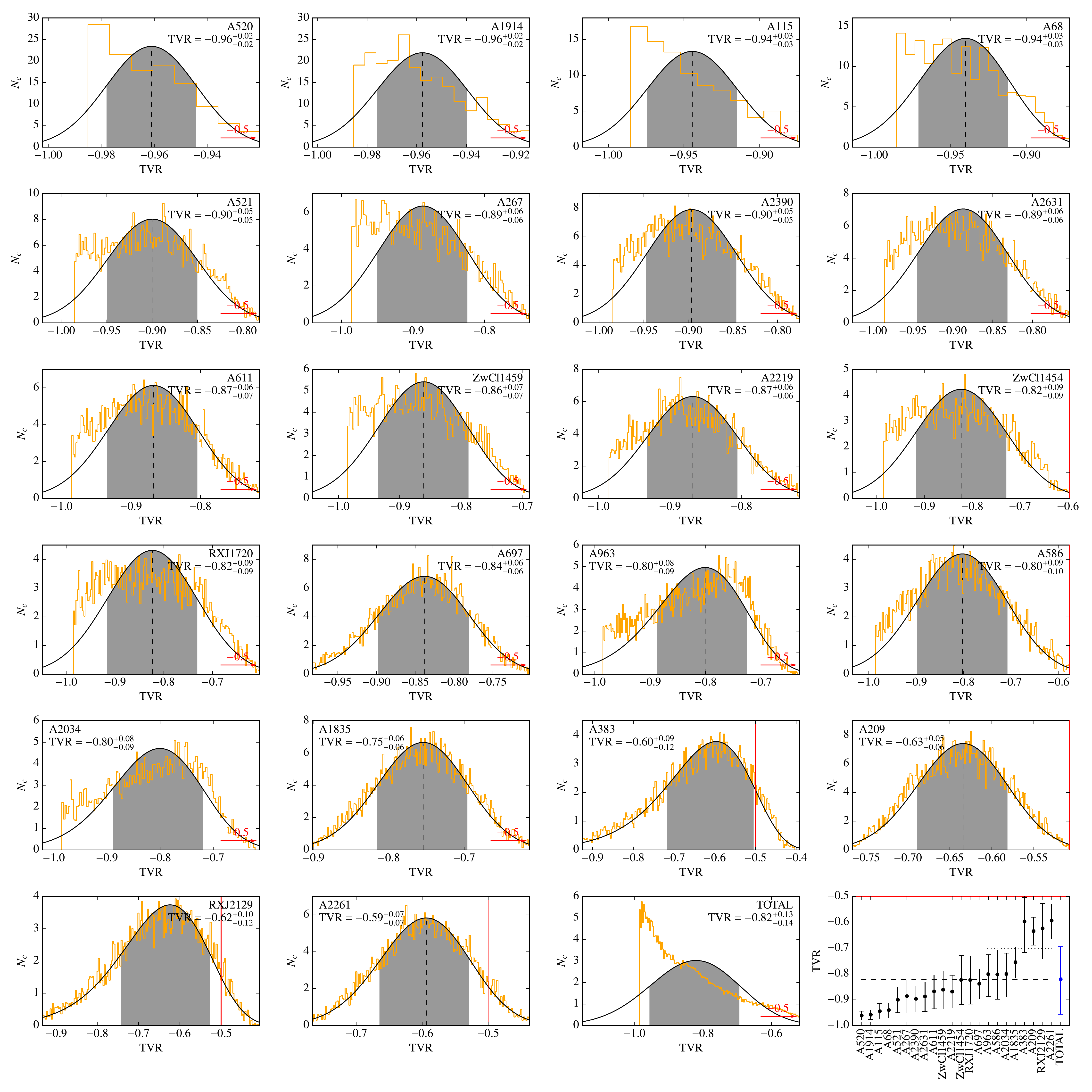}
        \caption{Histograms for the distributions of theoretical virial ratios
            with their log-normal fits for each cluster. The two lowest
            rightmost panel presents the cumulated histogram of all the clusters
            and the most probable \gls{tvr} values with error bars. The shaded
            areas marks the \sepercent\ \gls{cl}. The vertical dashed lines
            point the most probable values, while the solid red line, when
            shown, marks the position of the no interaction classic virial
            ratio. The mean value of all most probable values is also shown,
            represented by the horizontal dashed black line in the last panel.
            The two horizontal dotted lines indicate the means of the two
            groups.}
        \label{fig:allvrout6895}
    \end{figure}

    Armed with the results from the previous section, we compute with
    \eq{eq:virbalance} the \gls{tvr} each cluster would have at perfect
    equilibrium in the presence of interaction. 
    Figure~\ref{fig:allvrout6895} shows us their corresponding distributions.
    We keep conventions of shaded areas and error bars representing \sepercent\ \gls{cl}, the
    vertical dashed lines for the most probable values and the red solid line to
    indicate absence of interaction, the compounded distribution represented in
    the last panel by the blue error bar.
    As these histograms are very similar to the observed ones
    (section~\ref{ss:MCVR}), the following comments can be applied to both.
    The reasons for these similarities are discussed in
    section~\ref{subsec:outofbalance}.

    At this level, all the clusters exclude \num{-0.5} at $1\sigma$, which
    confirms the result from section~\ref{ss:MCVR} (one cluster did not
    exclude that value but only marginally).
    However, 16 out of the 22 clusters present log-normal fits that reflects
    poorly the underlying distributions: 
    A520, A1914, A115, A68, A521, A267, A2390, A2631, A611, ZwCl 1459, A2219,
    ZwCl 1454, RX J1720, A963, A586 and A2034.
    For all of those problematic distributions the log-normal fits break down
    for virial ratios in the proximity of \num{-1}.
    This is related to the singularity in \eq{eq:quadraticsolution}.
    In addition, the compounded \gls{tvr} points towards a single value of
    $-0.82^{+0.13}_{-0.14}$, which represents a detection at $2\sigma$, in
    contradiction with the results of the previous section.
    All this suggests a problem with our method that assumed small deviation
    from equilibrium, as previously pointed out.

    \subsection{The departure from equilibrium factors}
    \label{subsec:outofbalance}
\begin{figure}[tb]
        \centering
        \includegraphics[width=0.8\textwidth]{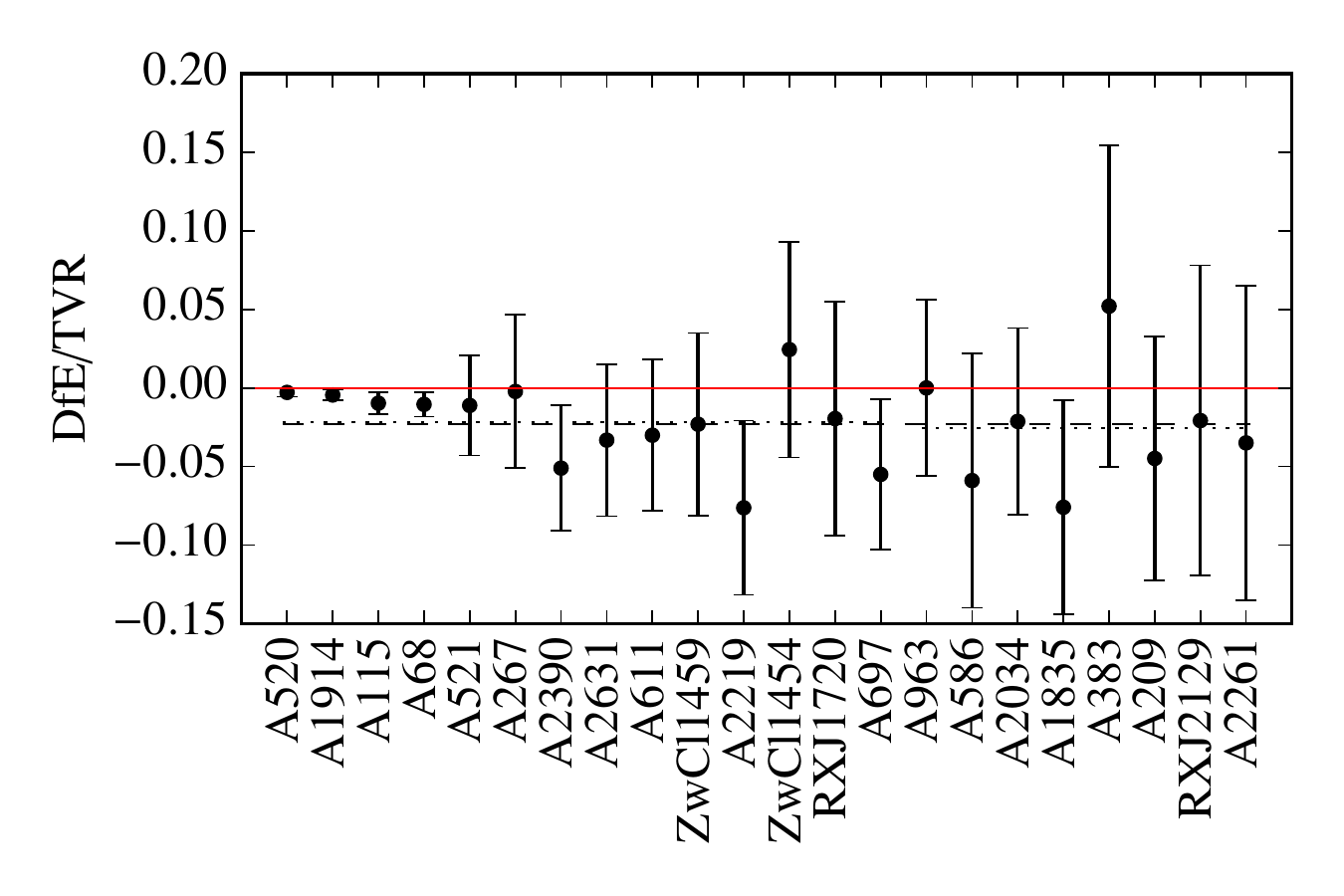}
        \caption{\gls{dfe} factors relative to the \gls{tvr} with confidence
            intervals for each cluster in our sample. Error bars give the
            \sepercent\ \gls{cl}. Dotted lines represent means for
            each group, while dashed line marks overall mean. The solid red line
            marks the absence of deviation.}
        \label{fig:OB}
    \end{figure}

    \Eq{eq:Out(xi)} with the results of section~\ref{sec:interaction} allows us
    to compute the \gls{dfe} factor for each cluster.
    The values of this factor relative to their \gls{tvr},
    \begin{align}
        \frac{\mathrm{DfE}}{\mathrm{TVR}} =
        - \frac{\left( 2 + 3 \zeta \right)^{-2}
            \dpot/H\pot}{(\kin/\pot)_{\mathrm{th}}} = \frac{\dpot / H
            \pot}{\left(1 + 3 \zeta \right) \left( 2 + 3 \zeta \right)},
    \end{align}
    are presented in figure~\ref{fig:OB}. 
    Except for A1914, A115, A68, A2390, A2219, A697 and A1835, all those relative
    departures are compatible with zero.
    We should note as well that ZwCl 1454 and A383 are the only clusters with
    positive \gls{dfe}.
    For this figure and for numerical reasons, we fit the distributions for each
    cluster with uniform distributions so as to evaluate the order of magnitude
    of those departures and produce the values displayed on figure~\ref{fig:OB}.
    Although not good fits, these uniform distributions enable us to show how
    small those values are, validating our hypothesis \eq{eq:cond7}.
    This explains the similarities between \gls{ovr} and \gls{tvr}, as seen in
    comparing figure~\ref{fig:MCVR_all} and the last panel of
    figure~\ref{fig:allvrout6895}. 

    \section{Discussion of the results}
    We analyzed the virial ratios of a set of clusters using a simple model
    based on the \LI\ equation
    \citep{Bertolami2007,Bertolami2009,Abdalla2009,Abdalla2010,He2010,Bertolami2012},
    using weak-lensing mass profiles and intracluster gas temperatures from
    optical and X-ray observations \citep{Okabe2010,Aghanim2012}.
    Our treatment involved assessing the virial balance of each cluster as well
    as their equilibrium state, using a Monte Carlo statistical analysis on the
    data.

    Our method, a first proof of concept for out of equilibrium virial
    evaluation, enabled us to find mild evidence for an interacting dark sector
    in the virial balance of those clusters, however yielding only small
    amplitudes of departure from equilibrium: although the compounded
    distribution of all clusters would accommodate $\zeta = 0$, a majority of
    the individual clusters, of their virial ratios and of the compounded
    evaluation of the virial ratio all point towards a positive interaction.
    The compounded estimates give us $\zeta = 0.29^{+2.25}_{-0.40}$, which is
    not a detection, but $(\kin/\pot)_{\mathrm{th}} = -0.82^{+0.13}_{-0.14}$,
    which is a detection at $2\sigma$. 
    This tension between the compounded results for the interaction strength and
    the \gls{tvr}, while the latter is constructed out of the former,
    points to the main problem in our results: despite the scatter in the values
    of virial ratios, the \gls{dfe} factors remain small, as imposed in our
    hypotheses.
    In addition, our method contains an unphysical singularity at $\kin/\pot =
    \num{-1}$ in \eq{eq:quadraticsolution}.
    These problems, in spite of encouraging results, call for follow-up work
    which should remove the small departure from equilibrium hypothesis, as well
    as the singularity we introduced in this work for $\kin = - \pot$.

    \chapter{Final considerations}
    \label{ch:final}
    We have studied dark energy interacting models and tested them against
    different types of large-scale structure observations.
    These works culminated in the production of two papers
    \citep{Marcondes2016,LeDelliouetal2015}, one of them already published,
    besides the development of a code for \gls{mcmc} parameter estimation and
    another one for obtaining equations of motion and other quantities in \gls{gr}
    from a given metric.
    Both codes are entirely written in python, with potential for a wide variety
    of uses, and will be made publicly available in the near future.
    In the first work here presented, we have derived the analytical expression
    for the growth rate of structures in terms of the growth index $\gamma$ in
    the presence of a \gls{de}-\gls{dm} interaction.
    The derivation was based on the expansion of the growth index and the
    \gls{de} equation of state in power series of the dark energy density
    parameter $\Omega_{\de}(z)$, which parametrizes the time evolution.
    We have proved the method to be successful in yielding an expression for
    $\gamma$ in terms of the \gls{eos} coefficients $w_0, w_1, \ldots$ and the
    coupling constant $\zeta$ when the interaction term in the energy
    conservation equation is written as $Q_{0}^{\dm} \propto \mathcal{H} \zeta
    \rho_{\de}$, i.e., proportional do the \gls{de} density.
    The growth rate is then written as
    \begin{align*}
        f(\Omega_{\dm}) \approx \left[ \Omega_{\dm} \right]^{\gamma_0 + \gamma_1
            \Omega_{\de}}
    \end{align*}
    with the growth index coefficients given by
    \begin{align*}
        \gamma_0 &= \frac{3 \left(1 - w_0 + 5 \zeta \right)}{5 - 6 w_0 - 6 \zeta}, \\
        \gamma_1 &= \frac{- \gamma_0^2 + \gamma_0 \left(1 + 12 w_1 + 18 \zeta \right) - 6 w_1 + 6 \zeta \left(5 - 6 w_0 + 6 \zeta \right)}{2 \left( 5 - 12 w_0 -12 \zeta\right)},
    \end{align*}
    up to first order in the \gls{de} density parameter.
    Since we consider only the two components, the \gls{dm} density parameter is
    $\Omega_{\dm} = 1 - \Omega_{\de}$ and
    \begin{align*}
        \Omega_{\de}(z) \approx \frac{\Omega_{\de,0} \left(1 + z \right)^{3
                \left(w_0 + \zeta \right)} \left[ 1 - \Omega_{\de,0} + \Omega_{\de,0} \left(1 +
                    z \right)^{3 \left( w_0 + \zeta \right)} \right]^{ \frac{
                    w_1 + \zeta}{ w_0 + \zeta}}}{1 - \Omega_{\de,0} + \Omega_{\de,0} \left(1 + z \right)^{3 \left( w_0 + \zeta \right)} \left[1 - \Omega_{\de,0} + \Omega_{\de,0} \left(1 + z \right)^{3 \left( w_0 + \zeta \right)} \right]^{\frac{w_1 + \zeta}{w_0 + \zeta}}},
    \end{align*}
    This is one of the main results of this thesis.
    We showed that the growth rate should be altered by a term proportional to
    the interaction coupling to make the continuity equation compatible with the
    non-interacting model assumed for the measurements, namely $\tilde f \equiv
    f + 3 \zeta \frac{1 - \Omega_{\de}}{\Omega_{\de}}$.
    The analytical expressions obtained then enabled us to
    compare the modified growth $\tilde f \sigma_8$ predicted by the
    interacting \gls{de} model with $f\sigma_8$ measurements from
    \glsentrylongpl{rsd}, with $\sigma_8$ given by
    \begin{align*}
        \sigma_{8}(z) \approx \sigma_{8,0} \left[
            \frac{\Omega_{\de}(z)}{\Omega_{\de,0}} \right]^{- \frac{1}{3
                \left(w_0 + \zeta \right)}} \exp \left\lbrace \frac{\varepsilon_1
                \left[ \Omega_{\de}(z) - \Omega_{\de,0} \right] + \varepsilon_2
                \left[ \Omega_{\de}^2(z) - \Omega_{\de,0}^2 \right]}{3 \left(
                    w_0 + \zeta \right)} \right\rbrace,
    \end{align*}
    with
    \begin{align*}
        \varepsilon_1 &\equiv \gamma_0 - \frac{w_0 - w_1}{w_0 + \zeta}, \\
        \varepsilon_2 &\equiv -\frac{\gamma_0^{2}}{4} + \frac{\gamma_0}{2}
        \left(\frac{1}{2} + \frac{w_0 - w_1}{w_0 + \zeta} \right) + \frac{\gamma_1}{2} -
        \frac{w_0^2 + w_1^2 - w_1 \left( w_0 - \zeta \right) - w_2 \left( w_0 + \zeta \right)}{2 \left( w_0 + \zeta \right)^2} .
    \end{align*}
    However, it was not possible to obtain tight constraints on the interaction strength
    due to the small number of measurements and their large uncertainties.
    The constraining power of these \gls{rsd} measurements is remarkable when
    combined with \gls{cmb} and other data, drastically reducing the region of
    the parameter space favored by observations.

    When considering the analytical evaluation of the growth rate, Hubble rate
    measurements and supernovae data are particularly interesting for combining
    with \gls{rsd}, since their predictions are made at the background level and
    can also be expressed analytically in terms of the model parameters.
    Their integration with the code developed for this work may be
    straightforward and is being considered for a follow-up work.
    We also plan to analyze the possibility of inclusion of the decaying mode of
    the matter perturbations, which could contribute more significantly
    in some cases, depending on the interaction.

    When the interaction is proportional do the \gls{dm} density, we find that
    the zero-th order terms of the expansion around $\Omega_{\de} = 0$ require
    $\zeta$ to be zero, thus forbidding the interacting cosmology.
    This is another important result, since the approximation $f \approx
    \Omega_{\matter}^{\gamma}$ is known to describe very well the growth rate in
    a wide variety of \gls{de} models and is widely adopted.

    We have thus shown that the growth index parametrization cannot account for
    this specific type of interacting model.
    Still, we have studied this interaction $Q_0^{\dm} \propto \mathcal{H}
    \zeta \rho_{\dm}$ but in a different scenario and with other type of
    observations.
    Galaxy clusters have been considered as dynamical systems whose equilibrium
    state---altered by the coupling between dark sectors as dictated by the \LI\
    equation---may reveal the existence of interaction.
    We find a hint of a positive interaction, although still compatible with
    zero at $1\sigma$ \gls{cl}:
    \begin{align*}
        \zeta = 0.29^{+2.25}_{-0.40}, \qquad \text{($1\sigma$ \gls{cl})}
    \end{align*}
    but also a $2\sigma$ detection in the compounded theoretical virial ratio
    \begin{align*}
        \left(\frac{\kin}{\pot}\right)_{\mathrm{th}} =
        -0.82^{+0.13}_{-0.14}. \qquad \text{($1\sigma$ \gls{cl})}
    \end{align*}

    Most observed clusters appear to be somewhat perturbed systems and are maybe
    still forming (accreting mass), which is expected in the current standard
    cosmological scenario.
    Our approach treating clusters as out of equilibrium systems is
    therefore natural, despite the measurement of such departure not following
    the observed wide variations in virial states.
    The tension between the results of our measured departure from the classic
    virial ratio and our measured interaction strength indicates that our
    method shows potential but also has room for improvement.
    We expect that accommodation for large departures will enable the use of
    much larger samples, statistically enhancing the significance of these
    results.

    \printbibliography[heading=bibintoc]
    %\bibliography{bib}

\end{document}

%% file: glossary.tex
\newacronym{cmb}{CMB}{cosmic microwave background}
\newacronym{firas}{FIRAS}{Far Infrared Absolute Spectrophotometer}
\newacronym{cobe}{COBE}{COsmic Background Explorer Satellite}
\newacronym{lcdm}{$\Lambda$CDM}{$\Lambda$-Cold Dark Matter}
\newacronym{bbn}{BBN}{Big Bang nucleosynthesis}
\newacronym{lss}{LSS}{large-scale structure}
\newacronym{sneIa}{SNe~Ia}{Type Ia supernovae}
\newacronym{wmap}{WMAP}{Wilkinson Microwave Anisotropy Probe}
\newacronym{ecmb}{eCMB}{extended cosmic microwave background}
\newacronym{act}{ACT}{Atacama Cosmology Telescope}
\newacronym{spt}{SPT}{South Pole Telescope}
\newacronym{bao}{BAO}{baryon acoustic oscillations}
\newacronym{2df}{2dFGRS}{Two-degree-Field Galaxy Redshift Survey}
\newacronym{6df}{6dFGS}{Six-degree-Field Galaxy Survey}
\newacronym{sdss}{SDSS}{Sloan Digital Sky Survey}
\newacronym{boss}{BOSS}{Baryon Oscillation Spectroscopic Survey}
\newacronym{rms}{RMS}{root mean square}
\newacronym{flrw}{FLRW}{Friedmann--Lema\^itre--Robertson--Walker}
\newacronym{rsd}{RSD}{redshift-space distortion}
\newacronym{fog}{FoG}{Fingers-of-God}
\newacronym{nfw}{NFW}{Navarro--Frenk--White}
\newacronym{cl}{CL}{confidence level}
\newacronym{pdf}{PDF}{probability density function}
\newacronym{cdf}{CDF}{cumulative distribution function}
\newacronym{camb}{CAMB}{Code for Anisotropies in the Microwave Background}
\newacronym{mcmc}{MCMC}{Markov Chain Monte Carlo}
\newacronym{cosmomc}{CosmoMC}{Cosmological Monte Carlo}
\newacronym{gr}{GR}{General Relativity}
\newacronym{efe}{EFE}{Einstein's Field Equations}
\newacronym{eos}{EoS}{equation of state}
\newacronym{de}{DE}{dark energy}
\newacronym{dm}{DM}{dark matter}
\newacronym{cde}{CDE}{coupled dark energy}
\newacronym{cpde}{CPDE}{coupled phantom-like dark energy}
\newacronym{cqde}{CQDE}{coupled quintessence-like dark energy}
\newacronym{wcqde}{$w$CQDE}{fixed-$w$ coupled quintessence-like dark energy}
\newacronym{dfe}{DfE}{departure from equilibrium}
\newacronym{ovr}{OVR}{observed virial ratio}
\newacronym{tvr}{TVR}{theoretical virial ratio}
\newacronym{eds}{EdS}{Einstein--de Sitter}
\newacronym{srf}{SRF}{scale reduction factor}
\newacronym{psrf}{PSRF}{potential scale reduction factor}
\newacronym{mpsrf}{MPSRF}{multivariate potential scale reduction factor}

%% file: symbols.tex
\DeclareSIUnit\jansky{Jy}
\DeclareSIUnit\parsec{pc}
\newcommand{\kB}{k_{\text{B}}} % Boltzmann constant
\newcommand{\Dirac}{\delta_{\mathrm{3D}}}%{three-dimensional Dirac delta function}

\newcommand{\nfwc}{c} %NFW concentration parameter
\newcommand{\rs}{r_{\mathrm{s}}}%{NFW scale radius}
\newcommand{\rhos}{\rho_{\mathrm{s}}}%NFW density parameter corresponding to \svar{rs}}
\newcommand{\mH}{m_{\mathrm{H}}}%{hidrogen mass}
\newcommand{\icgmu}{\mu} %mean intracluster gas molecular mass
\newcommand{\sigx}{\sigma_{\mathrm{X}}}%{galaxy velocity dispersion}
\newcommand{\sepercent}{\SI{68}{\percent}}
\newcommand{\nfpercent}{\SI{95}{\percent}}
\newcommand{\loc}{x_{\mathrm{loc}}}
\newcommand{\mvir}{M_{\mathrm{vir}}}
\newcommand{\cvir}{\nfwc_{\mathrm{vir}}}
\newcommand{\rvir}{r_{\mathrm{vir}}}
\newcommand{\tx}{T_{\mathrm{X}}}
\newcommand{\kin}{\rho_{\mathrm{kin}}}
\newcommand{\dkin}{\dot\rho_{\mathrm{kin}}}
\newcommand{\pot}{\rho_{\mathrm{pot}}}
\newcommand{\dpot}{\dot\rho_{\mathrm{pot}}}
\newcommand{\ppot}{\rho'_{\mathrm{pot}}}
\newcommand{\LI}{Layzer--Irvine}
\newcommand{\dmderatio}{\rho_{\dm}/\rho_{\de}}
\newcommand{\dm}{\mathrm{DM}}
\newcommand{\de}{\mathrm{DE}}
\newcommand{\indi}{A}
\newcommand{\indI}{\mathrm{I}}
\newcommand{\vac}{\mathrm{vac}}
\newcommand{\matter}{\mathrm{M}}
\newcommand{\rad}{\mathrm{rad}}
\newcommand{\tplus}{\text{+}}
\newcommand{\tminus}{\text{--}}
\newcommand{\critical}{\mathrm{cr}}
\newcommand{\baryons}{\mathrm{b}}
\newcommand{\ud}{\text{d}}
\newcommand{\vecnotation}[1]{\mathbf{#1}}

\newcommand{\vers}[1]{\mathbf{\hat{#1}}} % versor notation
\newcommand{\rmn}[1]{\mathrm{#1}}
\newcommand{\Nabla}{\nabla}

\newcommand{\eff}{\mathrm{eff}}
\newcommand{\gal}{\mathrm{g}}
\newcommand{\obs}{\mathrm{obs}}
\newcommand{\emit}{\mathrm{emit}}
\newcommand{\redspace}{\mathrm{s}}

\newcommand{\anl}{\mathrm{anl}}
\newcommand{\tnum}{\mathrm{num}}
\newcommand{\mcs}[1]{\multicolumn{5}{c}{#1}}
\newcommand{\mcf}[1]{\multicolumn{6}{c}{#1}}